\documentclass[12pt]{article}
\usepackage{dsfont}
\usepackage{multirow}
\usepackage{bm}
\usepackage{natbib}
\usepackage[pdftex]{graphicx}
\DeclareGraphicsExtensions{.png,.pdf,.jpg}
\usepackage[hmargin=1in,vmargin=1in]{geometry}
\usepackage{setspace}
\usepackage{verbatim}
\usepackage{amssymb, amsthm}
\usepackage{amsmath}
\usepackage{booktabs}

\usepackage{hyperref}

\usepackage{appendix}
\usepackage{tcolorbox}
\usepackage{xcolor}
\usepackage{longtable}
\usepackage{afterpage}
\usepackage{pdflscape}
\usepackage{capt-of}
\usepackage{lscape}
\usepackage{subcaption}
\usepackage{mathtools}
\usepackage[maxfloats=25]{morefloats}
\usepackage[export]{adjustbox}
\usepackage{float}

\DeclareMathOperator*{\argmax}{arg\,max}

\title{Discrete Autoregressive Switching Processes with Cumulative Shrinkage Priors for Graphical Modeling of Time Series Data
}
\author{Beniamino Hadj-Amar\thanks{Department of Epidemiology and Biostatistics, University of South Carolina, Columbia, SC},
Aaron M. Bornstein\thanks{Department of Cognitive Sciences, University of California, Irvine, CA},
Michele Guindani\thanks{Department of Biostatistics, UCLA Fielding School of Public Health, Los Angeles,CA}\\
and\\ Marina Vannucci\thanks{Department of Statistics, Rice University, Houston, TX}}

\begin{document}

\maketitle

\begin{abstract}
We propose a flexible Bayesian approach for sparse Gaussian graphical modeling of multivariate time series. We account for temporal correlation in the data by assuming that observations are characterized by an underlying and unobserved hidden discrete autoregressive process. We assume multivariate Gaussian emission distributions and capture spatial dependencies by modeling the state-specific precision matrices via graphical horseshoe priors. We characterize the mixing probabilities of the hidden process via a cumulative shrinkage prior that accommodates zero-inflated parameters for non-active components, and further incorporate a sparsity-inducing Dirichlet prior to estimate the effective number of states from the data.  For posterior inference, we develop a sampling procedure that allows estimation of the number of discrete autoregressive lags and the number of states, and that cleverly avoids having to deal with the changing dimensions of the parameter space. We thoroughly investigate performance of our proposed methodology through several simulation studies. We further illustrate the use of our approach for the estimation of dynamic brain connectivity based on fMRI data collected on a subject performing a task-based experiment on latent learning.\\

\noindent%
\end{abstract}

\section{Introduction}
\label{sec:introduction}
In this paper we consider the problem of estimating sparse Gaussian graphical models based on time series data. Time-changing dependencies and sparse structures are often encountered when investigating multi-dimensional physiological signals \citep{safikhani2022joint}, environmental and sensor data \citep{lam2012factor}, as well as macroeconomic and financial systems \citep{kastner2020sparse}. Among existing approaches, \citet{song2009time} introduced a time-varying dynamic Bayesian network for modeling the fluctuating network structures underlying non-stationary biological time series. \citet{kolar2010estimating} proposed a method for estimating time-varying networks based on temporally smoothed $l_1$-regularized logistic regression. \citet{danaher2014joint} and \citet{qiu2016joint} addressed the challenge of estimating multiple related Gaussian graphical models when observations belong to distinct classes, and \citet{Warnick2018b} and \citet{quinn2018task} employed Hidden Markov Models (HMMs) for the estimation of recurrent brain connectivity networks during a neuroimaging experiment.  Other procedures for modeling the temporal evolution of dynamic networks include change-point detection methods \citep{cribben2013detecting, xu2015dynamic} and time-varying parameter models \citep{lindquist2014evaluating, zhang2021bayesian}. Change-point techniques provide a data-driven approach for the temporal partitioning of the network structure into segments of adaptable length. However, these methods do not provide a system for identifying potentially recurring network patterns over time. Time-varying parametric methods offer a principled way of modeling dynamic correlations but are computationally intensive.

We propose a flexible Bayesian approach for sparse Gaussian graphical modeling of multivariate time series. In order to represent switching dynamics, we assume an unobserved hidden process, underlying the time series data, which at each time point exists in one of a finite number of states. We account for the temporal structure of this hidden process by assuming a Discrete Autoregressive (DAR) process of order $P$ \citep{biswas2009discrete}, which flexibly incorporates long-term dependencies by considering the $P$ previous lags of the process. Given the state of the latent process, we model the observations as conditionally independent of the observations and states at previous times and generated from state-specific multivariate Gaussian emission distributions. Under the multivariate Gaussian assumption, networks can be estimated by the graphical models induced by the state-specific inverse covariance matrices.  We capture these spatial dependencies by modeling the state-specific precision matrices via graphical horseshoe priors.

The DAR hidden process construction we adopt is reminiscent of higher-order HMMs, where the present state depends not only on the immediately preceding state but also on prior states further back in time. First-order HMMs, which constrain the temporal dynamics of the hidden state sequence to be Markovian, have been successfully applied in many scientific fields, including neuroimaging \citep{Warnick2018b,quinn2018task}, climate \citep{holsclaw2017} and animal behavior \citep{deruiter2017multivariate}, to cite a few.  While higher-order HMMs have been suggested \citep{cappe2005inference}, they require the estimation of transition probability matrices that grow exponentially in size as the order increases, making their estimation challenging \citep[see, for a discussion,][]{sarkar2019bayesian}. In our proposed model, the state-switching behavior of the process is captured by the time-varying mixing probabilities of the DAR process. To model these probabilities, we propose a nonparametric zero-inducing cumulative shrinkage prior.
Building upon the construction of the finite Dirichlet process \citep[DP; see][]{IshwaranJames2001}, the proposed prior accommodates zero-inflated parameters, to account for non-active components, and employs cumulative shrinkage \citep{legramanti2020bayesian} to handle increasing model complexity. This construction ensures that if a parameter in the DAR model is zero, then all subsequent lag parameters are also zero. This results in a flexible and computationally efficient framework for learning the time-varying mixing probabilities and the effective order of the process, as opposed to learning the entire transition matrix, as required in HMM modeling. Such reduction in the number of parameters leads to a substantial computational advantage. It also allows to learn the number of lags in a data-driven fashion. Related sparsity-inducing prior constructions have been developed by \citet{heiner2019structured} for the simplex model, and by \citet{tang2019zero} for zero-inflated generalized Dirichlet multinomial regression models. These constructions are specific to those models and less flexible than our approach, which models the ordering of the lags as the process evolves in time while promoting lower-order complexity. We complete our modeling framework with a sparsity-inducing Dirichlet prior that allows estimation of the effective number of hidden states in a data-driven manner. Drawing inspiration from the literature on overfitted finite mixture models  \citep{rousseau2011asymptotic, malsiner2016model}, we consider more states than strictly necessary, while employing a prior that effectively constrains the model’s complexity. This promotes sparsity while leading to more interpretable inferences. 

For posterior inference, we take a fully Bayesian approach and develop a sampling procedure that accommodates the multiple model selection problems, namely the number of DAR lags and the number of states, while cleverly avoiding having to deal with the changing dimensions of the parameter space. Specifically, we implement a Gibbs sampler that alternates between updating the DAR parameters, the sparse emission parameters, and the latent state sequence. To update the DAR probabilities, we leverage the stick-breaking construction of the DP by augmenting the space with auxiliary indicator variables and design a joint sampling scheme that alternates between adding or removing the sticks of the zero-inducing DP formulation. Our zero-inducing cumulative shrinkage prior significantly accelerates the proposed sampler, particularly in regard to the forward-backward algorithm for updating the latent state sequence.  Estimates of the number of hidden states and DAR order are determined based on the most frequently occurring number of active states and DAR order observed during MCMC sampling, respectively.

We thoroughly investigate performance of our proposed methodology through several simulation studies. We further illustrate the use of our proposed approach to estimate dynamic brain connectivity networks based on functional Magnetic Resonance Imaging (fMRI) data. Identifying the dynamic nature of brain connectivity is critical for understanding our current knowledge about human brain functioning.  In our application, we consider data collected on a subject performing an experiment aimed at understanding neural representations that are formed during latent learning. Inferred networks by our method identify distinct regimes of functional connectivity, that can be mapped onto cognitive interpretation. 

The rest of the paper is organized as follows. Section \ref{sec:the_model} introduces the proposed model, including the DAR process and the proposed prior structures, and the MCMC algorithm for posterior inference.  Section \ref{sec:simulation_studies} contains results from the simulation studies and Section \ref{sec:application} illustrates the application to fMRI data on latent learning. Section \ref{sec:conclusion} provides concluding remarks. The Julia software \texttt{sggmDAR}, which implements our proposed methodology is available on GitHub at \href{https://github.com/Beniamino92/sggmDAR}{https://github.com/Beniamino92/sggmDAR}.

\section{Sparse Modeling of Multivariate Time Series Data via Cumulative Shrinkage DAR} 
\label{sec:the_model}
In this Section, we describe the proposed latent variable approach for modeling sparse multi-dimensional time series. Let $\bm{y} = \big\{ \bm{y_t} \big\}_{\, t= 1}^{T}$,  $\bm{y}_t = (y_{t1}, \dots, y_{tD}) \in \mathbb{R}^{D}$, be the observed $D$-dimensional time series data, with $T$ indicating the number of time points. We envision an unobserved, latent hidden process underlying the observations and assume that, at each time point, the process assumes one of a finite number of states, represented as $\bm{\gamma} = \big\{ \gamma_t \big\}_{\, t = 1}^{T}$, with $\gamma_t \in \{1, \dots, M\}$ and $M$ denoting the (unknown) finite number of latent states. Given the value of $\gamma_t$, the observations $\bm{y}_t$ are assumed to be independent of both the observations and states at previous time points. We further assume that the state-specific emissions follow a $D$-variate Gaussian distribution
\begin{equation}
	\bm{y}_t \, |\,   \gamma_t, \, \bm{\mu}, \bm{\Omega} \sim   \sum_{j=1}^{M}\mathds{1}_{\{j\}}(\gamma_t) \, \mathcal{N}_D\, (\bm{y}_t \, | \, \bm{\mu}_j, \bm{\Omega}^{-1}_j),
	\label{eq:emissions}
\end{equation}
with state-specific means $\bm{\mu}_j$  and precision matrices $\bm{\Omega}_j$, $j=1, \ldots, M$, $t=1, \dots, T$. Here, $\mathds{1}_{\{j\}}(\gamma_t)$ denotes the indicator function, which equals 1 if $\gamma_t = j$, and 0 otherwise. Conditional dependencies can be inferred from the off-diagonal entries of the precision matrices. Specifically, for a given state $j$, if the entry $\omega_{j,il}$ is zero, the corresponding variables $y_{ti}$ and $y_{tl}$ are conditionally independent given the other variables. 


\subsection{State Dynamics via Discrete Autoregressive Processes}
\label{sec:DAR}
In order to learn the dependence structure between time points, represented by the sequence $\bm\gamma$, we design an approach that employs a discrete autoregressive process, with a cumulative shrinkage prior that enables a computationally efficient estimation of the order of the process.
More specifically, we assume that the evolution of the hidden state sequence $\gamma_t$ follows a Discrete Autoregressive (DAR) process of order $P$ \citep{biswas2009discrete}, a framework originally introduced in the context of multi-lag finite-state Markov chains by \citet{pegram1980autoregressive}, and subsequently adapted to categorical time series modeling by \citet{biswas2009discrete}. The DAR process allows the hidden sequence to incorporate long-term dependencies by considering the previous $P$ lags. Formally, the conditional distribution of $\gamma_t$ given the past values $\gamma_{t-1:t-P}$ is expressed as 
\begin{equation}
	p(\gamma_t | \gamma_{t-1:t-P}, \bm{\phi}, \bm{\pi}) = \phi_1 \mathds{1}_{\{\gamma_{t-1}\}}(\gamma_t) + \phi_2 \mathds{1}_{\{\gamma_{t-2}\}}(\gamma_t) + \ldots + \phi_P \mathds{1}_{\{\gamma_{t-P}\}}(\gamma_t) + \phi_0 \,  \pi_{\gamma_t},
	\label{eq:DAR}
\end{equation} where $\bm{\phi} = (\phi_0, \dots, \phi_P)$ and $\bm{\pi} = (\pi_1, \dots, \pi_M)$.  We denote with  $\{\phi_j\}_{j=0}^{P}$ the \emph{autoregressive probabilities}, with $\phi_0 = 1 - \sum_{j=1}^{P} \phi_j$, while the \emph{state innovation probabilities}  $\{ \pi_i \}_{i=1}^{M}$ are defined as  $\pi_i:=p(\gamma_t = i)$,  for $ i = 1, \dots, M$, and allow the process to transition to any of the $M$ states, including those not observed in the previous $P$ lags. Here, $\mathds{1}_{\{j\}}(i)$ is an indicator function equal to 1 if $i=j$ and 0 otherwise. According to Equation \eqref{eq:DAR}, at each time point \( t \), the model selects a latent state \( \gamma_t \in \{1, \dots, M\} \), and assigns probabilities based on how this state relates to the past \( P \) values of the sequence. Specifically, the probability of choosing a particular state \( i \in \{1, \dots, M\} \) is given by the sum of two components: an autoregressive component and an innovation component. The autoregressive probabilities \( \{\phi_j\}_{j=1}^{P} \) govern the recurrence structure, so that if \( \gamma_t = \gamma_{t-j} \) for any \( j \in \{1, \dots, P\} \), then state \( i \) receives mass \( \phi_j \). That is, the model assigns probability \( \phi_j \) to each state \( i \) that matches \( \gamma_{t-j} \). In this way, the latent sequence favors persistence or recurrence of recently visited states. Complementing this, the innovation probability \( \phi_0  \) allows for exploration beyond the observed history. With probability \( \phi_0 \), the model selects a state independently of \( \gamma_{t-1}, \dots, \gamma_{t-P} \), according to the innovation distribution \( \bm{\pi} \), which assigns nonzero probability to each of the \( M \) possible states. 

\vspace{1cm}

The transition probabilities in the DAR process can be represented by a multi-dimensional array. The dimensions of this array are determined by the number of autoregressive lags, $P$, and the number of hidden states, $M$. As an illustration, when $P = 2$, the transition probabilities are described by an $[M \times M \times M]$ array, say $\bm{\eta}$. The individual components of this array, denoted as $\eta_{\, {l,i,j}}$, represent the probability $p(\gamma_t = j | \gamma_{t-1} = i, \gamma_{t-2} = l)$ for $l, i, j \in \{1, \dots, M\}$, as defined in  \eqref{eq:DAR}. However, as the number of lags $P$ increases, the dimensionality of this array grows exponentially. Therefore, the DAR characterization simplifies inference by allowing us to focus only on making inferences on the $\phi$ parameters, as opposed to learning the entire transition matrix, which is the case with HMM models, for example. In fact, when dealing with higher order HMMs, the task involves estimating $M^{P}$ parameters for the transition arrays. In contrast, our proposed method streamlines this process by estimating $(M+P)$ parameters, resulting in a substantial computational advantage.

\subsubsection{Zero-inducing cumulative shrinkage prior for learning time dependence}
\label{sec:zero_inflated}
The time-varying mixing probabilities of the DAR model, denoted as $\phi_j$, characterize the state-switching behavior of the process. To model these probabilities, we propose a nonparametric zero-inducing cumulative shrinkage prior that accommodates zero-inflated parameters to account for non-active components, and that employs cumulative shrinkage \citep{legramanti2020bayesian} to handle increasing model complexity. This prior modifies the stick-breaking construction to allow for an increasing probability of setting $\phi_j=0$ as $j$ increases. In addition,  our formulation enforces that once $\phi_j$ becomes zero for a specific $j$, $j = 1, 2, \dots, P$, subsequent lags obey the condition $p(\phi_{j+k} = 0 \, | \, \phi_{j} = 0) = 1$, $k=1, \ldots, P-j$. To formally introduce our prior, we need to define a binary latent process, namely an ``active order" latent indicator, denoted as $z_{j}\in \{0,1\}$,  $j=1, \ldots, P$. If $z_j=0$, then $\phi_j$ is almost surely non-zero. However, when the first $j$ such that $z_{j}=1$ occurs, then $\phi_j=1-\sum_{l=1}^{j-1}\phi_l$ and $z_l=1$ almost surely for  $l=j+1, \ldots, P$.  More formally, the mixing probabilities $\phi_j$ are generated via a modified stick-breaking construction, 
\begin{equation}
	\phi_j = v_j \, \prod_{l=0}^{j-1}(1 - v_l), \quad \text{for} \, \, j = 1, \dots, {P},
	\label{eq:stick_break_1}
\end{equation}
with $\phi_0 = v_0$, where the stick-breaking weights $v_j$ are mixtures of a Beta distribution and a spike at one, 
\begin{equation}
v_{j} \mid z_{j} \sim\left(1-z_{j}\right) \, \operatorname{Beta}\left(a_{v}, b_{v}\right)+z_{j} \, \delta_{1},
\label{eq:stick_break_2}
\end{equation}
with $\delta_x$ denoting a point mass at $\{x\}$, $j=1, \ldots, P$, and by specifying $v_0 \sim \operatorname{Beta}(a_0, b_0)$.

For $z_j=0$,  \eqref{eq:stick_break_1}--\eqref{eq:stick_break_2} define the stick-breaking construction typical of the Dirichlet process. If at some point $z_j=1$ occurs, then $v_j=1$, and $\phi_j=\prod_{l=0}^{j-1}(1 - v_l)=1-\sum_{l=1}^{j-1} \, \phi_l$. For all remaining lags, our construction ensures  $\phi_l=0$, $l=j+1, \ldots, P$. More specifically, to enforce the desired behavior and promote lower order model complexity, we leverage the increasing shrinkage prior construction of  \citet{legramanti2020bayesian} and assign increasing probability mass to selecting the spike component as the order of the DAR grows.  In particular, we assume $z_{j} \, |\,  \bm{v}_{0:\,j-1} \,  \sim \text{Bern} \, (\xi_j)$ with probability $\xi_j =  \sum_{i=0}^{j-1}\phi_i$ increasing with the lag $j$,  where $z_1 | v_0 \sim \operatorname{Bern}(v_0)$. See also \citet{zhang2021bayesian}, where an increasing shrinkage prior is used in a VAR model. Our construction ensures that $p \, (z_{l} = 1 | z_{l-1} = 1) = 1$ and $p \, (\phi_{l} = 0 \, | \,  \phi_{l-1} = 0) = 1$.  We define the effective order of the DAR process as the random element $\hat{P}=\inf_{j\in \{1, \ldots, P\}}\{z_j=1\}$, that is the number of ``active'' lags of the DAR process. The proposition below demonstrates the aforementioned property. 

\textbf{Proposition 1.}  Let $\bm{\phi} = \{ \phi_j \in \Delta_\phi : j = 0, \dots, P \}$ with $\Delta_{\phi} = \{ \phi_l : 0 \leq \phi_l \leq 1, \sum_{l=0}^\infty \phi_l = 1\}$, be constructed according to \eqref{eq:stick_break_1}, and $\bm{v}= \{v_i \}_{i=0}^P$ and $\bm{z} = \{z_i \}_{i=1}^P$  be specified  as in \eqref{eq:stick_break_2}. Under these assumptions, the cumulative shrinkage DAR formulation implies that $p(z_{j+1} = 1 | z_j = 1) = 1$, for $j=1, \dots, P$.

\textit{Proof.}  Recall that $z_{j} \, |\,  \bm{v}_{0:\,j-1} \,  \sim \text{Bern} \, (\xi_j)$ with probability $\xi_j =  \sum_{i=0}^{j-1}\phi_i$, $j=1, \ldots P$. Therefore, for $j=1, \ldots, \hat{P}$, we can write
\begin{equation*}
\begin{split}
    p (z_{j+1} = 1 | z_j = 0, \bm{v}_{0:j}) = \sum_{i=0}^{j}\phi_i 
     =  \, v_0 +  \, v_1(1-v_0) + \dots + v_j \prod_{l=0}^{j-1}(1-v_l).
\end{split}
\end{equation*}
Thus,   
$
    p (z_{j+1} = 0 | z_{j} = 0, \bm{v}_{0:j}) = 1 - p (z_{j+1} = 1 | z_{j} = 0, \bm{v}_{0:j}) 
    = \prod_{l=0}^{j}(1-v_l).
$
For $j=\hat{P}$, since $v_{\hat{P}}=1$ a.s., we have  $\sum_{j=0}^{\hat{P}} \phi_j =\, v_0 +  \, v_1(1-v_0) + \dots + v_{\hat{P}} \prod_{l=0}^{\hat{P}-1}(1-v_l)=1$. Thus,  for $j=\hat{P}+1, \ldots, P-1$,  $p(z_{j+1} = 1 | v_{0:\hat{P}})= p(z_{j+1} | z_{j}=1) = 1$. \qedsymbol{}

Given the one-to-one relationship between the sequence $z_j$ and $\hat{P}$, the process can be alternatively defined in terms of the random quantity $\hat{P}$, which is computationally convenient, as we explain in Section \ref{sec:MCMC} below. We note that the previous characterization can also be extended to the case of $P=\infty$. However, for computational purposes, it is convenient to consider only a finite number of terms, say, $P_{max}$, and thus specify the autoregressive coefficients as $\bm{\phi} = (\phi_0, \dots, \phi_{P_{max}})$, where $\phi_{P_{max}} = 1 - \sum_{l=0}^{P_{max}-1} \phi_j$.  In implementations, this approach offers considerable versatility when $P_{max}$ is set to a moderately high upper bound, and it  is advisable to choose $P_{max}$ such that it exceeds the expected number of lags. 


\subsubsection{Sparsity-inducing Dirichlet prior to infer state transitions and space size}
\label{sec:sparse_finite_mixture}

As for the innovation probabilities $\bm\pi$, to facilitate a substantial reduction in the effective number of states compared to the maximum number, \(M=M_{\text{max}}\), we draw insights from recent literature on overfitted finite mixture models \citep{rousseau2011asymptotic, malsiner2016model}. Specifically, we assume a symmetric Dirichlet prior 
$\bm{\pi}=(\pi_1, \ldots, \pi_M) \sim \text{Dir}\,(\kappa_0, \dots, \kappa_0)$, 
where the concentration parameter \(\kappa_0\) is set at a very small value, so that the marginal densities of each $\pi_j$ are spiked around the values zero and one,  $j=1, \ldots, M$.  This approach results in estimating a reduced number of hidden states, denoted as \(\hat{M}\), which is significantly less than \(M\). Thus, unnecessary hidden states are effectively removed from the posterior distribution. The hyperparameter \(\kappa_0\) plays a crucial role. Here, we set \(\kappa_0 = 0.001\) following the recommendation by  \citet{malsiner2016model}. In Section \ref{sec:inference}, we propose to estimate the number of hidden states based on the most frequent number of active states during MCMC sampling. By setting a large value for $M$, our approach provides a simple and automated framework for estimating the number of hidden states, without relying on computations of marginal likelihoods, post-MCMC model selection criteria, or reversible-jump MCMC.

\subsection{Graphical Horseshoe Priors for the Precision Matrices}
\label{sec:GHS}
To induce prior sparsity in the state-specific precision matrices \(\bm{\Omega}_j\)'s, we employ the graphical horseshoe (GHS) prior proposed by \citet{li2019graphical}. This prior utilizes normal scale mixtures with half-Cauchy hyperpriors for the off-diagonal entries of the precision matrix while using uninformative priors for its diagonal elements. Specifically, 
\begin{align*}
    \begin{split}
        \omega_{j,ii} & \propto 1, \\ 
        \omega_{j,il \, : i<l} &\sim \mathcal{N}(0, 
        \lambda^{2}_{j,il} \tau^2_j), \\
        \lambda_{j,il \, :i<l} &\sim C^{+}(0, 1), \\
        \tau_j &\sim C^{+}(0, 1), 
    \end{split}
    \label{eq:GHS}
\end{align*}
for $i, l = 1, \dots, D$, and $j = 1, \dots, M$. The global shrinkage parameter \(\tau_j\) plays a crucial role in promoting sparsity across the entire matrix \(\bm{\Omega}_j\), by shrinking the estimates of all the off-diagonal values towards zero. On the other hand, the local shrinkage parameters \(\lambda_{jil:i<l}\) allow to preserve the magnitudes of the nonzero off-diagonal elements, ensuring that the element-wise biases do not become too large. This combination of global and local shrinkage enables the GHS prior to induce sparsity in the precision matrices while capturing the relevant dependencies between the elements.

We complete the prior specification on the emission distributions by assuming Gaussian priors on the state-specific means, that is,
$
p(\bm{\mu}_j) \sim \mathcal{N}(\bm{\mu}_0, \bm{R}_0^{\,-1}),
$
for $j=1, \dots, M$.

\subsection{Markov Chain Monte Carlo Algorithm} 
\label{sec:MCMC}

We now outline the MCMC algorithm we designed for posterior inference. For notational convenience, we collect all parameters except $\bm{\gamma}$ as the set $\bm{\theta} = \{ \bm{v}, \bm{z},  \bm{\pi}, \bm{\mu}, \bm{\Omega}, \bm{\tau}, \bm{\Lambda} \}$ with $\bm{\Lambda} =  \{ \bm{\Lambda}_j \}_{j=1}^M $ and $\bm{\Lambda}_j = \{\lambda^2_{j,il}\}$ the matrices of local shrinkage parameters in the GHS prior, $\bm{\tau} = (\tau_1, \dots, \tau_M)$ the global parameters, $\bm{\mu}=(\bm{\mu}_1, \ldots, \bm{\mu}_M)$, and $\bm{\Omega}=(\bm{\Omega}_1,\ldots, \bm{\Omega}_M)$.  We then write the posterior distribution of $\bm\theta$ conditional upon the current value of $\bm\gamma$ as
\begin{equation}
    p (\bm{\theta} \, | \, \bm{y}, \bm{\gamma}) \propto \mathcal{L}(\bm{\theta} ; \, \bm{y}, \bm{\gamma}) \, p (\bm{v}, \bm{z}) \, p(\bm{\pi}) \, p (\bm{\mu})   
    \, p  (\bm{\Omega}, \bm{\tau}, \bm{\Lambda}), 
    \label{eq:posterior}
\end{equation} 
where the conditional likelihood is factorized as 
\begin{equation}
    \mathcal{L}(\bm{\theta} ; \, \bm{y}, \bm{\gamma}) = \prod_{t=P+1}^{T}  p(\gamma_t 
    \, |
    \, \gamma_{t-1:t-P}, \bm{v}, \bm{z}, \bm{\pi}) \, p (\bm{y}_t \, | 
    \,  \gamma_t, \bm{\mu}, \bm{\Omega})
    \label{eq:likelihood}
\end{equation}
and where the joint prior $p (\bm{v}, \bm{z})$ of the indicator variables and the stick-breaking weights can be expressed as
\begin{equation}
    p(\bm{v}, \bm{z}) = p(v_0) \prod_{j = 1}^{\hat{P}-1} p(v_j|z_j) \prod_{j=0}^{\hat{P}} p (z_{j+1} | \bm{v}_{0:j}),
\end{equation}
with 
\begin{equation}
p ( v_j | z_j) \propto \text{Beta}(a_v, b_v)^{1 - z_j}, \qquad p (z_{j+1} | \bm{v}_{0:j}) \propto  \text{Bern}(1)^{z_j} \,  \text{Bern}(\xi_j)^{1-z_j},
\end{equation}
and the conditioning on $\bm{v}_{0:{j-1}}$ induced by the cumulative shrinkage parameter $\xi_j$. 

Since the posterior distribution is not available in closed form, we develop a Gibbs sampler that alternates between: (i) drawing the stick-breaking weights $\bm{v}$ and auxiliary indicators $\bm{z}$. For this, we design a Metropolis-Hastings algorithm similar to \citet{savitsky2011variable}, that cleverly avoids having to deal with the changing dimensions of the parameter space
via a joint update of the indicators and the weights; 
(ii) updating the innovation probabilities $\bm{\pi}$ related to the sparsity-inducing Dirichlet prior; (iii) sampling the multivariate sparse emission parameters,  i.e. the mean vectors in $\bm{\mu}$, the precision matrices in $\bm{\Omega}$ and the global and local shrinkage parameters $\bm{\tau}$ and $\bm{\Lambda}$; (iv) updating the latent state sequence $\bm{\gamma}$, through a forward-backward algorithm, which is significantly accelerated by the proposed zero-inducing cumulative shrinkage prior formulation. We now describe these updates in full detail.

\begin{itemize} 
\item {\textbf{Update} $\bm{z}$ and $\bm{v}$}: 
We perform a joint update of the indicators $\bm{z}$ and weights $\bm{v}$ by designing a Metropolis-Hastings sampler with \textit{birth} and \textit{death} moves, that increase or decrease the order of the DAR process by one. 
Formally,  let us define the current number of active components $\hat{P}^{\, curr}$, stick-breaking weights $\bm{v}^{\, curr}=(v_0, v_1, \dots, v_{\hat{P}^{\,curr}-1}, 1)$, and indicator variables  $\bm{z}^{\,curr} = (0, 0, \dots, 0, 1)$, of dimensions $\hat{P}^{\,curr} + 1$ and $\hat{P}^{\,curr}$, respectively; note that $\bm{z}^{\,curr} = 1$ when $\hat{P}^{curr} = 1$. A new vector of  indicators $\bm{z}$ is drawn by proposing at random one of the following two moves:

\begin{enumerate}
    \item[(i)] \textit{birth move:} Set $\hat{P}^{\,prop} =  \hat{P}^{\,curr} + 1$ and construct $\bm{z}^{\,prop}$ from  $\bm{z}^{\, curr}$ by adding a zero entry; for this move, the proposed vector of weights is constructed as $\bm{v}^{\,prop} = (v_0, v_1, \dots, v_{\hat{P}^{\,curr - 1}},v_{\hat{P}^{\,prop-1}}, 1)$ with $v_{\hat{P}^{\,prop}-1}$ drawn from the prior, i.e. $v_{\hat{P}^{\,prop}-1} \sim \text{Beta}(a_v, b_v)$, and $v_{\hat{P}^{\,prop}}$ set equal to one. This move is accepted or rejected with probability 

    \begin{equation}
        \alpha = \text{min} \bigg\{1, \dfrac{p ( \bm{v}^{prop}, \bm{z}^{prop} \,  | \bm{\gamma}, \bm{y}, \cdot )}{p ( \bm{v}^{curr}, \bm{z}^{curr} \,  | \bm{\gamma}, \bm{y}, \cdot )} \dfrac{1}{\text{Beta}(v_{\hat{P}^{\,prop}-1} | a_v, b_v)} \bigg\},
        \label{eq:MH_acceptance}
    \end{equation}
    where the joint posterior distribution $p (\bm{z}, \bm{v} | \cdot)$ is easily available by appropriate conditioning of the relevant variables in Eq. \eqref{eq:posterior} and \eqref{eq:likelihood}, i.e. 
\begin{equation}
p ( \bm{v}, \bm{z} \,  | \bm{\gamma}, \bm{y}, \cdot )  \propto  p (\bm{v}, \bm{z})  \prod_{t=P+1}^{T}  p(\gamma_t 
    \, |
    \, \gamma_{t-1:t-P}, \bm{v}, \bm{z}, \bm{\pi}),
    \label{eq:post_z_v}
\end{equation}
with the DAR probabilities $p(\gamma_t | \cdot)$ defined as in Eq. \eqref{eq:DAR}, recalling that $\bm{\phi}$ is a by-product of $\bm{v}$ and $\bm{z}$ using the formulation presented in Eq. \eqref{eq:stick_break_2}.  

    \item[(ii)] \textit{death move:} Set $\hat{P}^{\, prop} = \hat{P}^{\, curr} - 1$ and construct $\bm{z}^{\,prop}$ from  $\bm{z}^{\, curr}$ by removing a zero entry; here, $\bm{v}^{\, prop}$ is obtained from $\bm{v}^{\,curr}$ by replacing the component $v_{\hat{P}^{\,curr}-1}$ with a one and setting $v_{\hat{P}^{\,curr}}$ equal to zero, namely $\bm{v}^{\,prop} = (v_0, v_1, \dots, v_{\hat{P}^{curr}-2}, 1)$. This is move is accepted or rejected with probability the inverse of Eq \eqref{eq:MH_acceptance} with the appropriate change of labeling. 
    
\end{enumerate}


After each death/birth move, to enhance the mixing efficiency of the MCMC algorithm, we further update each component of the weight vector $\bm{v}$ using a one-at-a-time slice sampler \citep{neal2003slice}. Slice sampling is particularly advantageous for drawing samples from one-dimensional conditional distributions within a Gibbs sampling framework. Here, we focus on multivariate targets by iteratively sampling each variable. In particular, we obtain posterior samples from the target function $p(v_j \, | \, \bm{v}_{-j}, \cdot)$, for $j = 0, \dots, \hat{P}-1$, where $\bm{v}_{-j} = (v_0, \dots, v_{j-1}, v_{j+1}, \dots, v_{\hat{P}-1})$.

We remark here that the order $P$ of the DAR process is not modeled as a random variable, but rather inferred directly from $\bm{z}$ and $\bm{v}$, eliminating the necessity of employing a trans-dimensional MCMC sampler \citep{green1995}.  



\item {\textbf{Update $\bm{\pi}$:}} We update the components of \(\bm{\pi}\) with a one-at-a-time slice sampler, drawing samples from the target function $p(\pi_l \, | \, \bm{\pi}_{-l}, \cdot)$, for $j = 0, \dots, M_{max}-1$, where $\bm{\pi}_{-l} = (\pi_0, \dots, \pi_{l-1}, \pi_{l+1}, \dots, \pi_{M_{max}}$). Note that $\pi_{M_{max}}$ is automatically obtained from its simplex, i.e.  $\pi_{M_{max}} = 1 - \sum_{l=0}^{M_{max}-1} \pi_l$.


\item{\textbf{Update} $\bm{\Omega}_j$, $\bm{\Lambda}$, and $\bm{\tau}$:} 
We use the augmented block Gibbs sampler method proposed by \citet{li2019graphical}. We center the observations belonging to each state to its current value of the emission mean, \(\bm{\mu}_j\), and consider a modified set of observations denoted as \(\tilde{\bm{Y}}_j = \{\bm{y}_t - \bm{\mu}_j : \gamma_t = j\}\). By doing this, we can closely follow the scheme proposed by \citet{li2019graphical}, which assumes zero-mean multivariate normal distributions. We apply the Gibbs sampler independently for each state \(j\), from\(j=1\) to \(M_{\text{max}}\)
and subsequently update the global shrinkage parameter $\tau_j$ and its corresponding augmented parameter $\xi_j$. 
We refer to the reader to Algorithm 1 of \citet{li2019graphical}, for the details of the GHS sampler.


\item{\textbf{Update} $\bm{\mu}_j$:} We sample the mean vectors \(\bm{\mu}_j\) from the corresponding full conditional, as is typical in the context of Gaussian Bayesian regression settings  (see e.g. \citealt{gelman1995bayesian}). The posterior  distribution is given by  $\bm{\mu}_j|_{ \bm{\Omega}_{j}, \bm{y}, \cdot} \sim \mathcal{N}(\bm{\mu}_{\,j}^{\star}, \bm{\Omega}_{\,j}^{\star})$, where
\begin{equation}
    \bm{\Omega}^{\star \, -1}_{ \,j} = \bm{R_0} + N_j \bm{\Omega}_j, \quad \text{and} \quad \bm{\mu}_{\,j}^{\star} = \bm{\Omega}_{ \,j}^{\star}(\bm{R_0} \bm{\mu}_0 + N_j\bm{\Omega}_j\bm{Y}_j),
\label{eq:prova}
\end{equation}
and $\bm{Y}_j$ denotes the ($N_j \times D)$-dimensional matrix of observations assigned to state $j$, with $N_j$ the corresponding number of observations belonging to that regime.

\item{\textbf{Update} $\bm{\gamma}$:}  We update the sequence of latent states \(\bm{\gamma}\) with a block-wise approach that adapts the forward-backward procedure employed by \citet{fox2011sticky} and \citet{hadj2021identifying} to take into account temporal dynamics that extend beyond a simple Markovian structure.  Conditional upon $\bm{\phi}$, $\bm{\pi}$, $\bm{\mu}$ and $\bm{\Omega}$, we harness the dependence structure of the DAR and develop an iterative sampling scheme based on the following representation of the posterior distribution of the hidden states
\begin{eqnarray}
    p ( \bm{\gamma} \, | 
    \bm{y}, \, \cdot) = p ( \gamma_1 |  \bm{y}, \cdot) \, p ( \gamma_2 |  \gamma_1,   \bm{y},  \cdot) \dots p(\gamma_{\hat{P}} | \gamma_{1:\hat{P}-1},   \bm{y}, \cdot) \,  \prod_{t = \hat{P}+1}^{T} p(\gamma_{t} | \gamma_{t-1:t-\hat{P}},   \bm{y}, \cdot).
    \label{eq:sampling_states}
\end{eqnarray}
Under this factorization, we first sample $\gamma_1 \sim p ( \gamma_1 |  \bm{y}, \cdot)$, then, conditioning on the value of $\gamma_1$, we draw $\gamma_2 \sim p (\gamma_2 | \gamma_1,  \bm{y}, \cdot)$, and so on,  where we update $\gamma_t \sim p(\gamma_t | \gamma_{t-1:t-\hat{P}},  \bm{y}, \cdot)$, given the previous sampled states $\gamma_{t-1:t-\hat{P}}$. 
Assuming $M = M_{\text{max}}$, the general form for the conditional posterior distribution of the states in Eq. \eqref{eq:sampling_states} is given by 
\begin{equation}
    p(\gamma_t = j_0 | \gamma_{t-1}=j_1, \dots, \gamma_{t-\hat{P}} = j_{\hat{P}},  \bm{y}, \cdot ) \propto \eta_{\{\,j_{\hat{P}}, \dots, j_1, j_0\}} 
    \, p (\bm{y}_t | \gamma_t=j_0, \bm{\mu}, \bm{\Omega}) \, \beta_{t+1}(j_0),
    \label{eq:posterio_states}
\end{equation}
for $ t = \hat{P}+1, \dots, T$,
and $j_l \in \{1, \dots, M \}$, $l = 1, \dots {\hat{P}}$,
where $\eta_{\{\,j_{\hat{P}}, \dots, j_1, j_0\}}$ are the DAR probabilities of selecting state $j_0$, given previous values $j_1, \dots, j_{\hat{P}}$, as defined in Eq. \eqref{eq:DAR}, and $p (\bm{y}_t \, |\cdot )$ are the multivariate spiked Gaussian  emission densities specified in Eq. \eqref{eq:emissions}.   Here, we define the \textit{backward messages} $\beta_t(j_1) = p(\bm{y}_{t:T}| \gamma_{t-1} = j_1, \cdot)$, as the probability of the partial observation sequence from time $t$ to $T$ given the state $j_1$ at time $t-1$, conditioned on all the other parameters. These messages can be recursively expressed as follows 
(see Proposition 2, Supplementary Material)
\begin{equation}
    \beta_t(j_1) = \underbrace{\sum_{j_{\hat{P}}=1}^{M} \dots \sum_{j_{2}=1}^{M} \sum_{j_{0}=1}^{M}}_\textrm{$\hat{P}$ times} \eta_{\{\,j_{\hat{P}}, \dots, j_2, j_1, j_0\}}  p (\bm{y}_t | \gamma_t = j_0, \bm{\mu}, \bm{\Omega}) \beta_{t+1}(j_0), \quad t \leq T, 
    \label{eq:backward_msg}
\end{equation}
with $\beta_{T+1}(\cdot) = 1$. Our zero-inducing formulation for the DAR probabilities, described in Section \ref{sec:zero_inflated}, allows a significant speed-up of the proposed sampler, since in Eq. \eqref{eq:backward_msg} we restrict summations to the active DAR terms only, rather than using the entire multi-dimensional array $\bm{\eta}$. Additionally, we specify the initial DAR probabilities $\eta_{\,\{\cdot\}}$ in Eq. \eqref{eq:sampling_states} and \eqref{eq:posterio_states} to be uniformly distributed.
\end{itemize}

Following similar practices as in \citet{fox2011sticky}  and \citet{hadj2021identifying}, we update only emission parameters for those states that have at least 1\% of the assignments, while for those states that do not satisfy this condition we draw the corresponding  emission parameters from their priors. For the GHS prior, we draw the diagonal entries of the
precision matrix using a diffuse prior $\omega_{ii} \sim U(0, 100)$. 

We acknowledge that the proposed Bayesian procedure may be susceptible to the  \textit{label switching} problem  \citep{jasra2005markov} 
due to the invariance of the likelihood \eqref{eq:likelihood} under permutations of the mixture components' labeling.
To mitigate this issue, we adopt a post-processing approach using the Equivalence Classes Representatives (ECR) algorithm, initially introduced by \citet{papastamoulis2013convergence} and later improved by \citet{rodriguez2014label}. The core idea of the ECR algorithm is to categorize analogous allocation vectors as mutually exclusive solutions to the label switching problem. In this context, two allocation vectors are considered analogous if one can be obtained from the other merely by permuting its labels. The ECR procedure divides the allocation vectors into analogous categories and identifies a representative from each category. Consequently, during post-processing, the ECR algorithm identifies the permutation corresponding to each MCMC iteration. This permutation is then applied to reorder the matching allocation with the aim of aligning it with the representative of its category. 



\subsection{Posterior Inference} 
\label{sec:inference}
After obtaining the (possibly relabeled) MCMC output, we first estimate the number of active DAR components by computing the posterior probabilities $p\,(\hat{P} = p \, | \cdot)$,  $p = 1, \dots, P_{max}$ and then identify the posterior mode as the value of \(\hat{P}\) that maximizes such posterior probabilities. Similarly, to estimate the number of hidden states, we first calculate the posterior probabilities \(p\,(M = m \, | \cdot)\) for \(m = 1, \dots, M_{\text{max}}\) as
\begin{equation}
    P(M = m | \cdot) = \frac{1}{S}  \sum_{s=1}^{S}  \mathds{1} (\hat{M}^{(s)}=m),  \qquad \text{where} \quad \hat{M}^{\,(s)} = \sum_{j=1}^{M_{max}} \mathds{1} \big( N_j^{(s)} \neq 0 \big),
\end{equation}
with  \(N_j\) the number of observations assigned to state \(j\), and where the superscript \((s)\) indicates the MCMC iteration for \(s = 1, \dots, S\). We then calculate the posterior mode to obtain the final estimate of the number of hidden states, \(\hat{M}\). 
Next, conditional upon these estimates, we perform posterior inference on the model parameters  \(\hat{\bm{\phi}}\), \(\hat{\bm{\pi}}\), \(\hat{\bm{\mu}}\), and \(\hat{\bm{\Omega}}\) by averaging their sampled values across the MCMC iterations with number of hidden states $\hat{M}$ and DAR order $\hat{P}$.


As for inference on the sequence of latent states, we perform both global and local decoding.  \textit{Global decoding} refers to the determination of the most likely sequence of the entire vector of latent states \(\bm{\hat{\gamma}} = (\hat{\gamma}_1, \dots, \hat{\gamma}_T)\). We obtain such a maximum a posteriori (MAP) estimate by using a variant of the scheme described in equation \eqref{eq:sampling_states}. Given the estimated parameters \(\hat{\bm{\phi}}\), \(\hat{\bm{\pi}}\), \(\hat{\bm{\mu}}\), and \(\hat{\bm{\Omega}}\), we iteratively maximize the posterior distribution of the states, where at each time step \(t\), we compute \(\hat{\gamma}_t = \argmax_{j = 1, \dots, \hat{M}} p(\gamma_{t}=j | \hat{\gamma}_{t-1:t-\hat{P}},  \bm{y}, \cdot)\).
In contrast, \textit{local decoding} of the hidden state at time \(t\), \(p( \gamma_t = j\, | \bm{y}, \, \cdot)\) refers to the determination of that state which is most likely at that time. This is achieved using 
\begin{equation}
\begin{split}
    p( \gamma_t = j\, | \bm{y}, \, \cdot) \propto p(\gamma_t = j, \bm{y}_{1:t} | \, \cdot) p( \bm{y}_{t+1:T} | \, \gamma_t = j, \cdot) = \alpha_{t+1}(j) \beta_{t+1}(j),
    \end{split}
\end{equation}
where the backward messages are defined as $\beta_t(j) = p(\bm{y}_{t:T}| \gamma_{t-1} = j, \cdot)$ and the forward messages are expressed as  $\alpha_t(j) = p\,(\bm{y}_{1:t-1}, \gamma_{t-1} = j \,|\, \cdot)$. In order to leverage the recursive nature of these messages, we also defined the DAR-forward messages $\alpha_t(j_1, \dots, j_{\hat{P}}) = p
\,(\bm{y}_{1:t-1}, \gamma_{t-1} = j_1,\dots, \gamma_{t-\hat{P}} = j_{\hat{P}} \,|\, \cdot)$. Further details and the validity of these expressions are provided in the Supplementary Material. 
 
For inference on the graphs, since the GHS approach is a shrinkage procedure, and thus it does not estimate the entries  as exact zeros, we utilize posterior credible intervals to perform variable selection, as suggested by \citet{li2019graphical}.  Specifically, we use a 95\% interval from the estimated posterior distribution of each off-diagonal element of the precision matrices, so that if the interval corresponding to an entry includes zero, that entry is assessed as non active. Note that \citet{li2019graphical} employed a 50\% symmetric credible interval, arguing  that such a procedure would have conservative properties, and would reduce false negatives while controlling for false positive. However, in our experiments, a  95\% interval seemed to outperform the choice suggested by \citet{li2019graphical}.

\section{Simulation Studies} \label{sec:simulation_studies}
We investigate the performance of our proposed methodology using simulated data, aiming to assess its ability to recover the true DAR probabilities, emission parameters, and the correct number of autoregressive lags and hidden states. We further evaluate the robustness of the approach under model misspecification and its effectiveness on synthetic fMRI data.


\subsection{Data Generation}
\label{sec:study_structured_precision}
We first consider a simulation framework where the data were generated with underlying time-varying means and structured precision matrices. We generated 30 distinct datasets from model \eqref{eq:emissions}-\eqref{eq:DAR}, each consisting of $D=15$-dimensional time series of length $T = 2,000$, and assumed $M = 5$ latent states and a DAR of order $P = 2$. The autoregressive probabilities were set to $\bm{\phi} =(0.1, 0.75, 0.15)$ and the innovations to $\bm{\pi} = (0.6, 0.1, 0.1, 0.1, 0.1)$. For each state $j$, the emission means $\bm{\mu}_j$ were independently simulated from a multivariate Gaussian distribution with mean vector $ \bm{b}_0 = (-\frac{5}{D},  \dots, -\frac{1}{D}, 0, \frac{1}{D}, \dots, \frac{5}{D})$ and identity matrix as the covariance matrix, i.e. $  \bm{\mu}_j \sim \mathcal{N}(\bm{b}_0, \bm{\mathit{I}}_D)$, and where the simulated components of these vectors were randomly shuffled.
The state-specific precision matrices $\bm{\Omega}_j$ were assumed to be sparse with diagonal elements fixed to one and off diagonal elements constructed using the following five structures:

\indent{\it (i)} \textit{Identity graph}: this structure assumes that the components are independent, i.e. the off-diagonal elements are all set to zero.
    
\indent {\it (ii)} \textit{Star graph}: a configuration similar to the identity matrix, except for the first row and first column, whose elements are set to $\omega_{il} = -\frac{1}{D}$ if $i=1$ or $l=1$, and 0 otherwise.

\indent {\it (iii)} \textit{Hub graph}: this structure is organized into five blocks (hubs) of the same size. For any $l \neq i$ in the same block as $i$ we specify $\omega_{il} = \omega_{li} = -\frac{2}{\sqrt{D}}$, and 0 otherwise.

\indent {\it (iv)} \textit{AR(2) graph}: in this structure the precision matrix displays an autoregressive pattern of order two over the main diagonal. The entries are specified as  $\omega_{il} = \frac{1}{2}$ if $l=i-1, i+1$, $\omega_{il} = \frac{1}{4}$ if $l=i-2, i+2$, and 0 otherwise. 

\indent {\it (v)} \textit{Random sparse graph}: for this setting, the precision matrix is generated by randomly selecting $ \lfloor \frac{3}{2}D \rfloor $ off-diagonal entries, and drawing each $\omega_{jl}$  uniformly from the interval $[-1.0, -0.4] \cup [0.4, 1.0]$, while the diagonal elements are fixed at 1, and the other entries at 0. Each  of the off-diagonal element is then divided by the sum of the off-diagonal elements in its row, and then the matrix is averaged with its transpose, to produce a symmetric, positive definite, matrix. 

\begin{figure}
	\centering
\includegraphics[width=1\linewidth]{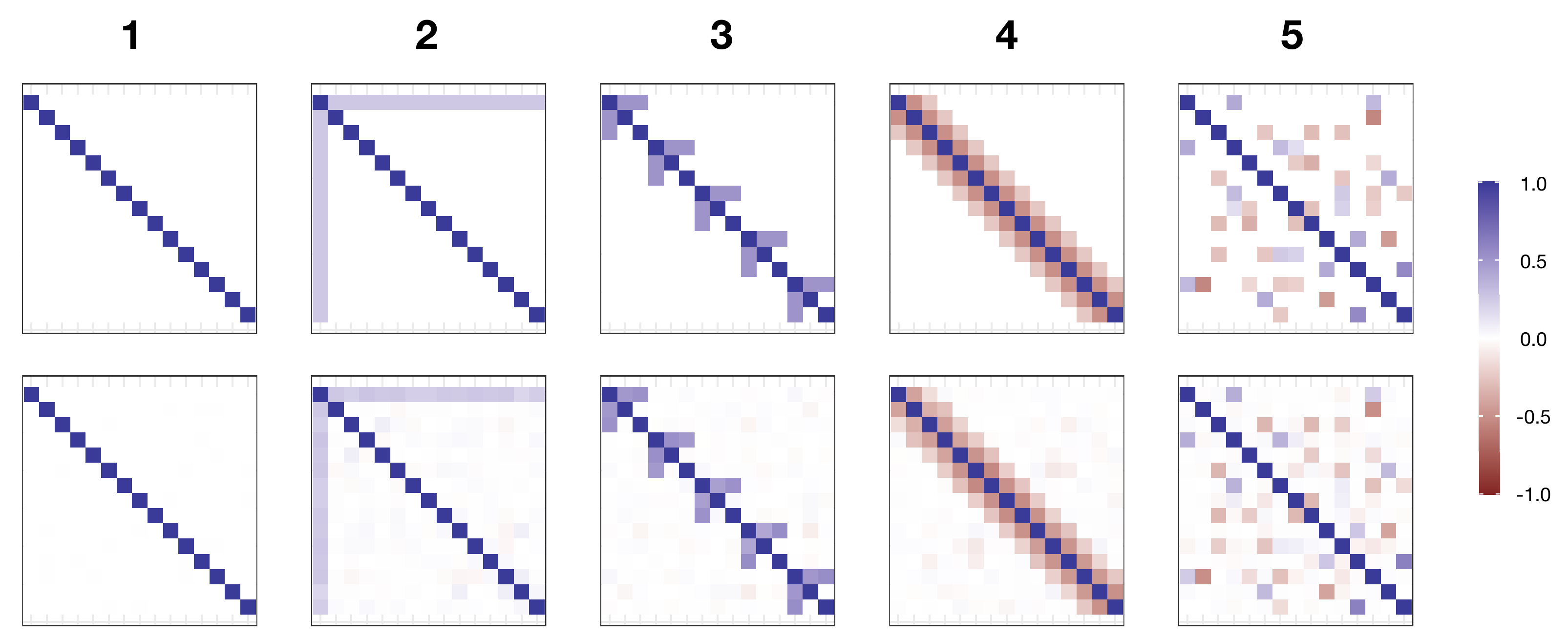}
	\caption{{\bf Simulation study.} (top) true state-specific partial correlation matrices; (bottom) estimated state-specific partial correlation matrices. These results are conditioned upon the estimated modal number of states and autoregressive order. }  
	\label{fig:partial_corr_simul}
\end{figure}

\begin{figure}[htbp]
	\centering
	\includegraphics[width=4.7in,height=1.5in]{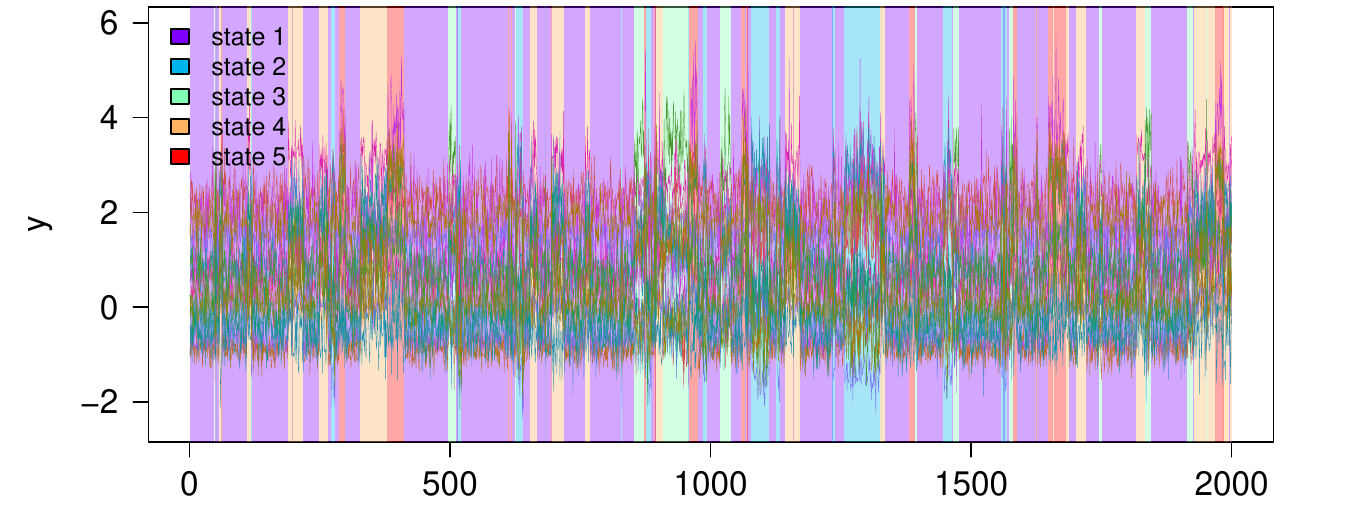}
     	\includegraphics[width=4.7in,height=1.5in]{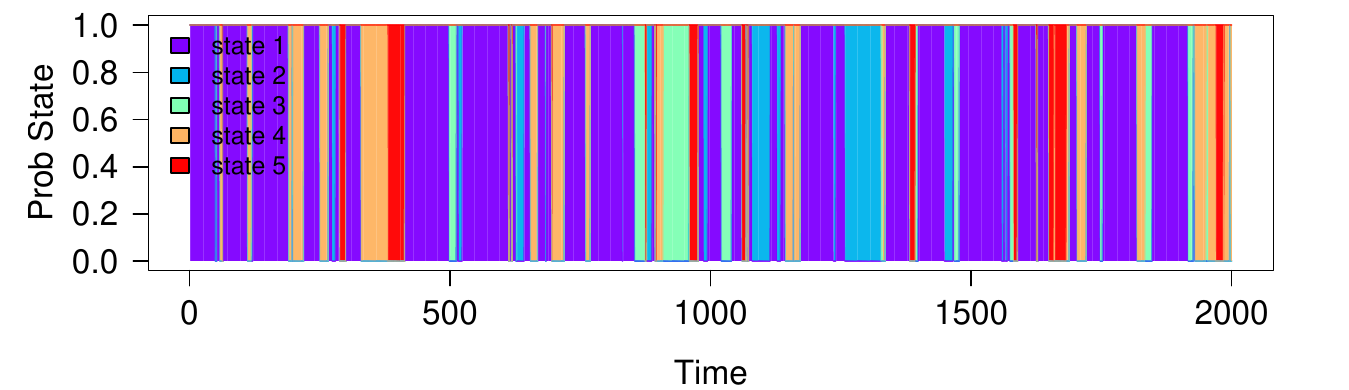}
	\caption{{\bf Simulation Study.} (top)  time series realization (lines), with each dimension represented by a different  line;  vertical bands represent the true underlying state sequence; (bottom)  estimated time-varying probability plot.} 
	\label{fig:data_and_predictive}
\end{figure}

Partial correlation matrices 
corresponding to these five scenario are displayed in Figure \ref{fig:partial_corr_simul} (top row).
A single realization from the simulation setting described in this section is shown in Figure \ref{fig:data_and_predictive} (top panel), with vertical bands representing the true underlying state sequence.  Here, we have further scaled the time series realization, independently for each dimension $d=1, \dots, D$, in such a way that the corresponding standard deviation of those observations $\{ y_{td} \}_{t=1}^{T}$ is equal to one. We note that the partial correlations are invariant under a change of scale and origin, allowing a meaningful comparison between true and estimated values of these matrices.


\subsection{Parameter Settings}
\label{sec:parameters}
Results reported below were obtained by fixing the maximum number of states to $M_{max} = 10$ and the maximum DAR order to $P_{max} = 5$. The DAR hyperparameters were chosen as $a_0 =1, b_0 = 10$, and $a_{\nu} = 10, b_{\nu} = 1$, so that the prior probabilities of innovation and autoregression were driven towards zero and one, respectively. The hyperparameters for the emission vector mean were specified as  $\bm{\mu}_0 = \bm{0}$ and $\bm{R}_0 = (1/10)\,  \bm{\mathit{I}}_D$, so that the mean components were a priori independent across different dimensions and with fairly large variance, hence reflecting weakly informative beliefs.

Initial values of the MCMC sampler were chosen as follows: the DAR parameters were sampled from the prior; the Gaussian emission means were fixed to the centers of a $k$-mean clustering and the covariance matrices were set to the identity. The GHS parameters were set to one. MCMC chains were run for 4,000 iterations, with 1,200 iterations discarded as burn-in. The algorithm took approximately 10 minutes to run, for each simulated time series, with a program written in Julia 1.6 on an Intel\textsuperscript{\textregistered} Core\textsuperscript{TM} i5 2GHz Processor 16GB RAM. We verified convergence of the MCMC sampler  by: (i) analyzing the trace plots of the parameters, e.g. the mean of the multivariate spiked Gaussian emissions, observing no pathological behavior; (ii) storing the values of the overall likelihood (Eq. \ref{eq:likelihood}) and plotting the corresponding trace, noting that it reached a stable regime; (iii) verifying the Heidelberger and Welch's convergence diagnostic \citep{heidelberger1981spectral} for the likelihood trace. We report some of the results in the Supplementary Material.

\subsection{Results}
\begin{figure}%
    \centering
    \subfloat{{\includegraphics[width=7.5cm]{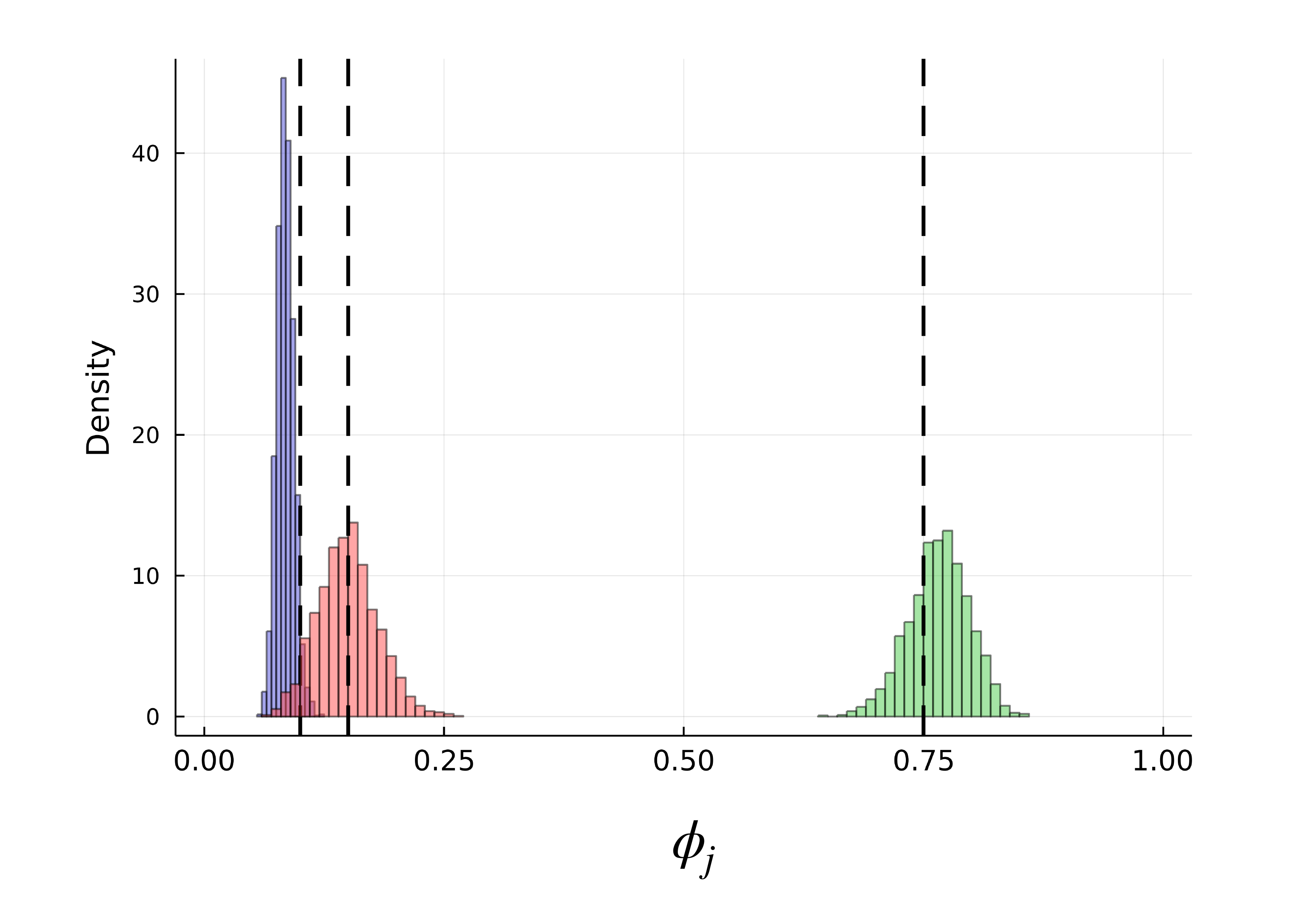} }}%
    \qquad
    \subfloat{{\includegraphics[width=7.5cm]{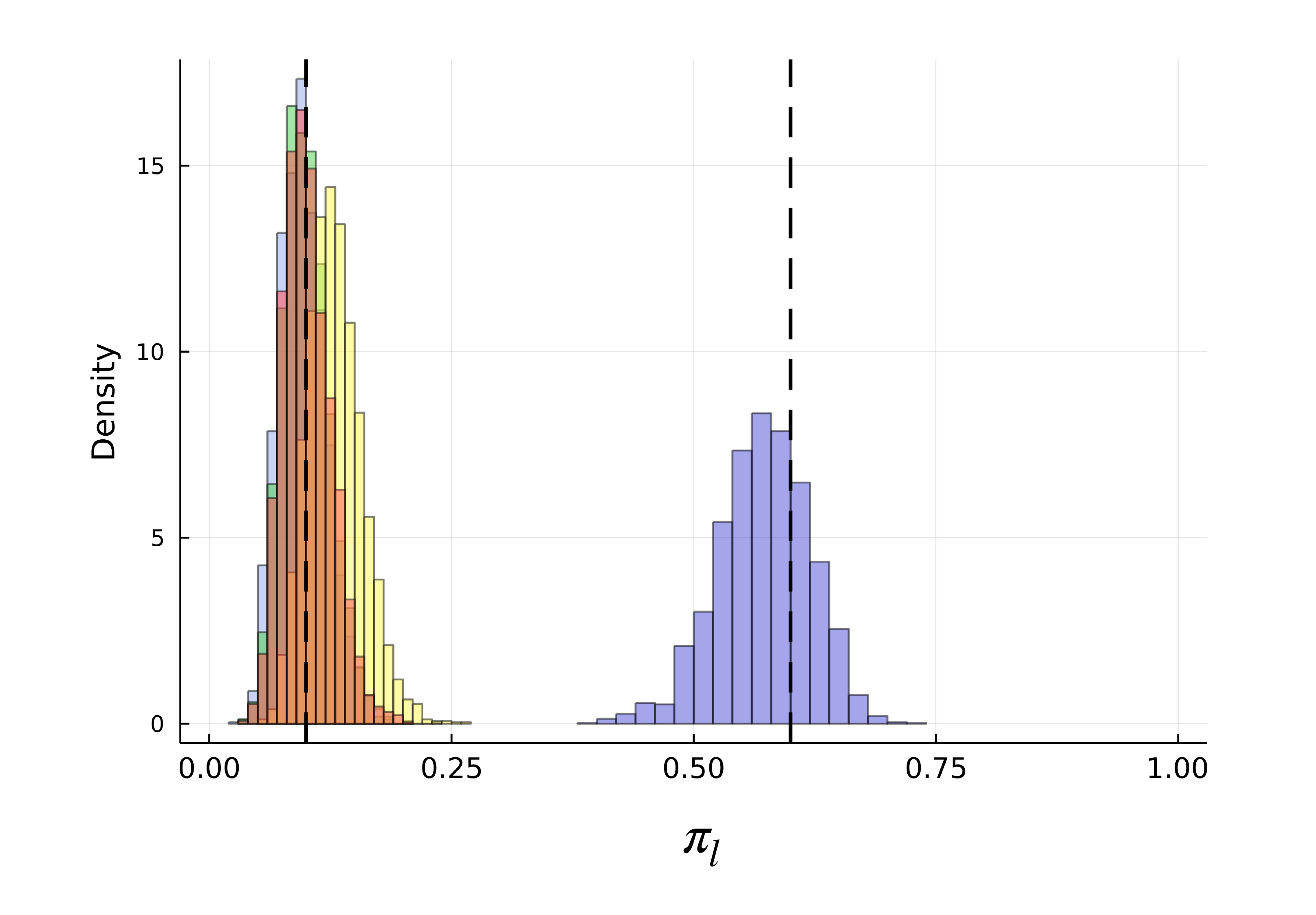} }}%
	\caption{{\bf Simulation Study.}  Posterior histograms of the DAR parameters. (left) autoregressive probabilities $\phi_l$, $l = 0, \dots \hat{P}$; (right) innovation probabilities $\pi_j$, $j = 1, \dots, \hat{M}$. Dotted vertical lines denote true parameters. These results are conditioned upon the modal number of states and autoregressive order. }
	\label{fig:histograms_DAR}
\end{figure}

Our approach consistently estimated the correct number of states $\hat{M}$ = 5 as the mode of the posterior distribution
and the number of active DAR probabilities as $\hat{P} = 2$ with high posterior probability, on all simulated replicates. 
For a single replicate, in Figure \ref{fig:partial_corr_simul} (bottom row) we show the estimates of the state-specific partial correlation matrices, conditioned upon the modal number of states and modal number of DAR parameters. 
Our approach successfully retrieves the distinct patterns of the true graphs. 
Figure \ref{fig:data_and_predictive} (bottom panel) displays a time-varying probability plot, namely the local decoding of the hidden state at time $t$, $p (\gamma_t = j \,| \,\bm{y}, \cdot)$, $j = 1, \dots, \hat{M}$, as described in Section \ref{sec:inference};  these plots are constructed by plotting the local probabilities  (which add to 1) cumulatively for each $t$. The proposed approach appears to correctly retrieve the true latent state sequence. Additionally,  Figure \ref{fig:histograms_DAR} displays the posterior histograms of autoregressive and innovation probabilities, with dotted vertical lines denoting the true generating values. Our proposed method appears to provide a good match between true and estimated values for the DAR parameters. In the Supplementary Material (Section C.1), we further analyze state-specific mean and variance values across different dimensions, revealing distinct patterns in both location and dispersion across states. These findings provide additional support for the model's ability to capture meaningful heterogeneity in the underlying data distribution.



Next, we investigated the performance of our proposed approach over the 30 replicated datasets and performed comparisons with alternative methods. We focused on the recovery of the state-specific precision matrices and compare the proposed methodology, which will be referred to as \texttt{sggmDAR}, to two alternative approaches. For the first approach, which we call  \texttt{mvHMM}, we fit a Bayesian multivariate HMM with Gaussian emissions, with a Normal inverse-Wishart prior on the state-specific emission parameters $(\bm{\mu}_j, \bm{\Sigma}_j) \sim NIW(\bm{\mu}_0, \bm{S}_0/\kappa_0; \nu_0, \bm{S}_0)$, where the hyper-parameters were specified in a weakly informative manner, i.e. $\bm{\mu}_0 = \bm{0}, \kappa_0 = 0.1, \nu_0 = D + 2, \bm{S}_0 = \bm{I}_D$. The number of states was set to five (i.e. the truth). The transition probabilities were assumed symmetric Dirichlet distributed, with concentration parameter equal to one. Since this HMM approach does not estimate precision entries as exact zeros, we once again  used 95\% posterior credible intervals to perform edge selection.
In the second approach, named \texttt{glassoSlide}, we followed \citet{allen2014tracking} and employed a sliding window to compute time-varying sparse inverse precision matrices via graphical lasso \citep{friedman2008sparse} using the R package \texttt{glasso}. In order to obtain an estimate of the latent state sequence, the windowed estimates of the precision matrices were then utilized as feature vectors in the $k$-means clustering algorithm. Finally, sparse state-specific precision matrices were estimated by applying graphical lasso to the MLE estimates of the covariances relative to the set of observations corresponding to each distinct state. The number of states was set to five (i.e. the truth), while the size of the sliding window and the magnitude of the penalization parameter were selected in such a way to maximize model selection performances averaged over the different states. 

\begin{table}[htbp]
\footnotesize
\begin{tabular}{clccccc}
  \hspace{-0.2cm}\\
  \hline
\multicolumn{1}{l}{}  &             &  Identity      &  Star          & Hub            & AR(2)          & Random         \\  \cmidrule{2-7}
\multirow{3}{*}{Acc}  & \texttt{sggmDAR}    & 1.0 (0.0)     & 0.995 (0.007) & 1.0 (0.002)    & 0.969 (0.021)  & 0.958 (0.027)  \\
                      & \texttt{mvHMM}       & 0.969 (0.032) & 0.933(0.039)  & 0.933 (0.088)  & 0.822 (0.138)  & 0.883 (0.110)  \\
                      & \texttt{glassoSlide} & 0.993 (0.018) & 0.835 (0.047) & 0.805 (0.053)  & 0.690 (0.031)  & 0.714 (0.031)  \\ \cmidrule{3-7}
\multirow{3}{*}{Spec} & \texttt{sggmDAR}    & 1.0 (0.0)     & 0.995 (0.007) & 1.0 (0.0)      & 0.995 (0.007)  & 0.997 (0.005)  \\
                      & \texttt{mvHMM}       & 0.969 (0.032) & 0.942 (0.022) & 0.933 (0.093)  & 0.840 (0.158)  & 0.924 (0.100)  \\
                      & \texttt{glassoSlide} & 0.993 (0.018) & 0.937 (0.052) & 0.879 (0.065)  & 0.899 (0.040)  & 0.869 (0.044)  \\ \cmidrule{3-7}
\multirow{3}{*}{MCC}  & \texttt{sggmDAR}    & -             & 0.979 (0.028) & 0.998 (0.011)  & 0.919 (0.055)  & 0.868 (0.087)  \\
                      & \texttt{mvHMM}       & -             & 0.734 (0.212) & 0.765 (0.210)  & 0.591 (0.297)  & 0.630 (0.365)  \\
                      & \texttt{glassoSlide} & -             & 0.140 (0.198) & -0.019 (0.040) & -0.017 (0.101) & -0.004 (0.073) \\ \cmidrule{3-7}
\multirow{3}{*}{F1}   & \texttt{sggmDAR}    & -     & 0.981 (0.025) & 0.998 (0.011)  & 0.936 (0.045)  & 0.883 (0.083)  \\
                      & \texttt{mvHMM}       & -     & 0.761 (0.194) & 0.747 (0.243)  & 0.683 (0.254)  & 0.683 (0.332)  \\
                      & \texttt{glassoSlide} & -    & 0.193 (0.189) & 0.081 (0.051)  & 0.125 (0.078)  & 0.152 (0.062)  \\ \cmidrule{3-7}
\multirow{3}{*}{Sens} & \texttt{sggmDAR}    & -     & 0.994 (0.020) & 0.996 (0.020)  & 0.895 (0.071)    & 0.807 (0.124)  \\
                      & \texttt{mvHMM}      & -             & 0.874 (0.235) & 0.927 (0.245)  & 0.772 (0.267)  & 0.732 (0.356)  \\
                      & \texttt{glassoSlide} & -     & 0.171 (0.174) & 0.103 (0.072)  & 0.089 (0.063)  & 0.126 (0.059)  \\ \cmidrule{3-7}
\multirow{3}{*}{RMSE} & \texttt{sggmDAR}    & 0.002 (0.001) & 0.034 (0.008) & 0.021 (0.007)  & 0.049 (0.011)  & 0.042 (0.011)  \\
                      & \texttt{mvHMM}       & 0.019 (0.012) & 0.073 (0.013) & 0.080 (0.035)  & 0.116 (0.057)  & 0.087 (0.041)  \\
                      & \texttt{glassoSlide} & 0.001 (0.004) & 0.097 (0.003) & 0.162 (0.002)  & 0.205 (0.002)  & 0.155 (0.004) \\ \hline  
\end{tabular}
\caption{{\bf Simulation Study.} Accuracy, specificity,  Matthew correlation coefficient (MCC), F1 score,  sensitivity, and residual mean squared error (RMSE) of precision matrix estimates, for each state $j=1, \dots, \hat{M}$. Standard deviations over the 30 simulations are displayed in parentheses. Results are reported for \texttt{sggmDAR}, \texttt{mvHMM} and \texttt{glassoSlide}. Results for \texttt{sggmDAR} are conditioned upon the modal number of states and autoregressive order. A hyphen is used for those metrics that cannot be computed due to the structure of the underlying truth (e.g. TP+FN = 0).}
\label{tab:accuracy_etc}
\end{table}
\normalsize

To assess model selection performances we computed accuracy, sensitivity, specificity, $F1$-score and Matthew correlation coefficient (MCC),  for each regime $j=1, \dots, \hat{M}$.
In addition, to evaluate estimation accuracy, we calculated residual mean squared error (RMSE) of state-specific off-diagonal entries of the precision matrices as RMSE$_{\,j}$ $= \sqrt{\frac{1}{D}\sum_{i<l} \big( \omega_{jil} - \hat{\omega}_{jil} \big)^{2}}$. Results from these measures are summarized in Table \ref{tab:accuracy_etc}. Note that MCC, F1, and sensitivity  for the Identity state are not presented since these metrics cannot be computed due to the structure of the underlying truth (e.g. TP+FN = 0).
Overall, \texttt{sggmDAR} produced the best results both in estimation accuracy and model selection performances. Though accuracy and specificity of \texttt{glassoSlide} are somewhat high, this frequentist method is by far the worse, as illustrated by very low MCC scores. We remark that, while both \texttt{mvHMM} and \texttt{glassoSlide} need to specify the number of states in advance, our proposed approach produces an estimate of this parameter. In the Supplementary Material, we further investigate the performance of our proposed methodology for data-generating emissions with zero-mean, i.e. $\bm{\mu}_j = \bm{0}$, for $j = 1, \dots, M$. 
The results confirm the superiority of our approach over both \texttt{mvHMM} and \texttt{glassoSlide} in terms of estimation and model selection accuracy.
Additionally, the Supplementary Material contains a sensitivity analysis study that focuses on examining the impact of the hyperparameters $a_v, b_v$ associated with the zero-inducing cumulative shrinkage prior \eqref{eq:stick_break_2}. Results show that, for small and moderate $T$, different combinations of the hyperparameters may yield varying dynamics of the process. However, as $T$ increases, such differences are not noticeable.

\subsection{Simulations for Varying $T$ and $D$}
\label{sec:high_dim}

\begin{figure}[t!bp]%
    \centering
    \subfloat{{\includegraphics[width=7.2cm]{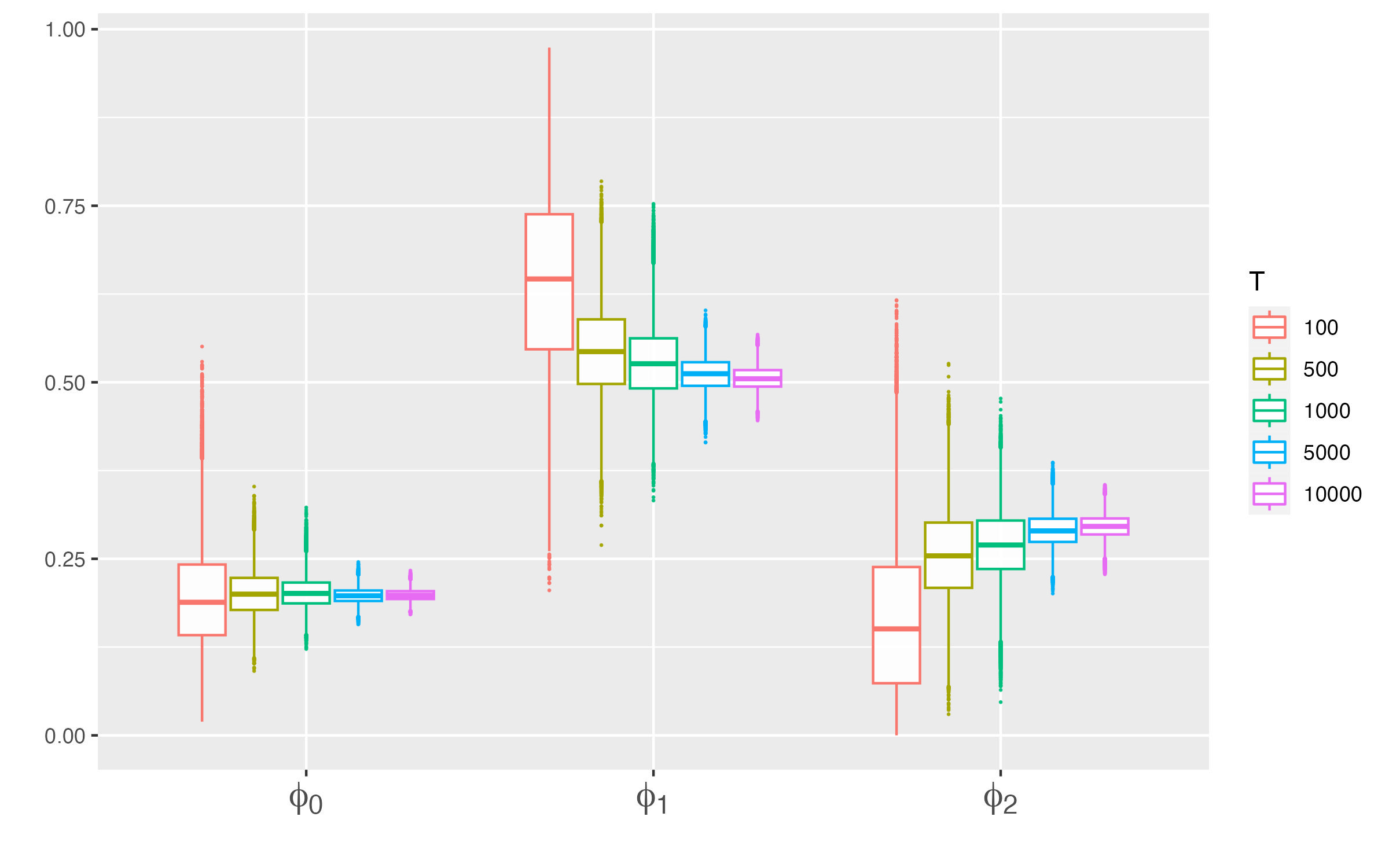} }}
    \qquad
    \subfloat{{\includegraphics[width=7.2cm]{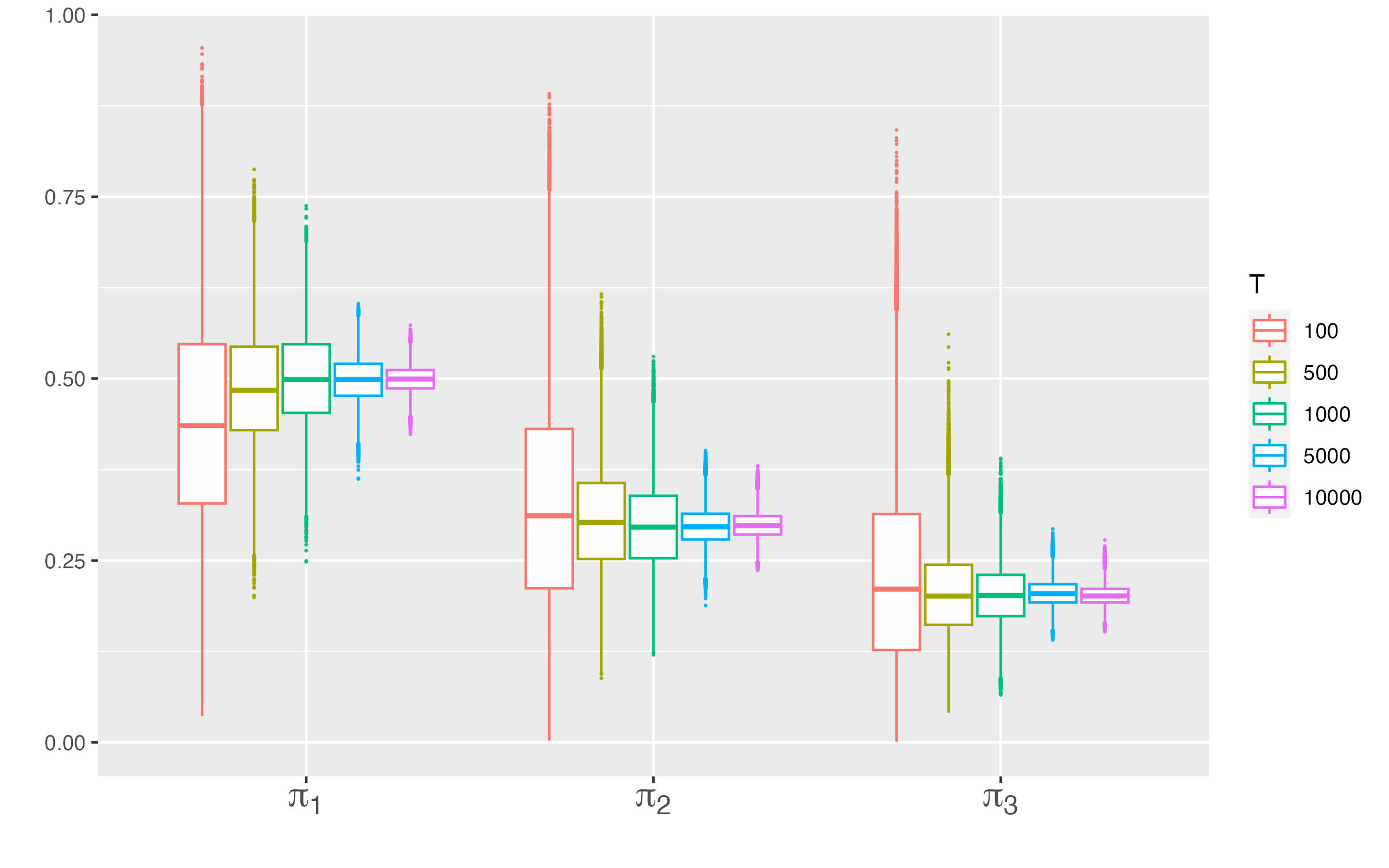} }}
    	\caption{{\bf Simulation Study.} Boxplots over 30 simulations of posterior distributions for (left) autoregressive probabilities $\phi_l$, $l = 0, \dots \hat{P}$, and (right) innovation probabilities $\pi_j$, $j = 1, \dots, \hat{M}$, for different sample size $T = 100, 500, 1000, 5000, 10000$.  These results are conditioned on the modal number
of states and autoregressive order.  }
   \label{fig:example}
   \end{figure}

Next, we investigated performance for different values of the sample size $T$.
For this, we generated 30 distinct time series for different sample sizes, $T = 100, 500, 1000, 5000, 10000$, assuming $M=3$ states and DAR order $P=2$, with autoregressive probabilities specified as $\bm{\phi} =(0.2, 0.5, 0.3)$ and innovations set to $\bm{\pi} = (0.5, 0.3, 0.2)$. The emission means were generated as in the main simulation above, whereas the precision matrices were constructed using patterns (i), (iii) and (v) from Section \ref{sec:study_structured_precision}. Here, to perform Bayesian inference we fixed the maximum number of states to $M_{max} = 3$ and maximum DAR order to $P_{max} = 2$. The hyperparameters were specified as in Section \ref{sec:study_structured_precision}.
Figure \ref{fig:example} displays boxplots over the 30 simulations of the posterior distributions of $\bm{\phi}$ and $\bm{\pi}$, for the different values of $T$, conditioned upon the the modal number of states and autoregressive order. As it was to be expected, estimates for both $\phi_j$ and $\pi_j$ showed larger variability for small sample sizes. Conversely, as $T$ increases, the generating DAR dynamics became more apparent, and our inference procedure is indeed able to retrieve the correct parameters more accurately, for most cases. 

\normalsize

Finally, we explored the performance of our approach in a scenario where the dimension $D$ of the  data is large. Here, we focused on assessing the ability of our proposed method in recovering the number of states, number of DAR parameters, and true sparse precision matrices.  We simulated 30 time series, each consisting of $D = 100$-dimensional time series of length $T= 2000$, with $M=3$ and $P =2$, and with the emissions generated as in Section \ref{sec:study_structured_precision} and the precision matrices for the three states specified as Identity, Hub (with four blocks) and Random, respectively. The innovations were set to $\bm{\pi} = (0.6, 0.2, 0.2)$, while the rest of the data-generating parameters were set as in Section \ref{sec:study_structured_precision}. Here we report results obtained by specifying the hyper-parameters as described in Section \ref{sec:study_structured_precision} and by running MCMC chains for 4,000 iterations, with 1,200 iterations discarded as burn-in.


As in the previous simulations, the correct number of states and DAR order were identified as those with the highest posterior probability for all replicated datasets.
In the Supplementary Material, we report model selection and estimation accuracy performances for the off-diagonal component of the precision matrices, for \texttt{sggmDAR}, \texttt{mvHMM}, and \texttt{glassoSlide}. 
The MCC scores  highlight the advantage of choosing our proposed method in large settings. Indeed, the number of parameters for each individual state is substantial, as there are 4950 distinct off-diagonal coefficients to be inferred for each precision matrix.

\subsection{Additional Simulations}

We conducted further simulations to assess the robustness of our model under two complementary settings: (i) a model misspecification scenario, where data are generated from a stationary vector autoregressive (VAR) process; and (ii) a realistic synthetic fMRI dataset designed to mimic experimental BOLD signals under structured noise conditions. These simulations provide insight into how the model behaves when its assumptions are violated and how well it performs in capturing task-driven neural dynamics. Full details are provided in the Supplementary Material (Sections F and G).

\subsubsection{Misspecified Model}

VAR is a popular model for both task and resting state fMRI, commonly used to capture temporal dependencies in neural time series. To explore the model's robustness under model misspecification, we generated multivariate time series from a stationary first-order VAR process with either sparse or dense autoregressive coefficient matrices. We summarize here the major findings. A complete description of the simulation setup and additional diagnostic plots are provided in the Supplementary Material (Section~F).

In the sparse setting, where dependencies among dimensions are limited, the model consistently favored a single latent regime, with an estimated number of states $\hat{M} = 1$. This outcome reflects the relative homogeneity of the time series structure under sparsity, which does not induce detectable changes in statistical properties over time. In contrast, under a dense VAR structure, where most components influence one another, the model identified multiple regimes, with $\hat{M} = 3$. Despite the absence of true regime-switching in the data-generating process, the inferred latent states captured subtle shifts in the local mean and variance, as well as changes in the estimated connectivity patterns. Residual diagnostics, including standardized residuals and Q-Q plots, confirmed the adequacy of the model fit, with deviations confined to minor departures in the tails.
These results highlight the model’s capacity to adapt to a variety of temporal structures, effectively extracting interpretable latent states even under substantial model misspecification.

\subsubsection{Simulated fMRI Data}

To further assess the performance of our model in a realistic neuroimaging context, we generated synthetic fMRI data using the \texttt{neuRosim} package in \textsf{R} \citep{welvaert2011neurosim}, which provides a principled framework for simulating blood-oxygen-level-dependent (BOLD) signals under controlled experimental and noise conditions. This simulation mimics a block-design experiment in which an external stimulus is applied every 80 seconds and persists for 40 seconds. The BOLD response is constructed using a canonical double-gamma hemodynamic response function (HRF), and several structured noise components—namely, temporal drift, physiological oscillations, and scanner-induced artifacts—are independently added to each brain region. The resulting signal is standardized to reflect typical preprocessing steps in fMRI analysis.

Figure~\ref{fig:BOLD_signal_generation} displays this process: panel (a) shows the final simulated multivariate fMRI time series; panel (b) presents the clean BOLD signal prior to noise addition; and panel (c) shows the posterior state probability plot obtained from our model. In this setting, the proposed approach correctly infers $\hat{M} = 2$ latent states, with state transitions that align closely with the stimulus schedule—despite the unsupervised nature of the analysis. This confirms the model's ability to recover task-related structure in high-dimensional noisy time series data.

\begin{figure}[htbp]
    \centering
    \includegraphics[width=0.65\linewidth]{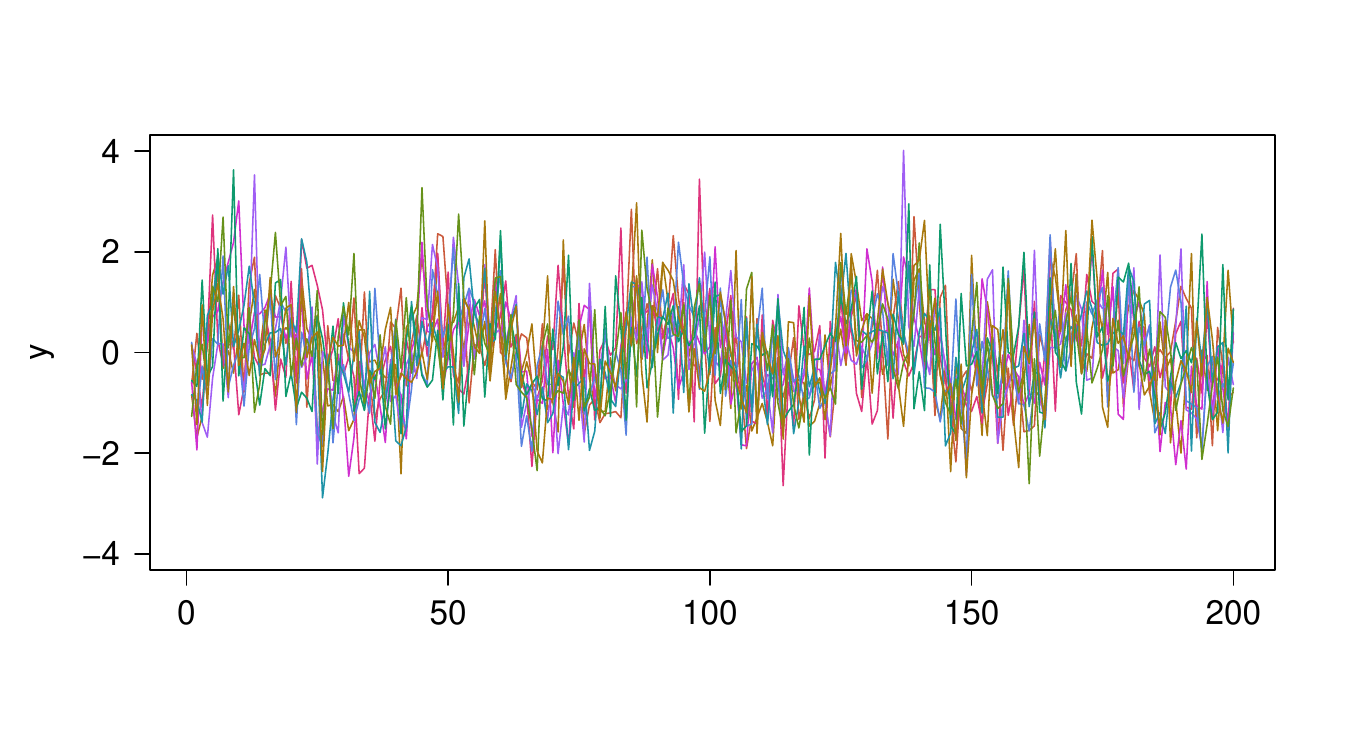}
    \includegraphics[width=0.65\linewidth]{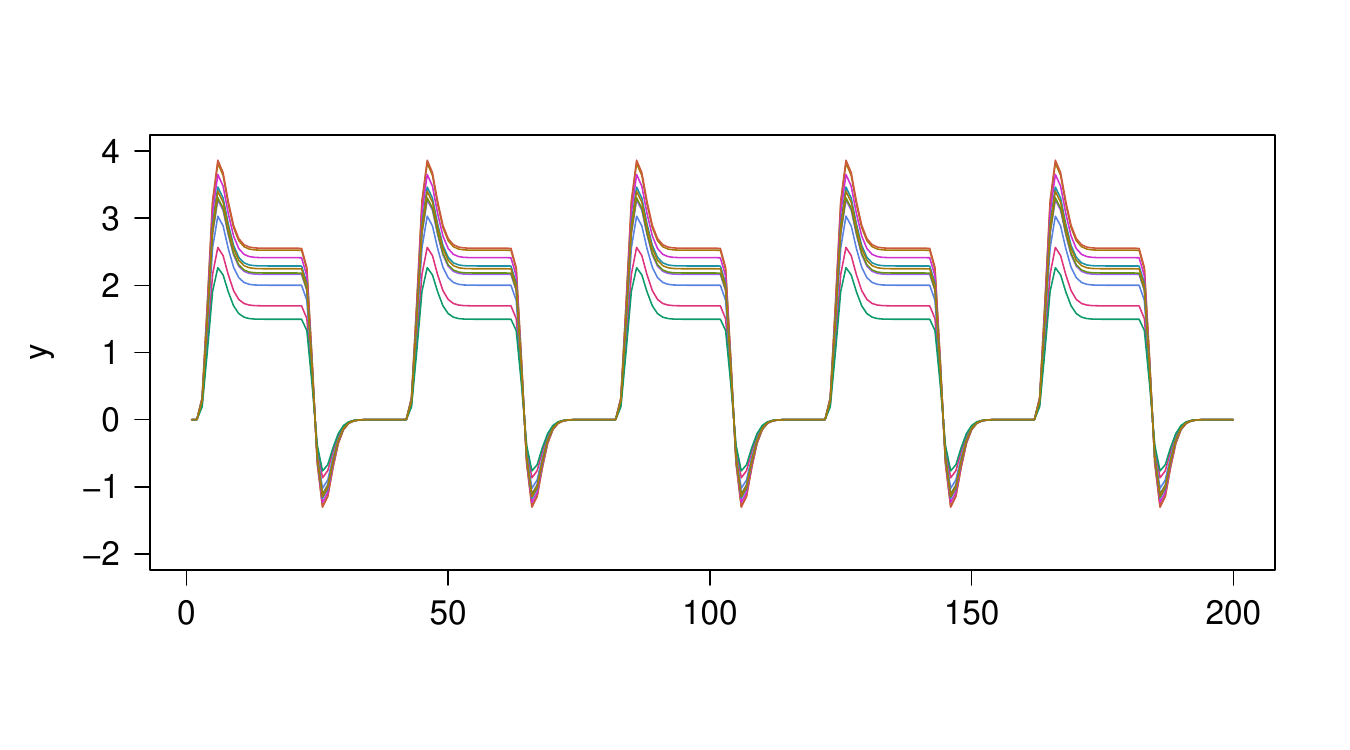}
    \includegraphics[width=0.65\linewidth]{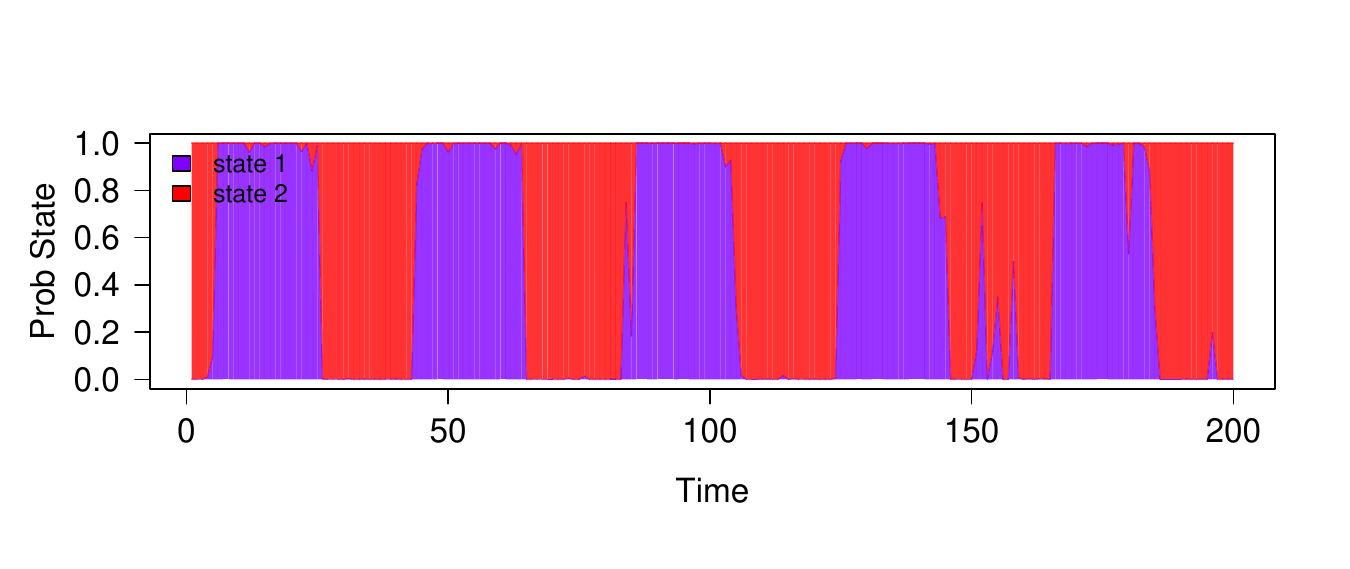}
    \caption{\textbf{Simulated fMRI Data.} (a) Simulated fMRI time series obtained by summing the clean BOLD signal with structured noise components and standardizing each region; (b) True underlying BOLD signal convolved with a canonical double-gamma HRF; (c) Estimated time-varying state probability plot.}
    \label{fig:BOLD_signal_generation}
\end{figure}

The model’s estimated state-specific mean and variance profiles (reported in the Supplementary Material, Section G) highlight interpretable differences between the two inferred states. One state corresponds to higher overall activation across regions, while the other reflects baseline or resting activity. These findings demonstrate the method’s ability to isolate functionally relevant brain states. Additional diagnostics, including standardized residuals and Q-Q plots, further confirm the model’s goodness-of-fit, suggesting that the proposed framework effectively characterizes stimulus-driven neural dynamics even under complex noise conditions.

\section{Application to fMRI Data} \label{sec:application}
Identifying the dynamic nature of brain connectivity is critical for understanding and advancing our current knowledge about human brain functioning. Functional magnetic resonance imaging (fMRI), which measures brain activity by detecting changes associated with blood flow, has become a successful and effective instrument for studying how the brain functions. Here we consider the problem of estimating brain connectivity, i.e., statistical dependence between fMRI time series in distinct regions of the brain. Recent evidence has shown that the interactions among brain regions vary over the course of an fMRI experiment, suggesting that brain connectivity is a dynamic process \citep{cribben2013detecting,lindquist2014evaluating,xu2015dynamic, vidaurre2017brain, vidaurre2018discovering, Warnick2018b}.  In this context, our proposed modeling framework is particularly well suited to address two key challenges in the analysis of fMRI data. First, the number of distinct cognitive or connectivity states that manifest during the course of the task is unknown and likely to vary across individuals or experimental conditions. Rather than assuming a fixed number of states, our model infers the number of latent regimes in a data-driven fashion using a sparsity-inducing prior. Second, the temporal structure governing transitions between these regimes may involve dependencies that extend beyond first-order Markovian assumptions. Employing a DAR process of unknown order allows for more flexible and biologically plausible switching dynamics. Together, these features enable the identification of meaningful patterns in brain connectivity that evolve over time, offering insight into latent cognitive processes.

We apply the proposed model to fMRI data from a subject performing a task-based experiment 
where the interest is to identify the neural representations that are formed during latent learning of predictive sequences \citep{bornstein2012dissociating}. In this experiment, participants carried out a task in which they observed a sequence of black-and-white natural scene images. They were instructed only to press a keyboard key (`d', `f', `j', `k') that they had previously been trained to associate with each image. Throughout the trials, the series of pictures were generated according to a first-order Markov process, though the participants were not aware of this structure. This form of task has been used to examine the cognitive and neural architecture of latent learning and the use of learned representations in support of predictive lookahead, a core computational process supporting decision-making in humans and animals \citep{strange2005information,harrison2006encoding,bornstein2013cortical,morris2018goal,rmus2022humans,hunter2018common,yoo2023cognitive}. Here, response times indicated the degree to which the participant implicitly expected the currently presented image, on the basis of the previously presented one. A consistent finding in these tasks is that participants implicitly learn the sequential structure, and that neural regions signal the degree to which they anticipate the upcoming image in the sequence. Several studies have identified more than one representation of sequential structure, each of which has an influence on behavior as estimated across the entire experiment. However, it is unclear how these multiple representations are arbitrated among to influence behavior -- e.g., as a weighted mixture at the single-trial level \cite{wang2022mixing,khoudary2022precision,nicholas2022uncertainty}, in alternation according to regimes of task statistics \cite{daw2005uncertainty,poldrack2001interactive,lengyel2007hippocampal,yoo2023cognitive}, or as a fixed proportion that varies according to individual differences such as in memory encoding precision \cite{noh2023memory}. Full details of the experiment are provided in \citet{bornstein2012dissociating}. 

The scanning session proceeded with four blocks consisting of 275 fMRI acquisitions. For the analyses of this paper we concatenated the four blocks and subtracted the mean of each block. $D=18$ lateralized regions of interest (ROIs) were selected on the basis of prior findings using this task \citep{bornstein2012dissociating,bornstein2013cortical} as those most sensitive to one or more of the identified representations of sequential structure (dorsal and ventral striatum, hippocampus), to the degree of conflict between the representations (anterior cingulate cortex), or to the stimulus content (scene images; ventral visual stream regions). The observed time series correspond to the preprocessed and standardized blood-oxygen-level-dependent (BOLD) signals recorded from each of the selected brain regions, serving as proxies for local neural activity across time. We scaled each dimension $d=1, \dots, D$, to have variance one. 
In our analysis, we seek to identify distinct regimes of functional connectivity that can be mapped onto cognitive interpretation -- specifically, to identify the manner in which multiple representations combine to control behavior. Since we do not have prior information on the cognitive states that are manifested during the experiment, we assumed that the number of the states is unknown, as well as the latent learning structure driving the switching of regimes. We set the hyperparameters of the model as described in Section \ref{sec:study_structured_precision} and ran MCMC chains with 4000 iterations, 1200 of which were discarded as burn-in.

\begin{figure}[t!bp]
	\centering
 \includegraphics[width=0.7\linewidth]{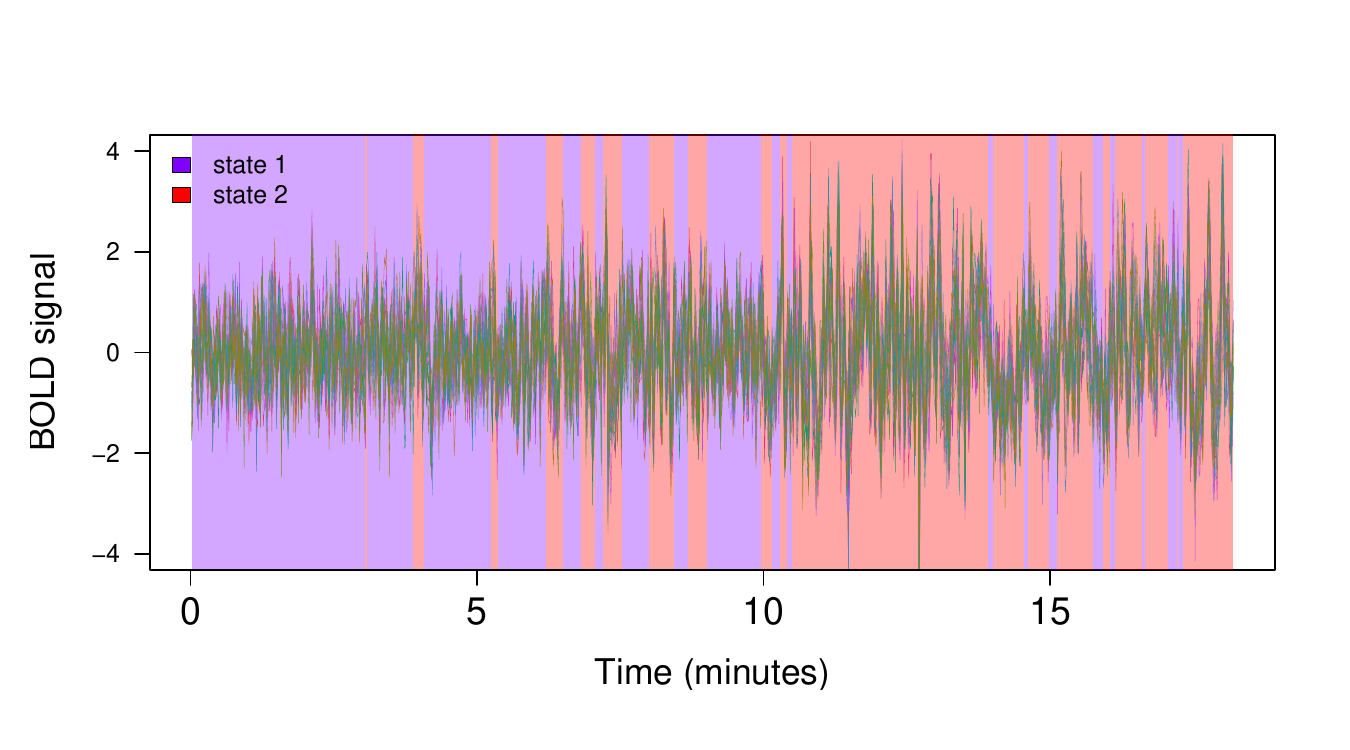}
 \includegraphics[width=4.7in,height=0.8in]{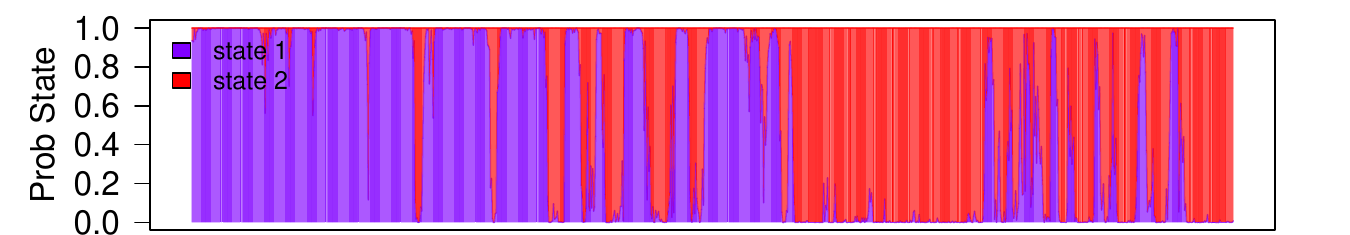}
 \caption{\textbf{Application to FMRI data}. (top) BOLD time series data from one participant, with each brain region
represented by a different line; vertical  bands represent the estimated state sequence. (bottom) estimated time-varying state probability plot.}
\label{fig:subject2}
\end{figure}

The  BOLD time series data for the $D=18$ selected ROIs are shown in the top panel of Figure \ref{fig:subject2}, along with the estimated latent state sequence. Our model identified a mode at $\hat{K} = 2$ distinct states and a DAR order $\hat{P} = 2$, with estimated values of innovations and autoregressive parameters $\hat{\bm{\pi}} = [0.50, 0.50]$ and $\hat{\bm{\phi}} = [0.08,0.87,0.05]$, respectively. The bottom panel of Figure \ref{fig:subject2} displays the time-varying probability plot, namely the local decoding of the hidden state at time $t$, $p (\gamma_t = j \, | \, \bm{y}, ·)$, for $t=1, \dots, T$, as described in Section \ref{sec:inference}. These probabilities are represented with a different color for each of the two inferred states, and they cumulatively add to one for each $t$. The state probability plot displays a clear transition from state 1 to state 2, approximately halfway through the task.

\begin{figure}[t!bp]
	\centering
  \includegraphics[width=0.7\linewidth]{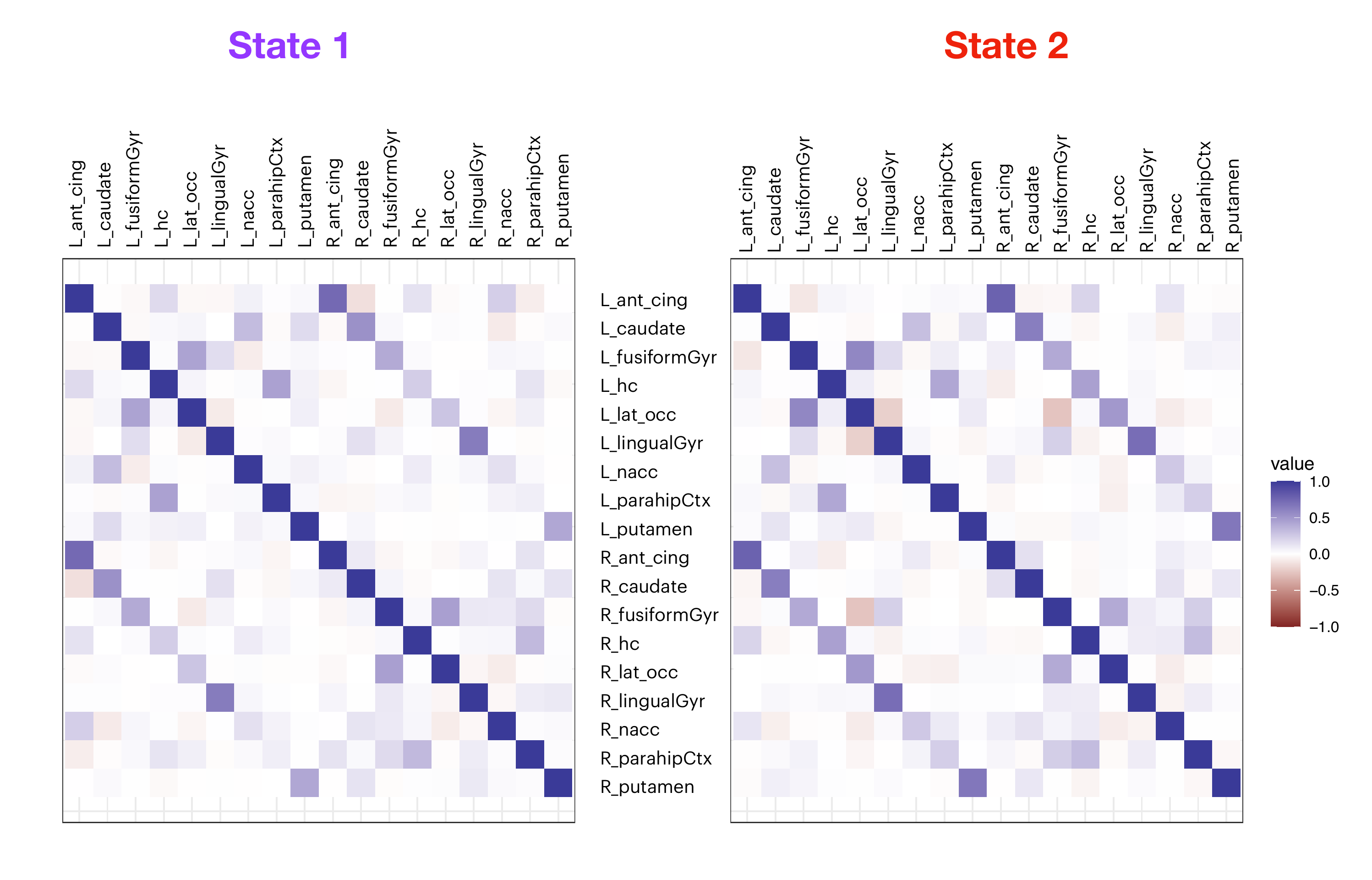}
 \includegraphics[width=0.7\linewidth]{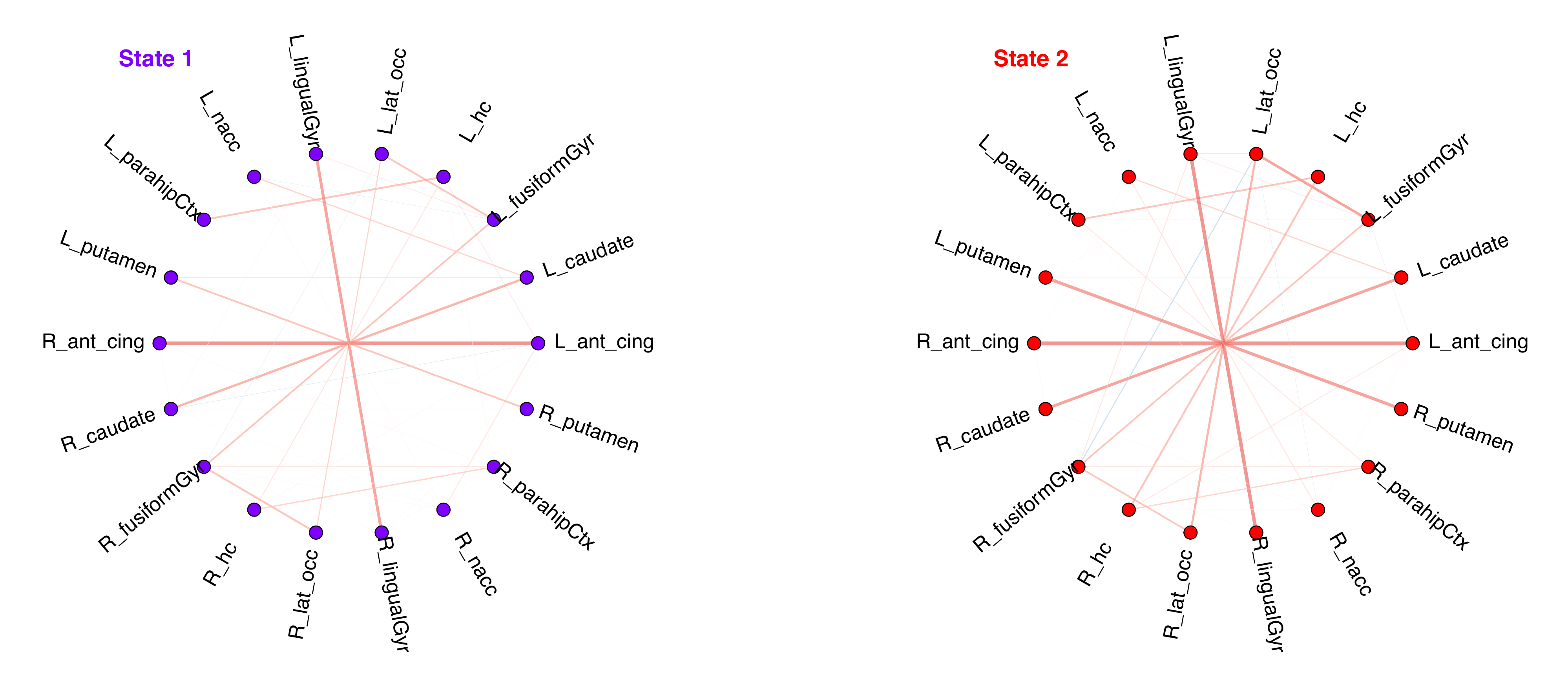}
 \caption{\textbf{Application to FMRI data}. (top) estimated partial correlation matrices, for each of the two inferred states. (bottom) estimated state-specific connectivity graphs.}
\label{fig:subject2_connectivity}
\end{figure}

The top panel of Figure \ref{fig:subject2_connectivity} shows the estimated state-specific partial correlation matrices, for the two estimated states, and the bottom panel the corresponding estimated connectivity graphs, with edges identified through the procedure described in Section \ref{sec:GHS}. State 1 has relatively stronger connectivity between hippocampus (HC) and anterior cingulate cortex (ACC), with mean difference between states across ROI pairs equal to $.028$; whereas State 2 shows stronger connectivity between Caudate and ACC, with mean difference between states across ROI pairs equal to $.048$. Across all ROI pairs, the average difference in partial correlation values between states is $.002$.  In the Supplementary Material, Section C.2, we provide additional analysis of the fMRI data, including state-specific mean and variance estimates, standardized residual diagnostics, and ACF/PACF plots to further assess model fit and temporal structure. These results show that while mean activation remains largely stable across states, variability differs meaningfully, suggesting distinct dynamic regimes. Residual diagnostics, including standardized residuals and Q-Q plots, further support model adequacy, indicating no major misspecification and approximate normality of the residuals. Together, these findings suggest that our model effectively captures the dominant temporal structure of the data, achieving a balance between flexibility and interpretability in the context of dynamic brain connectivity.

These observations are consistent with findings in the literature that at least two distinct networks mediate expectations in this task: one centered on hippocampus and thought to encode stimulus-stimulus predictive relationships (e.g. ``cognitive maps"), and the other centered on striatum and thought to encode response-response sequences \citep{bornstein2012dissociating,bornstein2013cortical}. Each has different dynamics with regard to the predictiveness of the learned sequences: activity in the hippocampal network scales with increasing uncertainty about the next item in the sequence, consistent with its proposed role in ``pre-fetching" upcoming states in support of decision-making \citep{johnson2007neural}; separately, activity in the striatal network \textit{decreases} with uncertainty about the next item in the sequence, consistent with observations that this structure is more strongly activated by highly predictive associations \citep{smith2016habit}. The observation that the network regime corresponds to shifts in its connectivity with anterior cingulate cortex is consistent with theoretical accounts of this region as signaling the ``expected value of control", mediating the influence of internal representations on behavior \citep{shenhav2013expected}. The transition between hippocampal and striatally-mediated regimes is consistent with extensive empirical findings that these regions ``trade-off" in control of behavior across highly repeated tasks, with hippocampus driving responses early on and striatum taking over when sequences are more well-practiced \citep{poldrack2001interactive,lengyel2007hippocampal}.  


\section{Concluding Remarks}
\label{sec:conclusion}
We have presented a flexible Bayesian approach for estimating sparse Gaussian graphical models based on time series data.
In order to represent switching dynamics of the time series data, we have assumed an unobserved hidden process underlying the data, with observations generated from state-specific multivariate Gaussian emission distributions. We have modeled the temporal structure of the hidden state sequence based on a DAR process, as a flexible approach to incorporate temporal dynamics that extend beyond simple Markovian structures. We have modeled the time-varying mixing probabilities capturing the state-switching behavior of the DAR process via a cumulative shrinkage non-parametric prior that accommodates zero-inflated parameters for non-active components. The proposed formulation ensures that if a parameter in the DAR model is zero, then all subsequent lag parameters are also zero, yielding a flexible and computationally efficient modeling framework for estimating the time-varying mixing probabilities as well as the effective order of the process. This considerably speeds up the posterior sampler, especially in regard to the forward-backward scheme for updating the latent state sequence. 
We have additionally integrated a sparsity-inducing Dirichlet prior to estimate the effective number of states in a data-driven manner. At the network level, we have assumed a graphical horseshoe prior to induce sparsity in the state-specific precision matrices.
We have thoroughly investigated the performance of our methods through simulation studies and performed comparisons with competing approaches. We have further illustrated our proposed approach for the estimation of dynamic brain connectivity based on fMRI data collected on a subject performing a task-based experiment on latent learning.

\section*{Acknowledgments}
B.H.-A. completed part of this work as a postdoctoral fellow in the Department of Statistics at Rice University, Houston, TX.

\section*{Disclosure Statement}
No potential competing interest was reported by the authors.


\bigskip
\begin{center}
{\large\bf SUPPLEMENTARY MATERIAL}
\end{center}

\begin{description}
\item[Supplement:] We include further details about backward and forward messages for our sampling algorithm. We also report results from additional simulations, sensitivity analyses and convergence diagnostics of the MCMC. 

\item[Software:] \texttt{sggmDAR} - a Julia software implementing the methodology outlined in the paper, accompanied by a comprehensive tutorial designed to guide users through replicating the findings detailed in the article. The software \texttt{sggmDAR} is also available on GitHub at \href{https://github.com/Beniamino92/sggmDAR}{https://github.com/Beniamino92/sggmDAR}.
\end{description}

\bibliographystyle{apalike}
\bibliography{bib/biblio}

\newpage
\bigskip
\begin{center}
{\large\bf SUPPLEMENTARY MATERIAL}
\end{center}



\newcommand{\blind}{0}
\renewcommand*{\thesection}{\Alph{section}.}
\renewcommand{\theequation}{S.\arabic{equation}}
\setcounter{figure}{0}
\setcounter{section}{0}
\setcounter{equation}{0}

This document provides supplemental material for the article ``Discrete Autoregressive Switching Processes in Sparse Graphical Modeling of Time Series.'' Sections \ref{sec:S_backwards} and \ref{sec:S_forward} present additional details on backward and forward messages, along with the validity of these expressions. Section \ref{sec:suppl_additional} includes supplementary results for the simulation study and fMRI application. Section \ref{sec:S_simulzeromean} investigates performance under zero-mean emissions, while Section \ref{sec:S_largedsetting} examines the approach in high-dimensional settings. Section \ref{sec:suppl:misspecified} evaluates model performance under misspecification, and Section \ref{sec:suppl_simulatedfMRI} assesses the model’s behavior on synthetic fMRI data. Finally, Sections \ref{sec:suppl_sensitivity} and \ref{sec:S_convergence} provide sensitivity analysis results and MCMC convergence diagnostics.

\section{Backward messages} \label{sec:S_backwards}
\noindent \textbf{Proposition 2}. Let consider $\eta_{\{\,j_{\hat{P}}, \dots, j_1, j_0\}}$, i.e. the DAR probabilities of selecting state $j_0$, given previous values $j_1, \dots, j_{\hat{P}}$, as defined in Eq. \eqref{eq:DAR}, and let $p (\bm{y}_t \, |\cdot )$ be the multivariate Gaussian  emission densities specified in Eq. \eqref{eq:emissions}. Then, the backward messages $\beta_t(j_1) = p(\bm{y}_{t:T}| \gamma_{t-1} = j_1, \cdot)$ can be recursively expressed as in Eq. \eqref{eq:backward_msg}.

\noindent  \textit{Proof:} Let $M = M_\text{max}$, and let $P = \hat{P}$ be the number of active DAR parameters. Then the proof proceeds as follows 
\begin{align*}
        \beta_t(j_1) &= p(\bm{y}_{t:T} | \gamma_{t-1} = j_1, \cdot ) \\
        &= \sum_{j_0=1}^{M} p (\bm{y}_{t:T}, \gamma_t = j_0 | \gamma_{t-1} = j_1, \cdot) \\
        &= \sum_{j_{P}=1}^{M} \dots \sum_{j_2=1}^{M} \sum_{j_0=1}^{M} p (\bm{y}_{t:T}, \gamma_t = j_0 | \gamma_{t-1}=j_1, \dots, \gamma_{t-P} = j_{P}, \cdot) \\
        &=\sum_{j_{P}=1}^{M} \dots \sum_{j_2=1}^{M} \sum_{j_0=1}^{M}
        p(\gamma_{t-1}=j_1, \dots, \gamma_{t-P} = j_{P}) \, p(\bm{y}_{t:T} | \gamma_t = j_0, \cdot)\\
        &= \sum_{j_P=1}^{M} \dots \sum_{j_2=1}^{M} \sum_{j_0=1}^{M} \eta_{\{\,j_P, \dots, j_1, j_0\}} \, p(\bm{y}_t | \gamma_t = j_0, \bm{\mu}, \bm{\Omega}) \, \beta_{t+1}(j_0).
\end{align*}

\section{Forward Messages for Local Decoding} \label{sec:S_forward}

The \textit{forward messages} $\alpha_t(j_1) = p
\,(\bm{y}_{1:t-1}, \gamma_{t-1} = j_1 \,|\, \cdot)$ utilized to conduct local decoding (Section \ref{sec:inference}) are defined as 
\begin{equation}
\alpha_t(j_1)  = \sum_{j_2 =1}^{\hat{M}} \dots  \sum_{j_{\hat{P}}=1}^{\hat{M}} \alpha_t(j_1, j_2,  \dots, j_{\hat{P}}), 
\end{equation}
for $j_l \in \{1, \dots, \hat{M}\}$, $l = 1, \dots {\hat{P}}$, and the DAR-\textit{forward messages}  $\alpha_t(j_1, \dots, j_{\hat{P}}) = p
\,(\bm{y}_{1:t-1}, \gamma_{t-1} = j_1,\dots, \gamma_{t-\hat{P}} = j_{\hat{P}} \,|\, \cdot)$ are described as the probability of the partial observations sequence $\bm{y}_{1:t-1}$, and states $\gamma_{t-1:t-\hat{P}}$ at time $t-1$, given all the other parameters. These messages can be recursively computed as follows
\begin{equation}
    \alpha_t(j_1, \dots, j_{\hat{P}}) = \sum_{j_{\hat{P}+1}=1}^{\hat{M}} \eta_{\{\,j_{\hat{P}+1}, \dots, j_2, j_1\}} \, p (\bm{y}_{t-1} | \gamma_{t-1}=j_{1},\bm{\mu}, \bm{\Omega}) \, \alpha_{t-1}(j_2, \dots, j_{\hat{P}+1}), 
    \label{eq:DAR-forward_msg}
\end{equation} 
as shown in the following Proposition. 

\noindent \textbf{Proposition 3.} Let  $\alpha_t(j_1, \dots, j_{\hat{P}}) = p
\,(\bm{y}_{1:t-1}, \gamma_{t-1} = j_1,\dots, \gamma_{t-\hat{P}} = j_{\hat{P}} \,|\, \cdot)$ be the  DAR-\textit{forward messages}, described as the probability of the partial observations sequence $\bm{y}_{1:t-1}$, and states $\gamma_{t-1:t-\hat{P}}$ at time $t-1$, given all the other parameters. Then, these messages can be recursively computed as  in Eq. \eqref{eq:DAR-forward_msg}.

\noindent  \textit{Proof:}  Let $M = \hat{M}$, and let $P = \hat{P}$ be the number of active DAR parameters. Then the proof proceeds as follows 
\begin{align*}
         \alpha_t(j_1, \dots, j_{P}) &= 
p\,(\bm{y}_{1:t-1}, \gamma_{t-1} = j_1, \dots \gamma_{t-\hat{P}} = j_{P} \,|\, \cdot) \\
&= \sum_{j_{P+1} = 1}^{M} p\,(\bm{y}_{1:t-1}, \gamma_{t-1} = j_1, \dots, \gamma_{t-(P+1)} = j_{P+1} \,|\, \cdot) \\ 
&= \sum_{j_{P+1} = 1}^{M} p (\bm{y}_{1:t-2}, \gamma_{t-2} = j_2, \dots,  \gamma_{t-(P+1)} = j_{P+1}) \,  p (\gamma_{t-1} = j_1 | \gamma_{t-2} = j_2, \dots, \gamma_{t-(P+1)} = j_{P+1}) \\[-1em]
& \qquad \qquad p (\bm{y}_{t-1} | \gamma_{t-1}=j, \bm{\mu}, \bm{\Omega}) \\
&= \sum_{j_{P+1} = 1}^{M}  \eta_{\{\,j_{P+1}, \dots, j_1 \}}  p (\bm{y}_{t-1} | \gamma_{t-1}=j, \bm{\mu}, \bm{\Omega}) \alpha_{t-1}(j_2, \dots, j_{P+1}).
\end{align*}

\section{Additional Results} \label{sec:suppl_additional}

In this section, we provide supplementary results for the simulation study presented in Section \ref{sec:simulation_studies} and the application to fMRI data discussed in Section \ref{sec:application}. 

\subsection{Simulation Study under Time-Varying Means}

\begin{figure}[htbp]
    \centering
    \includegraphics[width=0.5\linewidth]{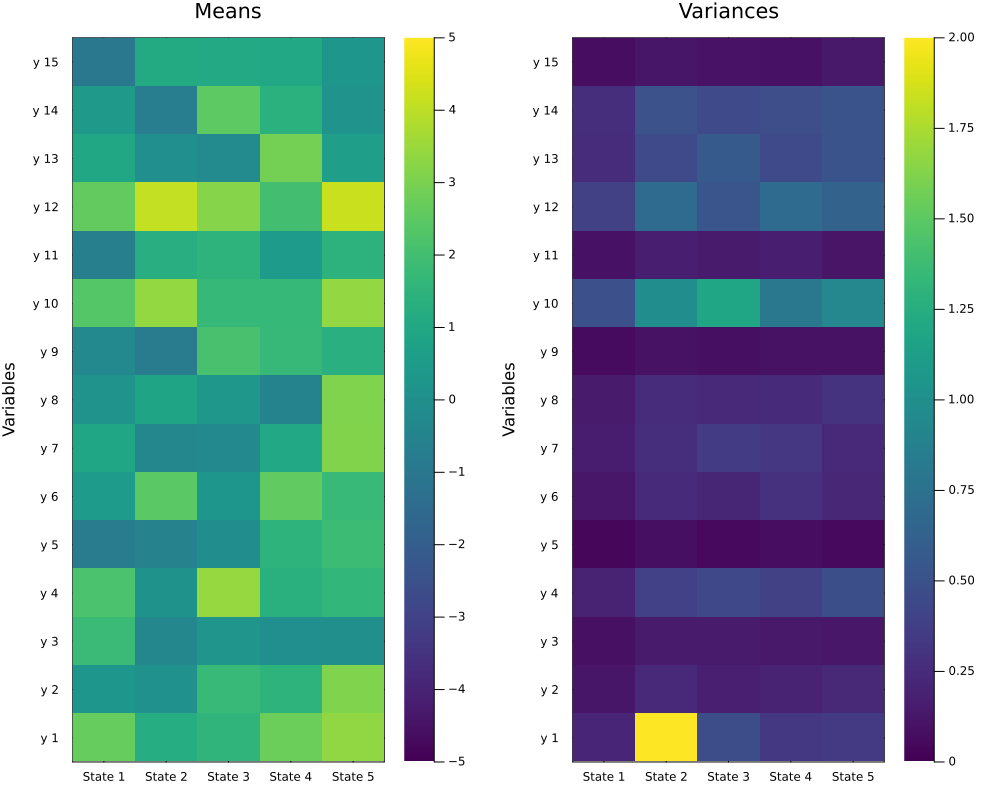}
    \caption{\textbf{Simulation Study.} State-specific mean (left) and variance (right) values across different dimensions. }
    \label{fig:mean_variance_ILLUSTRATIVE}
\end{figure}

As part of the simulation setting in Section~\ref{sec:simulation_studies}, beyond analyzing the estimated state-specific connectivity matrices, it is also informative to examine the state-wise mean and variance values across different dimensions.Figure \ref{fig:mean_variance_ILLUSTRATIVE} presents the estimated means (left panel) and variances (right panel) for each state and each variable. The heatmap of mean values reveals distinct patterns across states, with certain variables exhibiting strong shifts in activation levels between states. This suggests that the identified states effectively capture meaningful variations in the underlying data distribution. On the other hand, the variance heatmap indicates that some dimensions exhibit higher variability in specific states, potentially reflecting differences in signal stability or noise contributions across brain regions. Notably, variable \( y_1 \) in State 2 shows a particularly high variance, which may correspond to transient fluctuations in activity or an outlier effect.

\subsection{Application to fMRI Data}

\begin{figure}[htbp]
    \centering
    \includegraphics[width=0.5\linewidth]{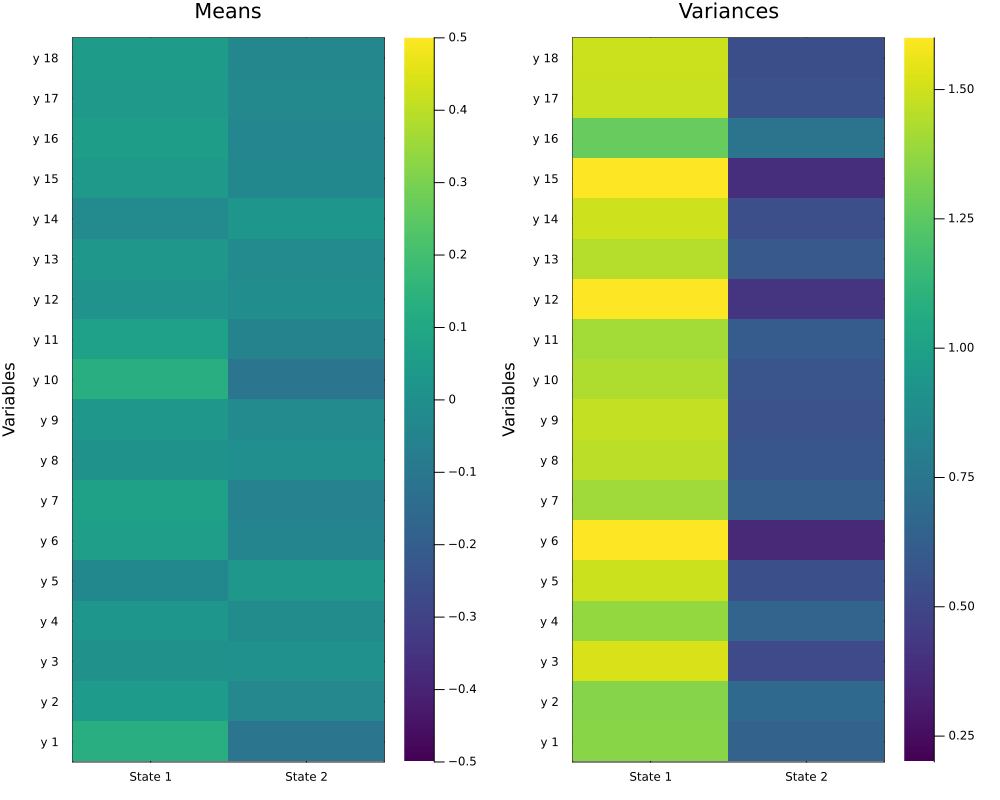}
    \caption{\textbf{Application to FMRI data.} State-specific mean (left) and variance (right) values across different dimensions. }
    \label{fig:mean_variance_realData}
\end{figure}

As part of the fMRI analysis in Section~\ref{sec:application}, we report state-specific mean and variance estimates across dimensions in Figure~\ref{fig:mean_variance_realData}. The left panel of the figure shows the estimated means, which remain relatively similar across the two states. In contrast, the right panel illustrates the variance estimates, where State 1 generally exhibits higher variability compared to State 2. This suggests that while the overall activation levels remain stable, the fluctuations in neural activity differ between states.

\begin{figure}[htbp]
    \centering
    \begin{subfigure}{0.35\textwidth} 
        \centering
        \includegraphics[width=\textwidth]{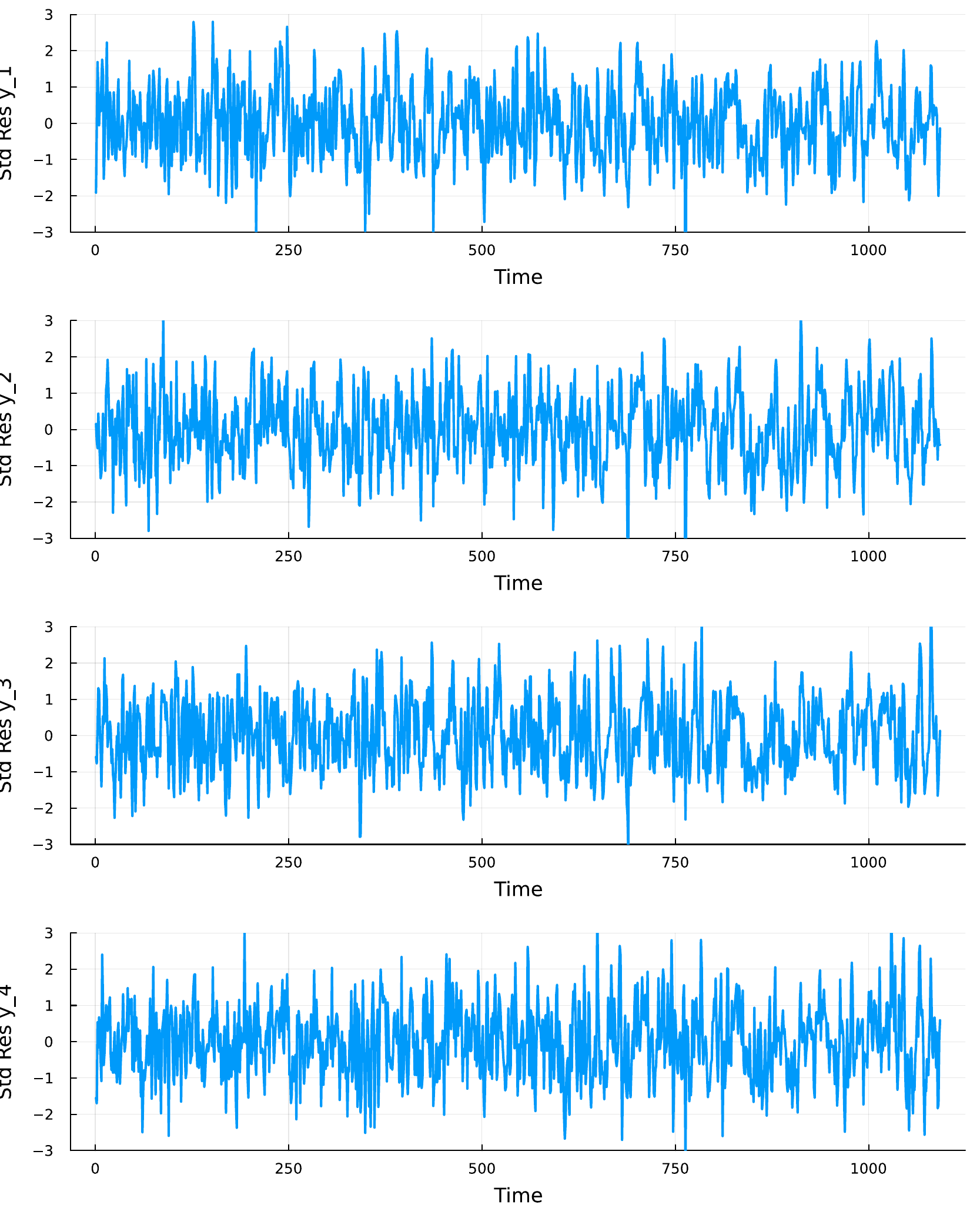}
        \caption{}
        \label{fig:a}
    \end{subfigure}
    \hspace{1em} 
    \begin{subfigure}{0.35\textwidth} 
        \centering
        \includegraphics[width=\textwidth]{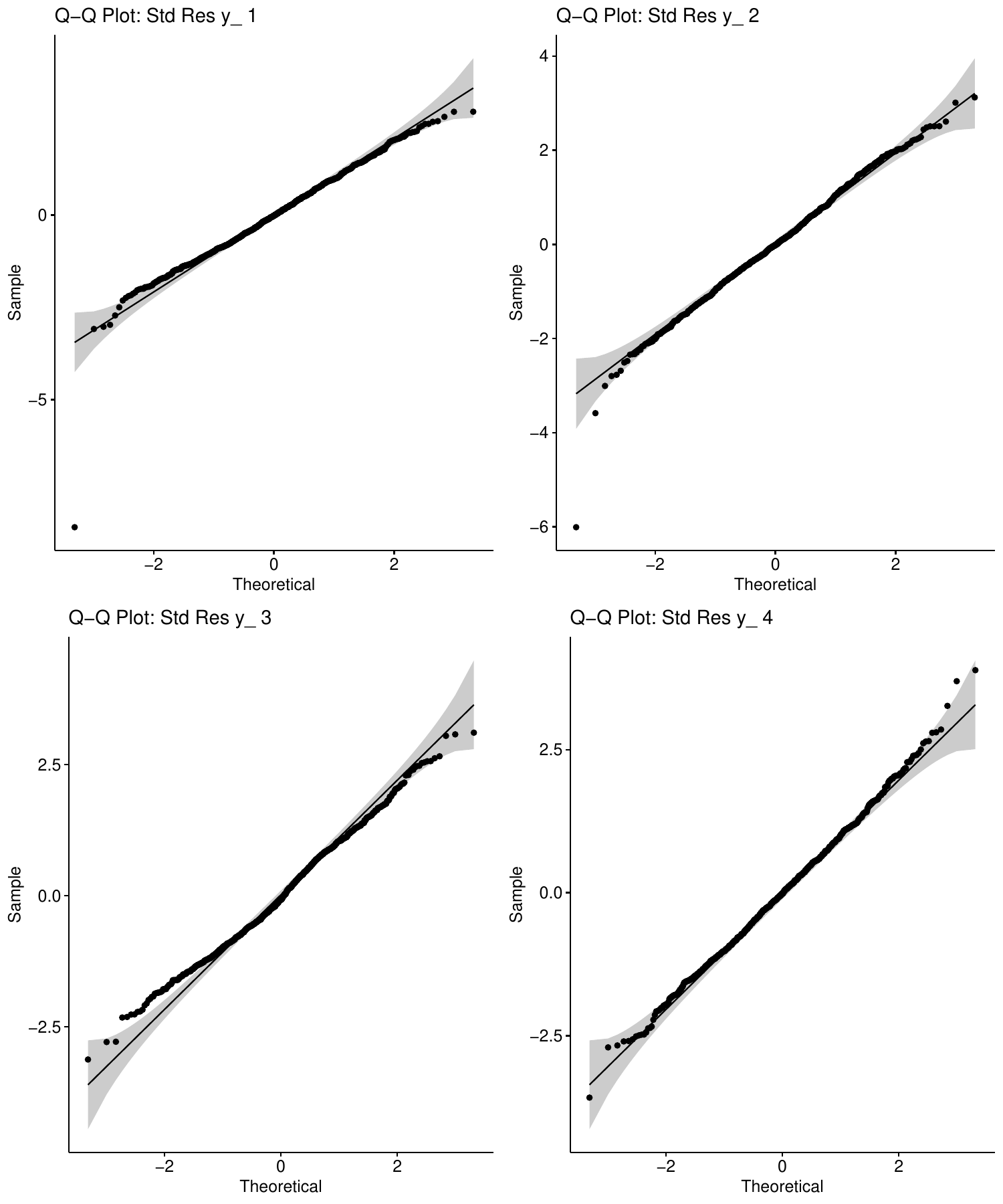}
        \caption{}
        \label{fig:b}
    \end{subfigure}

    \caption{\textbf{Application to fMRI Data}. (a) Standardized residuals; (b) Q-Q plots of standardized residuals. First four dimensions, representing representative brain regions.}
    \label{fig:residual_realData}
\end{figure}

To investigate the accuracy of the proposed approach in modeling the fMRI data, we display in Figure \ref{fig:residual_realData} standardized residuals and corresponding QQ-plots for four representative brain regions. The standardized residuals (left panel) show no obvious pattern, indicating a good fit, while the QQ-plots (right panel) suggest that the residuals closely follow a normal distribution, with only minor deviations in the tails. These results confirm that the model effectively captures the underlying structure of the data.

\begin{figure}[htbp]
    \centering
    \includegraphics[width=0.5\linewidth]{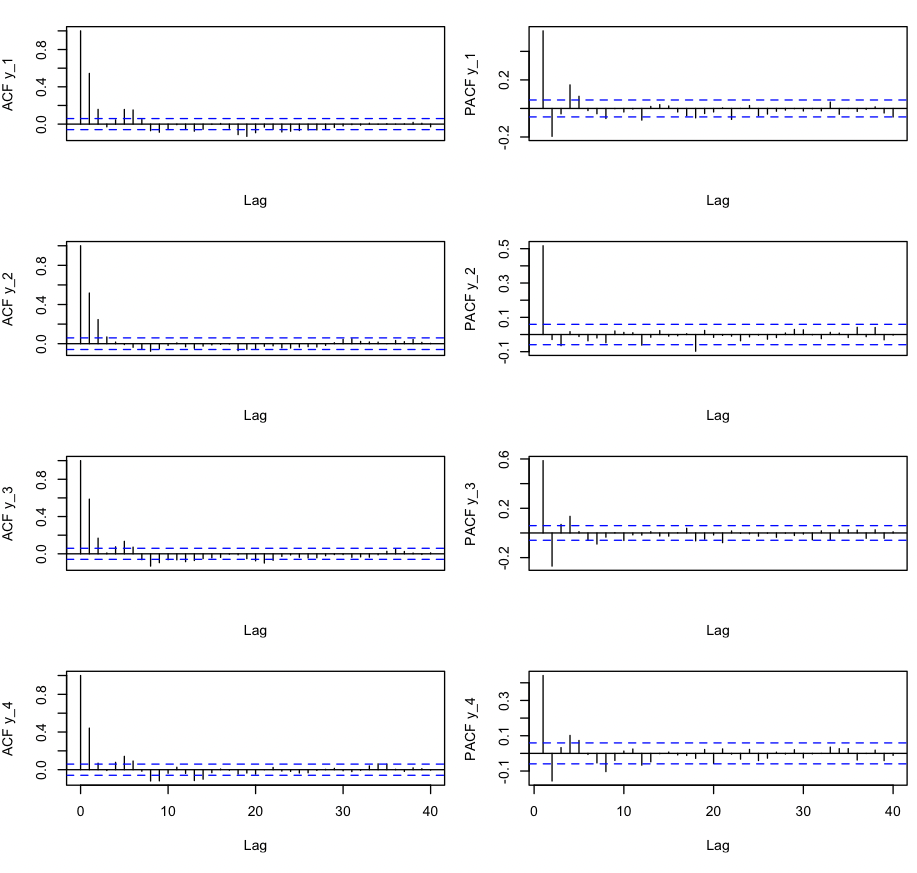}
\caption{\textbf{Application to fMRI Data.} ACF and PACF plots for four representative brain regions, illustrating the temporal dependencies in the observed time series.}

    \label{fig:PACF_ACF_realData}
\end{figure}

Finally, we display ACF and PACF plots for the same four representative brain regions in Figure \ref{fig:PACF_ACF_realData}. These plots provide insight into the temporal structure of the observed fMRI time series, highlighting short-term dependencies that may not always be fully captured by the proposed model. While our approach assumes a discrete autoregressive structure for the evolution of the latent states and independent Gaussian emissions at the observational level, its primary goal is to characterize connectivity patterns through the state-specific precision matrices. Future work could explore extensions that incorporate additional temporal dependencies while preserving the interpretability of the inferred functional connectivity.

\section{Simulated scenario with zero-mean observations} 
\label{sec:S_simulzeromean}

\begin{figure}[htbp]
	\centering
\includegraphics[width=0.7\linewidth]{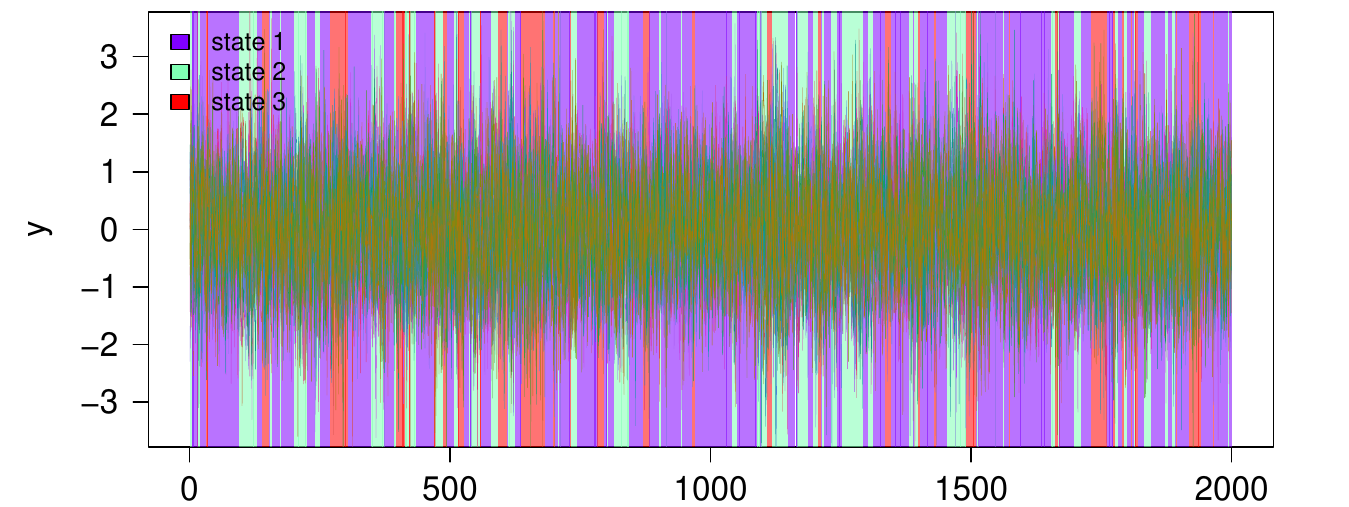}

\includegraphics[width=0.7\linewidth]{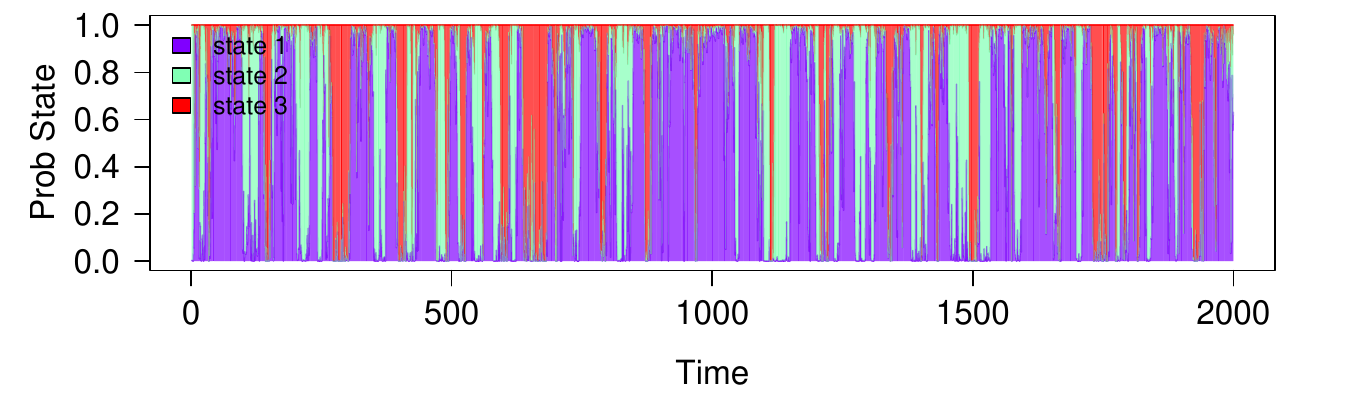}
	\caption{\textbf{Simulation with zero mean}. (top) time series realization (lines) where each dimension is represented by a different colored line;  vertical colored bands represent the true underlying state sequence; (bottom) estimated time-varying probability plot.  }

	\label{fig:data_and_predictive_Zeromean}
\end{figure}

Here, we investigate the performance of our approach when the data-generating emissions are zero-mean, i.e. $\bm{\mu}_j = \bm{0}$, for $j = 1, \dots, M$. We generated 30 distinct dataset, each consisting of $D=15$-dimensional $T=2000$, in a similar fashion as in Section \ref{sec:study_structured_precision}. We assumed $M=3$ states and DAR order $P=2$, where the autoregressive probabilities and innovations were specified as $\bm{\phi} =(0.2, 0.5, 0.3)$ and $\bm{\pi} = (0.5, 0.3, 0.2)$. The precision matrices were constructed using patterns  (i), (iv) and (v) from Section \ref{sec:study_structured_precision}. A single realization from this simulation setting is shown in Figure \ref{fig:data_and_predictive_Zeromean} (top panel), where vertical colored bands represent the true underlying state sequence. We chose $M_{max} = 6$, and we set the rest of the  hyperparameters as in Section \ref{sec:study_structured_precision}. 
 


Our approach consistently estimated  the correct number of states  $\hat{M} = 3$ as the mode of the posterior distribution and the number of active DAR probabilities $\hat{P} = 2$ with high posterior probability, on all simulated replicates. Figure \ref{fig:data_and_predictive_Zeromean} (bottom panel) displays a time-varying probability plot, namely the local decoding of the hidden state at time $t$, $p (\gamma_t = j \,| \,\bm{y}, \cdot)$, $j = 1, \dots, \hat{M}$, as described in Section \ref{sec:inference};  these plots are constructed by plotting the local probabilities  (which add to 1) cumulatively for each $t$, where each
state is associated with a different color. It is evident that our proposed approach correctly retrieves the true latent state sequence. We assessed the model selection performance of our approach in Table \ref{tab:accuracy_etc_zeromean}, showing accuracy, sensitivity, specificity, $F1$-score and Matthew correlation coefficient (MCC).  To evaluate estimation accuracy we report RMSE of the state-specific off-diagonal entries of the precision matrices. As in Section \ref{sec:simulation_studies}, our proposed methodology is compared with \texttt{mvHMM} and \texttt{glassoSlide}. These  results from our proposed approach are  conditioned on the modal number of states and autoregressive order. As in the investigation carried out in Section \ref{sec:study_structured_precision}, our approach seems to be superior to  \texttt{mvHMM} and \texttt{glassoSlide}, for what concerns both estimation accuracy and model selection performances.

\begin{table}[htbp]
\small
\centering
\begin{tabular}{clccc} \hspace{-0.2cm}\\
\multicolumn{1}{l}{}  &             & Identity      & AR(2)         & Random        \\ \cmidrule{2-5}
\multirow{3}{*}{Acc}  & \texttt{sggmDAR}    & 1.0 (0.0)     & 0.994 (0.007) & 0.987 (0.011) \\
                      & \texttt{mvHMM}       & 0.929 (0.072) & 0.953 (0.026) & 0.933 (0.057) \\
                      & \texttt{glassoSlide} & 0.999 (0.004) & 0.837 (0.049) & 0.869 (0.069) \\ \cmidrule{3-5}
\multirow{3}{*}{Spec} & \texttt{sggmDAR}    & 1.0 (0.0)     & 0.992 (0.01)  & 0.996 (0.006) \\
                      & \texttt{mvHMM}       & 0.929 (0.072) & 0.936 (0.035) & 0.947 (0.039) \\
                      & \texttt{glassoSlide} & 0.999 (0.004) & 0.916 (0.062) & 0.936 (0.047) \\ \cmidrule{3-5}
\multirow{3}{*}{MCC}  & \texttt{sggmDAR}    & -             & 0.985 (0.018) & 0.959 (0.034) \\
                      & \texttt{mvHMM}       & -             & 0.892 (0.055) & 0.865 (0.076) \\
                      & \texttt{glassoSlide} & -             & 0.564 (0.126) & 0.580 (0.267) \\ \cmidrule{3-5}
\multirow{3}{*}{F1}   & \texttt{sggmDAR}    & -     & 0.989 (0.013) & 0.967 (0.029) \\
                      & \texttt{mvHMM}       & -    & 0.918 (0.042) & 0.802 (0.278) \\
                      & \texttt{glassoSlide} & -     & 0.652 (0.092) & 0.646 (0.231) \\ \cmidrule{3-5}
\multirow{3}{*}{Sens} & \texttt{sggmDAR}    & -     & 0.999 (0.007)     & 0.949 (0.051) \\
                      & \texttt{mvHMM}       & -             & 1.0 (0.0)     & 0.877 (0.300)   \\
                      & \texttt{glassoSlide} & -     & 0.607 (0.153) & 0.617 (0.248) \\ \cmidrule{3-5}
\multirow{3}{*}{RMSE} & \texttt{sggmDAR}    & 0.003 (0.002) & 0.028 (0.005) & 0.030 (0.005) \\
                      & \texttt{mvHMM}       & 0.039 (0.012) & 0.045 (0.006) & 0.069 (0.036) \\
                      & \texttt{glassoSlide} & 0.0 (0.001)   & 0.165 (0.019) & 0.111 (0.019) \\ \hline
\end{tabular}
\caption{\textbf{Simulation with zero-mean.} Accuracy, sensitivity, specificity, F1 score, Matthew correlation coefficient (MCC), and residual mean squared error (RMSE) of precision matrix estimates, for each regime $j=1, \dots, \hat{M}$. Standard deviation over the 30 simulations are displayed in brackets. Results are reported for our \texttt{sggmDAR} , \texttt{mvHMM} and \texttt{glassoSlide}. The results of our approach are conditioned on the modal number of states and autoregressive order. A hyphen is used for those metrics that cannot be computed due to the structure of the underlying truth (e.g. TP+FN = 0).} 
\label{tab:accuracy_etc_zeromean}
\end{table}
\normalsize

\section{Large $D$ setting} \label{sec:S_largedsetting}

Here, we explore the performance of our approach in a scenario where the dimension $D$ of the  data is large, as discussed in Section \ref{sec:high_dim}. We focus on assessing the ability of our proposed method in recovering the number of states, number of DAR parameters, and true sparse precision matrices.  Table \ref{tab:large_d_setting} displays  model selection and estimation accuracy performances for the off-diagonal component of the high-dimensional precision matrices, for \texttt{sggmDAR}, \texttt{mvHMM}, and \texttt{glassoSlide}. 
The MCC scores  highlight the advantage of choosing our proposed method in high-dimensional settings. Indeed, the number of parameters for each individual state is substantial, as there are 4950 distinct off-diagonal coefficients to be inferred for each precision matrix.

 We report  model selection and estimation accuracy performances for the off-diagonal component of the high-dimensional precision matrices, for \texttt{sggmDAR}, \texttt{mvHMM}, and \texttt{glassoSlide}. 
The MCC scores  highlight the advantage of choosing our proposed method in high-dimensional settings. Indeed, the number of parameters for each individual state is substantial, as there are 4950 distinct off-diagonal coefficients to be inferred for each precision matrix.  

\begin{table}[htbp]
\centering
\footnotesize
\begin{tabular}{clccc}
\hspace{-0.2cm}\\ \hline
\multicolumn{1}{l}{}  &             & Identity      & Hub         & Random        \\ \cmidrule{2-5}
\multirow{3}{*}{Acc}  & \texttt{sggmDAR}     & 1.0 (0.0)     & 1.0 (0.0)     & 0.997 (0.001) \\
                      & \texttt{mvHMM}       & 0.923 (0.017) & 0.862 (0.045) & 0.861 (0.054) \\
                      & \texttt{glassoSlide} & 0.997 (0.008) & 0.963 (0.029) & 0.932 (0.025) \\ \cmidrule{3-5}
\multirow{3}{*}{Spec} & \texttt{sggmDAR}     & 1.0 (0.0)     & 1.0 (0.0)     & 1.0 (0.0)     \\
                      & \texttt{mvHMM}       & 0.923 (0.017) & 0.861 (0.043) & 0.861 (0.050) \\ 
                      & glassoSlide & 0.997 (0.008) & 0.981 (0.030) & 0.959 (0.027) \\ \cmidrule{3-5}
\multirow{3}{*}{MCC}  & \texttt{sggmDAR}     & -             & 0.999 (0.002) & 0.952 (0.02)  \\
                      & \texttt{mvHMM}       & -             & 0.301 (0.099) & 0.352 (0.148) \\
                      & \texttt{glassoSlide} & -             & 0.006 (0.063) & 0.015 (0.020) \\ \cmidrule{3-5}
\multirow{3}{*}{F1}   & \texttt{sggmDAR}     & -     & 0.999 (0.002) & 0.953 (0.02)  \\
                      & \texttt{mvHMM}       & -     & 0.214 (0.064) & 0.293 (0.104) \\
                      & \texttt{glassoSlide} & -     & 0.013 (0.033) & 0.044 (0.022) \\ \cmidrule{3-5}
\multirow{3}{*}{Sens} & \texttt{sggmDAR}     & -     & 1.0 (0.0)     & 0.911 (0.036) \\
                      & \texttt{mvHMM}       & -             & 0.903 (0.209) & 0.862 (0.256) \\
                      & \texttt{glassoSlide} & -      & 0.018 (0.061)        & 0.057 (0.043) \\ \cmidrule{3-5}
\multirow{3}{*}{RMSE} & \texttt{sggmDAR}     & 0.001 (0.0)   & 0.004 (0.001) & 0.008 (0.001) \\
                      & \texttt{mvHMM}       & 0.031 (0.003) & 0.063 (0.018) & 0.067 (0.025) \\
                      & \texttt{glassoSlide} & 0.001 (0.002) & 0.028 (0.001) & 0.059 (0.002) \\ \hline
\end{tabular} 
\caption{{\bf Simulation Study.} Large $D$ setting.  Accuracy, sensitivity, specificity, F1 score, Matthew correlation coefficient (MCC), and residual mean squared error (RMSE) of precision matrix estimates, for each regime $j=1, \dots, \hat{M}$. Standard deviation over the 30 simulations are displayed in brackets. Results are reported for our \texttt{sggmDAR} , \texttt{mvHMM} and \texttt{glassoSlide}. The results of our approach are conditioned on the modal number of states and autoregressive order. A hyphen is used for those metrics that cannot be computed due to the structure of the underlying truth (e.g. TP+FN = 0).}
\label{tab:large_d_setting}
\end{table}

\section{Misspecified Model} \label{sec:suppl:misspecified}
 We evaluate the performance of our proposed approach under model misspecification. Specifically, we generate data from a stationary vector autoregressive (VAR) process of order 1, defined as  

\begin{equation*}
\bm{y}_t = \bm{\mu} +  \bm{A} \bm{y}_{t-1} + \bm{\varepsilon}_t, \qquad \bm{\varepsilon}_t \sim N_D(\bm{0}, \Sigma_\varepsilon),
\end{equation*}

where \(\bm{y}_t \in \mathbb{R}^D\) represents the \(D\)-dimensional time series at time \(t\), and \(\bm{\mu} \in \mathbb{R}^D\) is an intercept term that captures the mean level of the process. Here, we generated \(T = 500\) observations with dimensionality \(D = 5\). The matrix \(\bm{A} \in \mathbb{R}^{D \times D}\) contains the autoregressive coefficients, governing both the temporal dependencies within each variable and the interactions across variables. The innovation term \(\bm{\varepsilon}_t\) follows a multivariate normal distribution with zero mean and covariance matrix \(\Sigma_\varepsilon\), capturing the variability and cross-correlations in the residual errors.  To ensure stationarity, we assume that the eigenvalues of \(\bm{A}\) lie within the unit circle, i.e., all eigenvalues of \(\bm{A}\) have modulus strictly less than one. This condition prevents explosive behavior and ensures that past influences decay over time, leading to a stable time series \citep{lutkepohl2005new}.  We consider two scenarios for the autoregressive coefficient matrix \(\bm{A}\): (i) a sparse VAR structure, where only a small subset of entries in \(\bm{A}\) are nonzero, reflecting limited and localized interactions among variables. Such sparsity is characteristic of high-dimensional applications like functional MRI (fMRI) data, where neuronal activity exhibits structured but selective dependencies. (ii) A dense VAR structure, where most entries in \(\bm{A}\) are nonzero, representing a system with widespread dependencies across variables. This setting corresponds to scenarios where all variables exert direct or indirect influences on each other, such as in economic or climatological models. By analyzing both cases, we assess the robustness of our approach under varying levels of dependency structure, from highly localized interactions to fully connected systems.

\vspace{0.1cm}

\begin{figure}[htbp]
    \centering
    \begin{subfigure}{0.32\textwidth}
        \includegraphics[width=\textwidth]{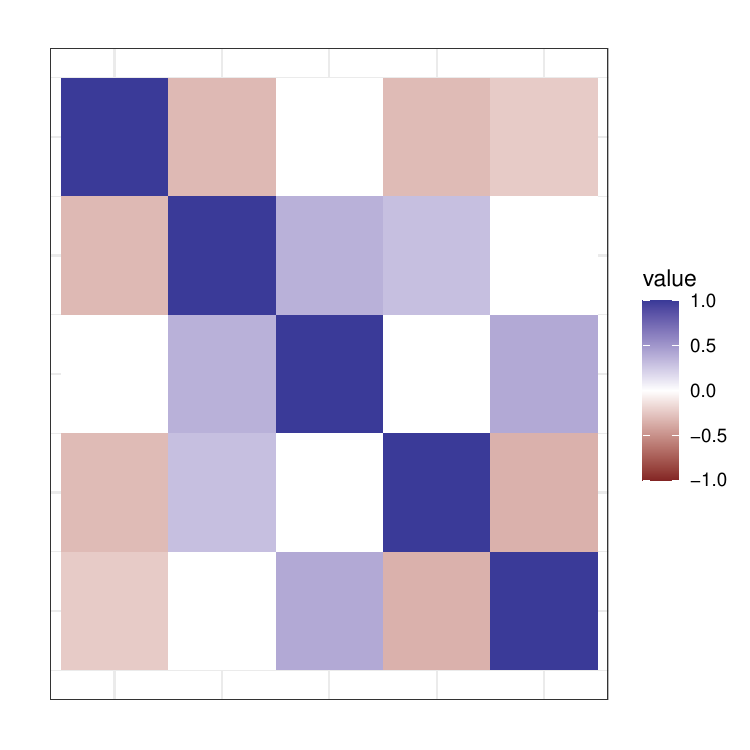}
        \caption{ }
        \label{fig:a}
    \end{subfigure}
    \hfill
    \begin{subfigure}{0.32\textwidth}
        \includegraphics[width=\textwidth]{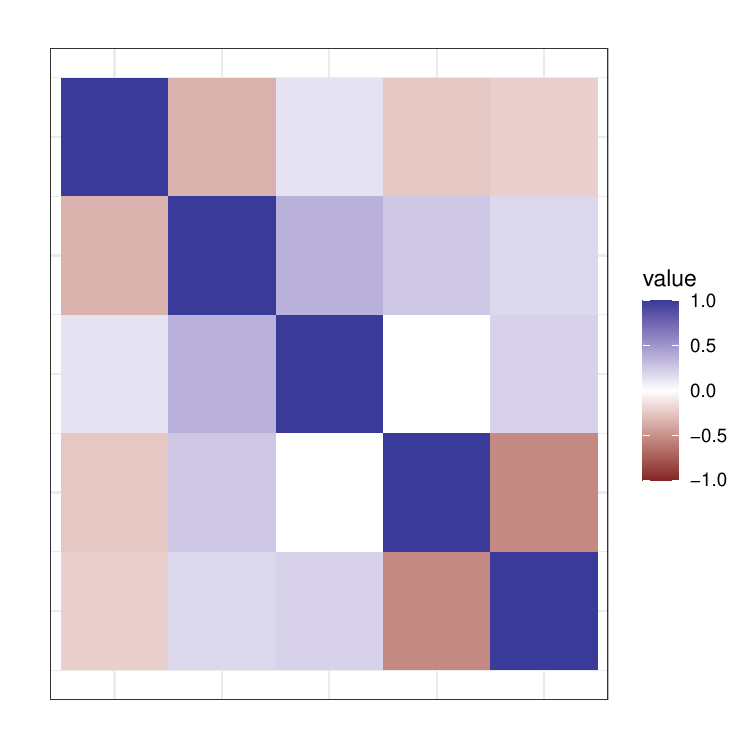}
        \caption{}
        \label{fig:b}
    \end{subfigure}
    \hfill
    \begin{subfigure}{0.32\textwidth}
        \includegraphics[width=\textwidth]{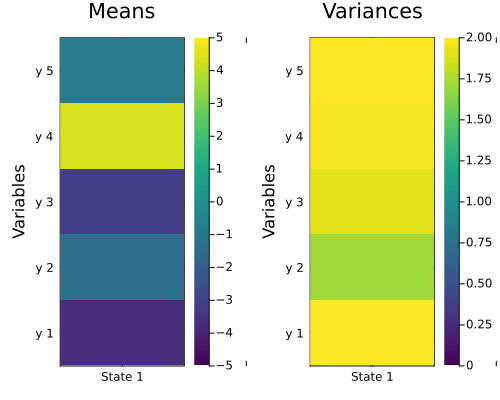}
        \caption{}
        \label{fig:c}
    \end{subfigure}
    
    \vspace{0.5cm}
    
    \begin{subfigure}{0.49\textwidth}
        \includegraphics[width=\textwidth]{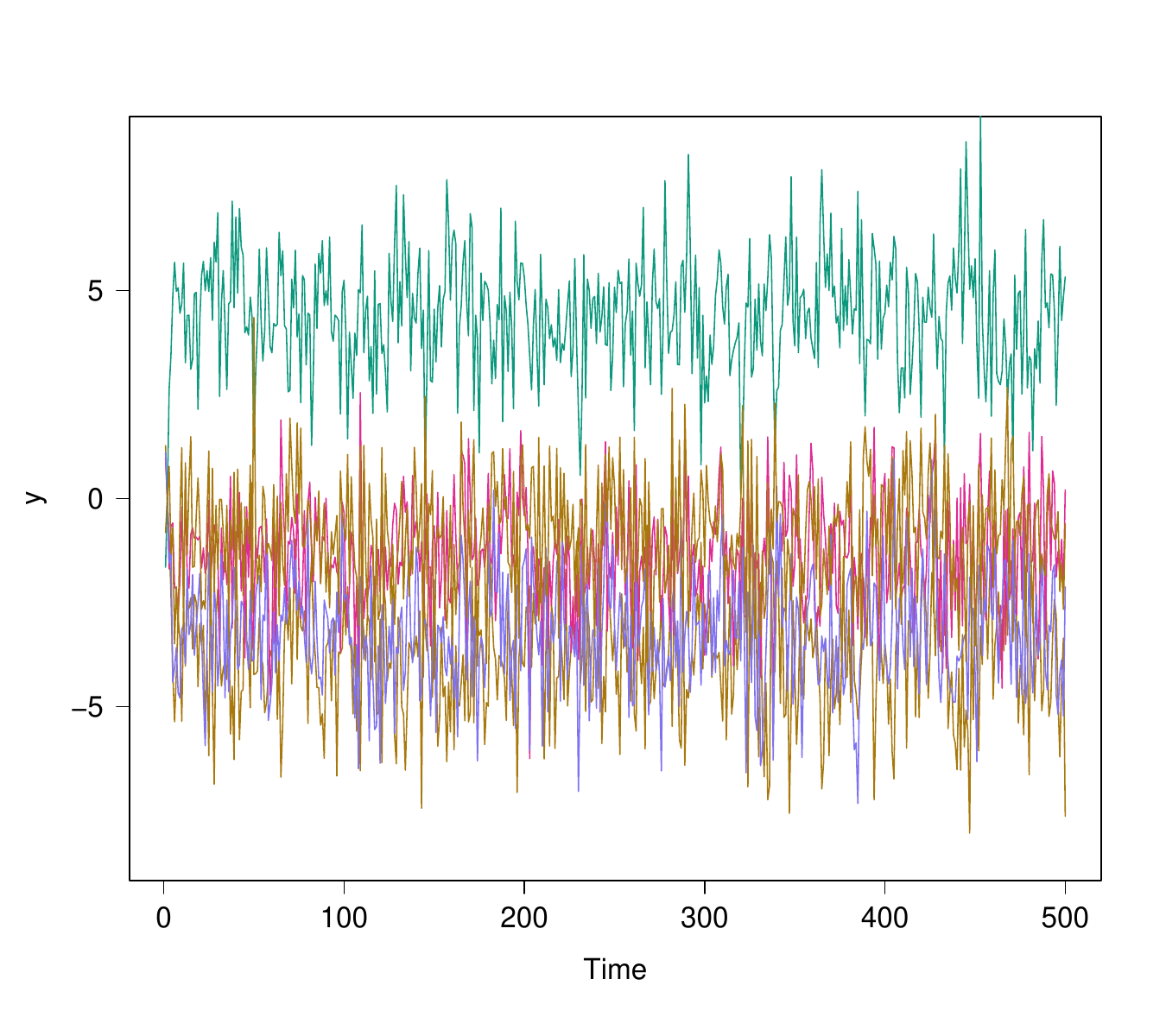}
        \caption{}
        \label{fig:d}
    \end{subfigure}
    \hfill
    \begin{subfigure}{0.49\textwidth}
        \includegraphics[width=\textwidth]{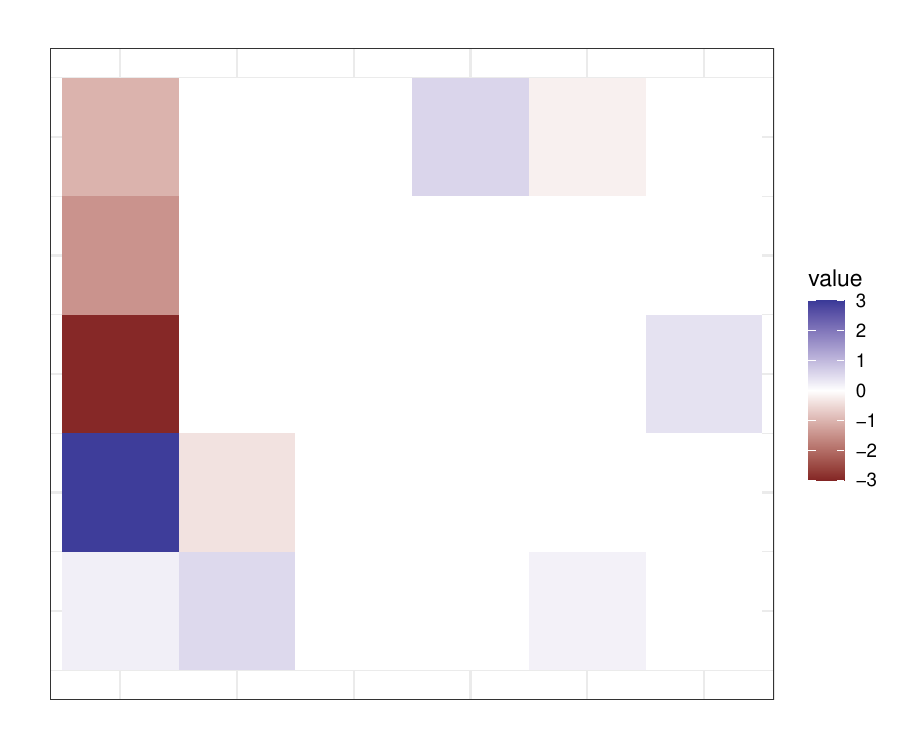}
        \caption{}
        \label{fig:e}
    \end{subfigure}

    \caption{\textbf{Misspecified Model - Sparse VAR.} (a) true partial correlation matrix; (b) estimated partial correlation matrix; (c) estimated mean and variances; (d) time series realization; (e) generated VAR matrix. }
    \label{fig:all_sparseVAR}
\end{figure}

\textbf{\textit{Sparse Autoregressive Matrix}}: The autoregressive coefficient matrix \(\bm{A}\) is constructed to be sparse, with a sparsity level of \(\rho = 0.7\), meaning that only 30\% of its entries are nonzero. The nonzero coefficients are drawn uniformly from the interval \([-0.8, -0.1] \cup [0.1, 0.8]\), ensuring a mix of positive and negative dependencies while avoiding near-zero values. The intercept terms \(\bm{\mu}\) are sampled from a uniform distribution on \([-3,3]\), while the covariance matrix \(\bm{\Sigma}_\varepsilon\) is randomly generated as described in Section \ref{sec:study_structured_precision}.  A realization of the generated time series is shown in Figure \ref{fig:all_sparseVAR}(d), illustrating the dynamical behavior induced by the sparse structure. Figure \ref{fig:all_sparseVAR}(a) and Figure \ref{fig:all_sparseVAR}(e) display the corresponding partial correlation matrix  and the sparse autoregressive coefficient matrix, respectively. Notably, the first column of \(\bm{A}\) incorporates the intercept \(\bm{\mu}\).

We apply our proposed sampler with a maximum allowable number of states set to \(M_{\max} = 6\), while keeping the remaining parameter settings as described in Section \ref{sec:simulation_studies}. The posterior distribution for the number of states indicates a preference for a single regime, with \(P(M = 1 \mid \cdot) = 0.65\) and \(P(M = 2 \mid \cdot) = 0.35\). Despite the model misspecification, our approach correctly identifies a dominant regime, yielding an estimated mode at \(\hat{M} = 1\).   Figure \ref{sec:study_structured_precision}(b) displays the estimated partial correlation matrix, demonstrating strong agreement with the true generating structure. Additionally, Figure \ref{sec:study_structured_precision}(c) presents the estimated means and variances for each dimension, further supporting the accuracy of our inference.  These results highlight the robustness of our method in capturing the underlying dependency structure, even when the data-generating process deviates from the assumed model. Notably, in the case of a sparse autoregressive matrix, the method predominantly identifies a single latent state, suggesting that sparsity in the autoregressive structure naturally reduces regime complexity.

\begin{figure}[htbp]
	\centering
\includegraphics[width=0.7\linewidth]{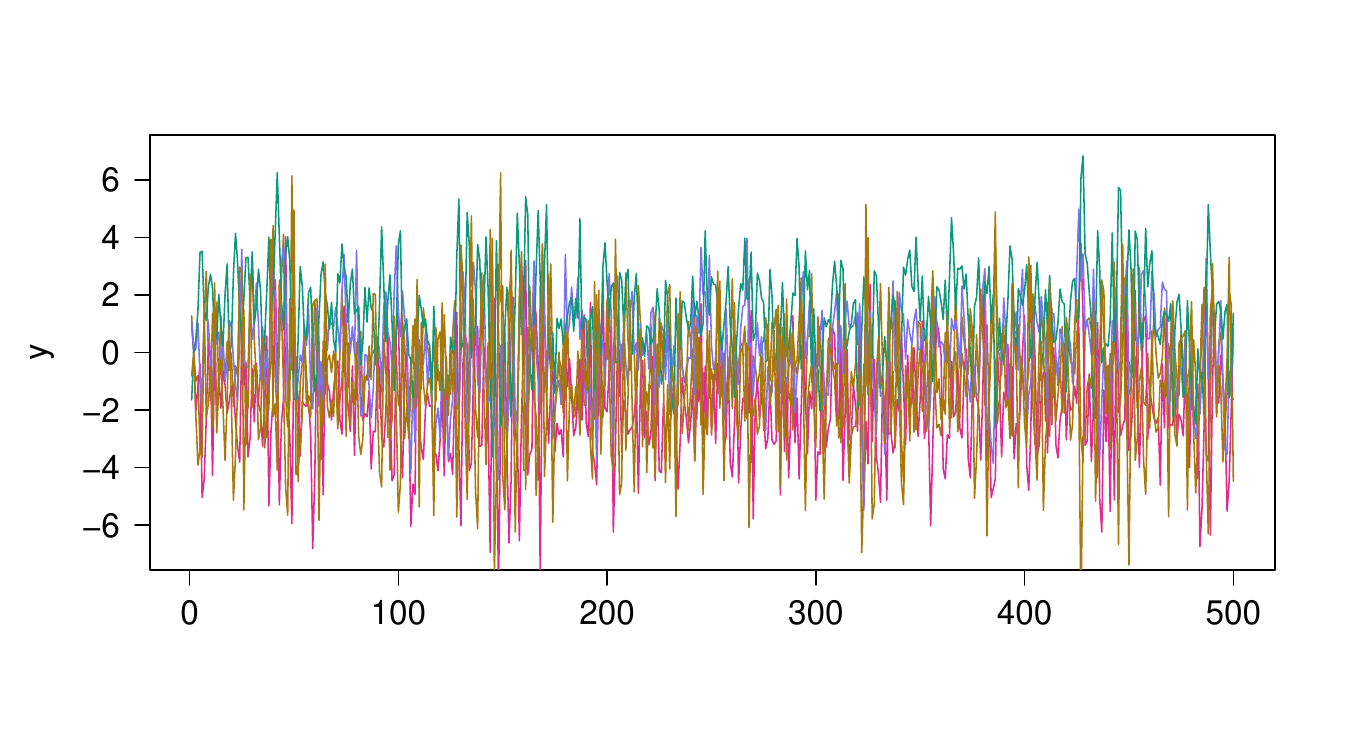}

\includegraphics[width=0.7\linewidth]{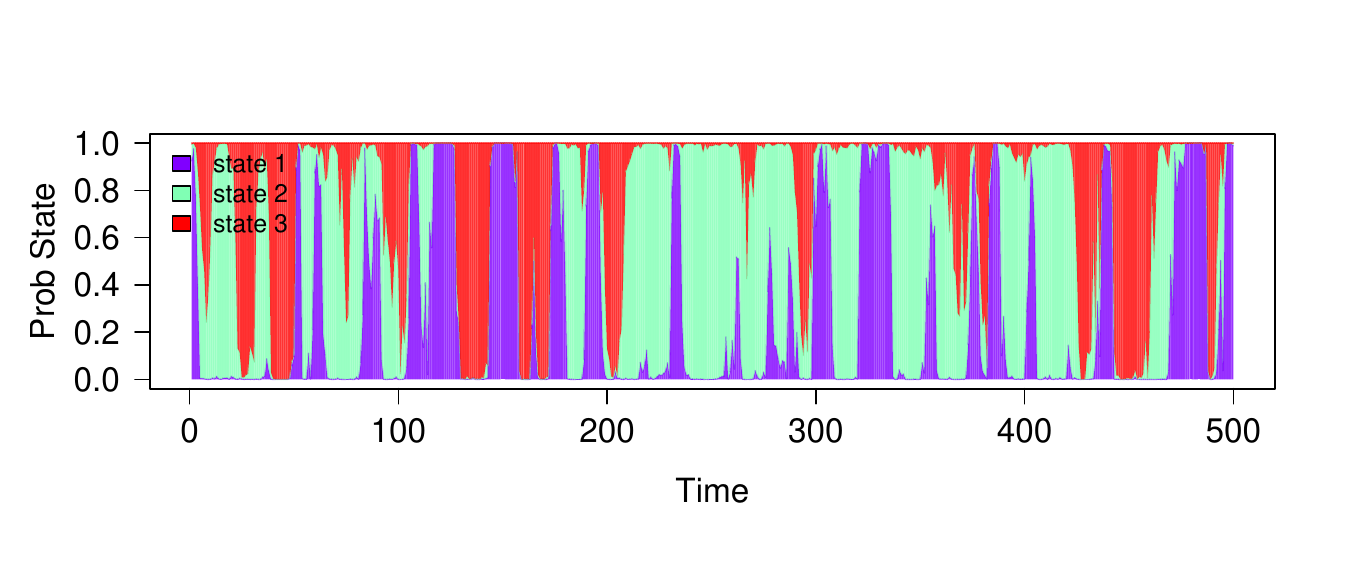}
	\caption{\textbf{Misspecified Model - Dense VAR}. (top) time-series realization (lines) where each dimension is represented by a different colored line;  (bottom) estimated time-varying probability plot. }

	\label{fig:data_and_state_probs_DENSE}
\end{figure}

\begin{figure}[htbp]
    \centering
    \begin{subfigure}{0.8\textwidth}
        \centering
        \includegraphics[width=\textwidth]{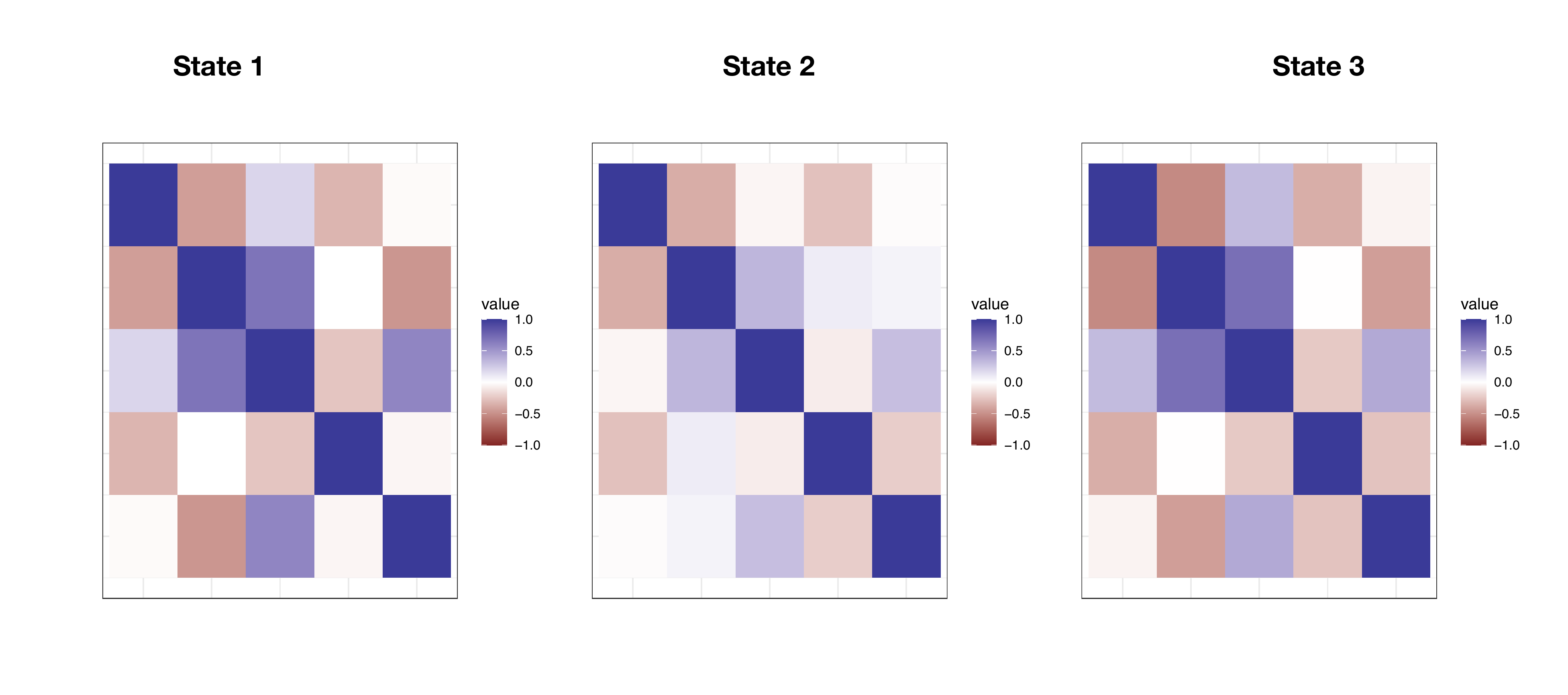}
        \caption{}
        \label{fig:a}
    \end{subfigure}
    
    \vspace{1em} 

    \begin{subfigure}{0.28\textwidth} 
        \centering
        \includegraphics[width=\textwidth]{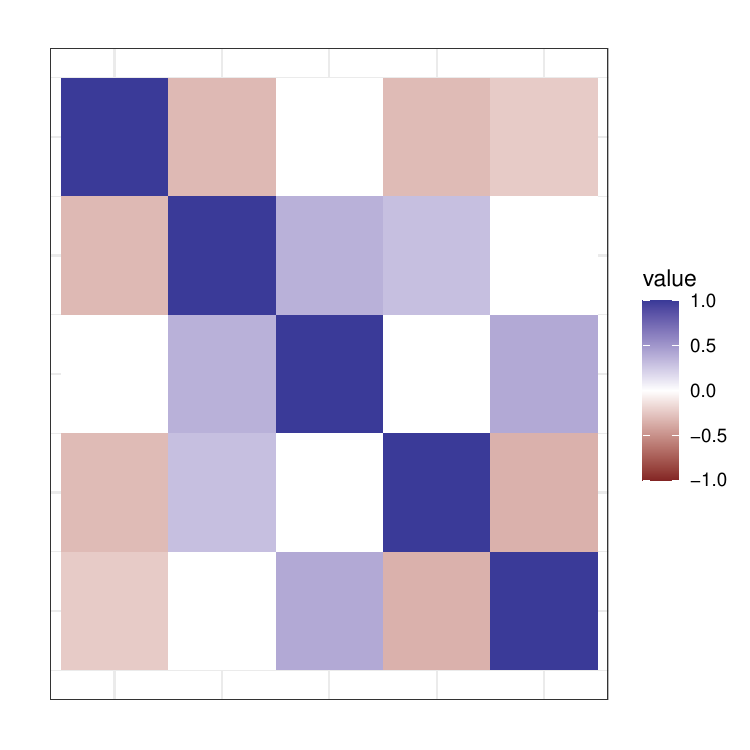}
        \caption{}
        \label{fig:b}
    \end{subfigure}
    \hspace{0.5em} 
    \begin{subfigure}{0.28\textwidth} 
        \centering
        \includegraphics[width=\textwidth]{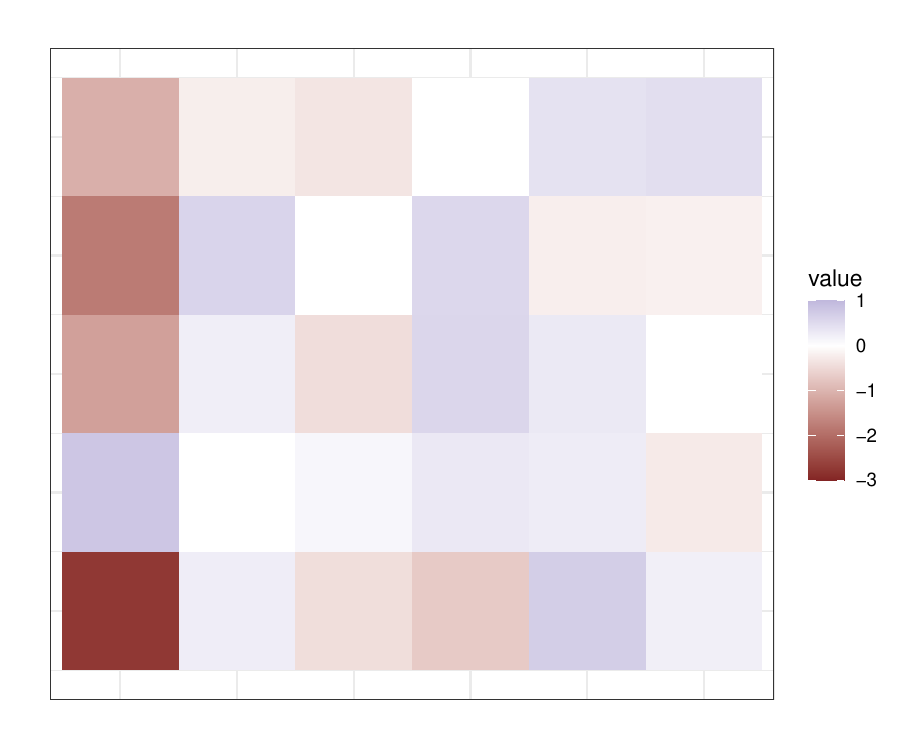}
        \caption{}
        \label{fig:c}
    \end{subfigure}
    \hspace{0.5em} 
    \begin{subfigure}{0.28\textwidth} 
        \centering
        \includegraphics[width=\textwidth]{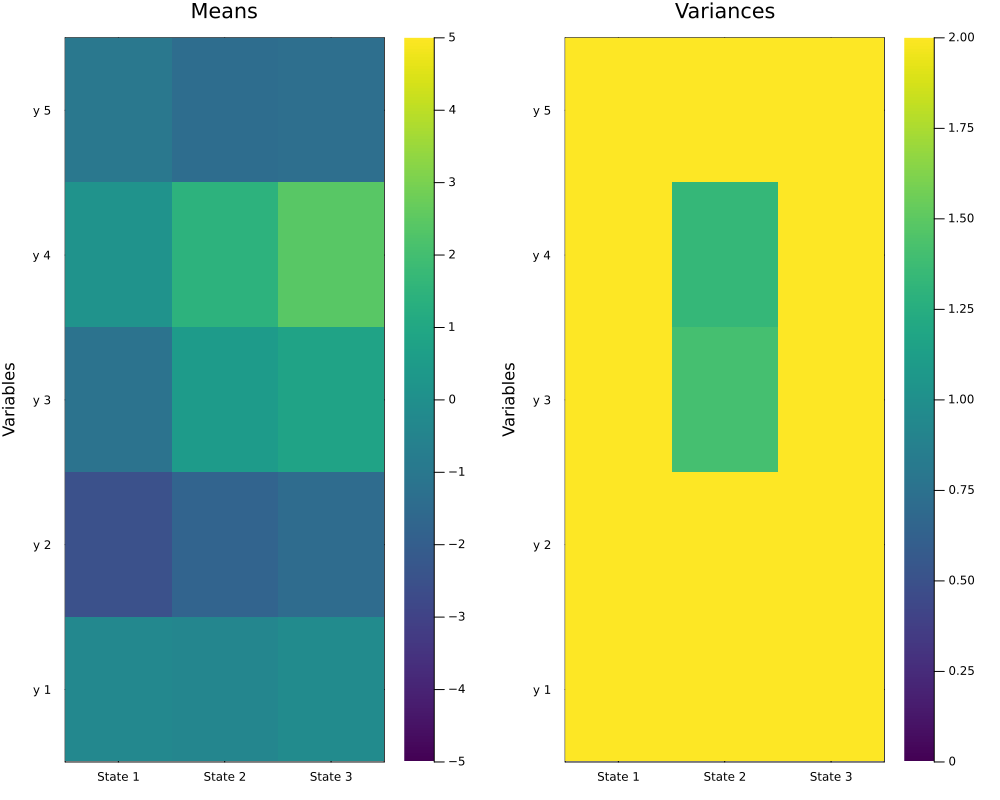} 
        \caption{}
        \label{fig:d}
    \end{subfigure}

    \caption{\textbf{Misspecified Model - Dense VAR} (a) estimated state-specific partial correlation matrices; (b) true partial correlation matrix; (c) generated VAR matrix; (d) estimated mean and variances.}
    \label{fig:matrixes_DENSE}
\end{figure}

\textbf{\textit{Dense Autoregressive Matrix}}:  
We now consider the case of a dense autoregressive coefficient matrix \(\bm{A}\), where the sparsity level is set to \(\rho = 0.1\), meaning that 90\% of its entries are nonzero. The nonzero coefficients and the intercept term \(\bm{\mu}\) are generated following the same procedure as in the sparse case. A realization of the generated time series is shown in Figure \ref{fig:data_and_state_probs_DENSE} (top), illustrating the more intricate dependencies introduced by the dense structure. Figures \ref{fig:matrixes_DENSE}(b) and \ref{fig:matrixes_DENSE}(c) display the corresponding partial correlation matrix and dense autoregressive coefficient matrix, respectively. The remaining parameterization for the proposed model is kept identical to the sparse case.  

\begin{figure}[htbp]
    \centering
    \begin{subfigure}{0.42\textwidth} 
        \centering
        \includegraphics[width=\textwidth]{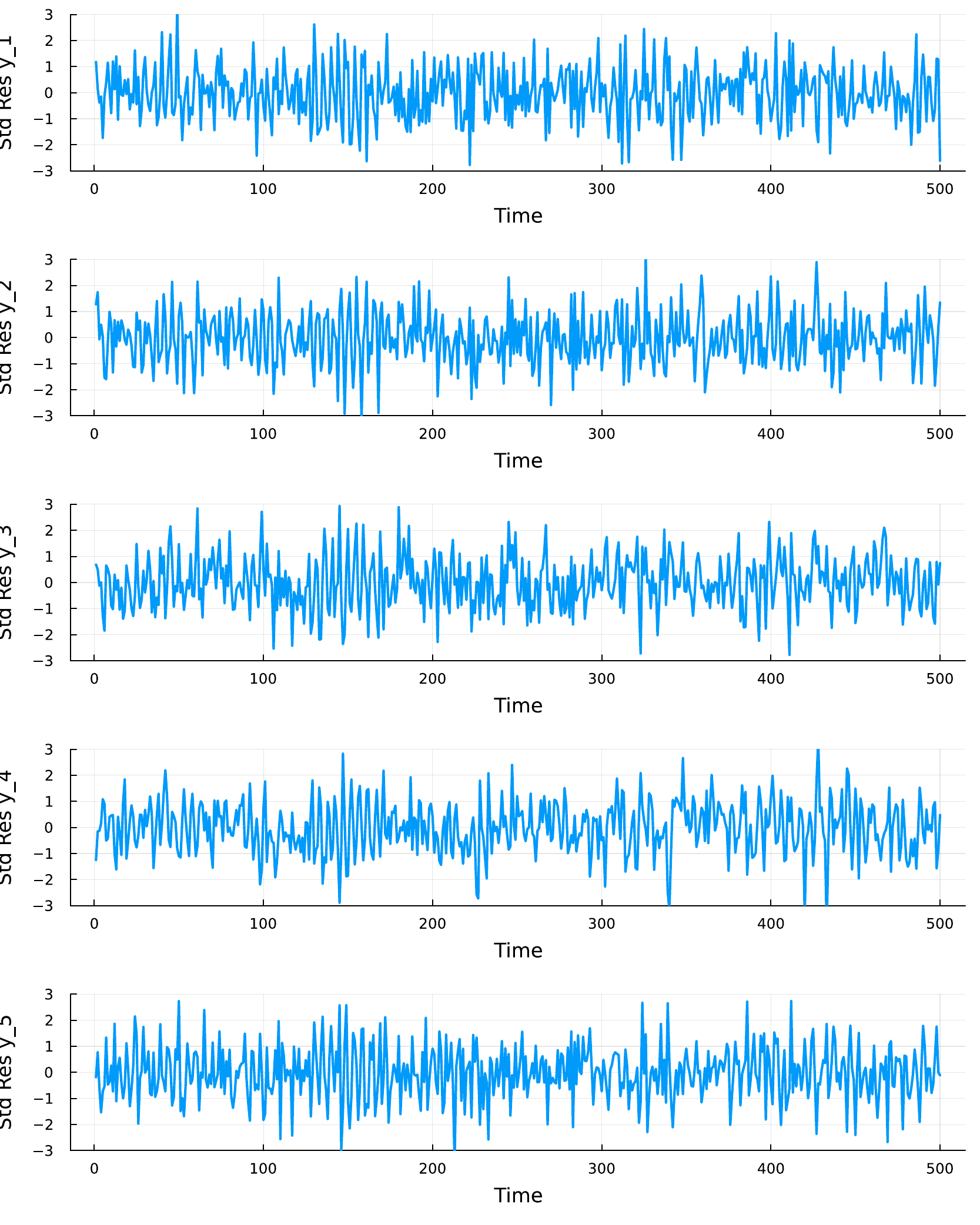}
        \caption{}
        \label{fig:a}
    \end{subfigure}
    \hspace{1em} 
    \begin{subfigure}{0.42\textwidth} 
        \centering
        \includegraphics[width=\textwidth]{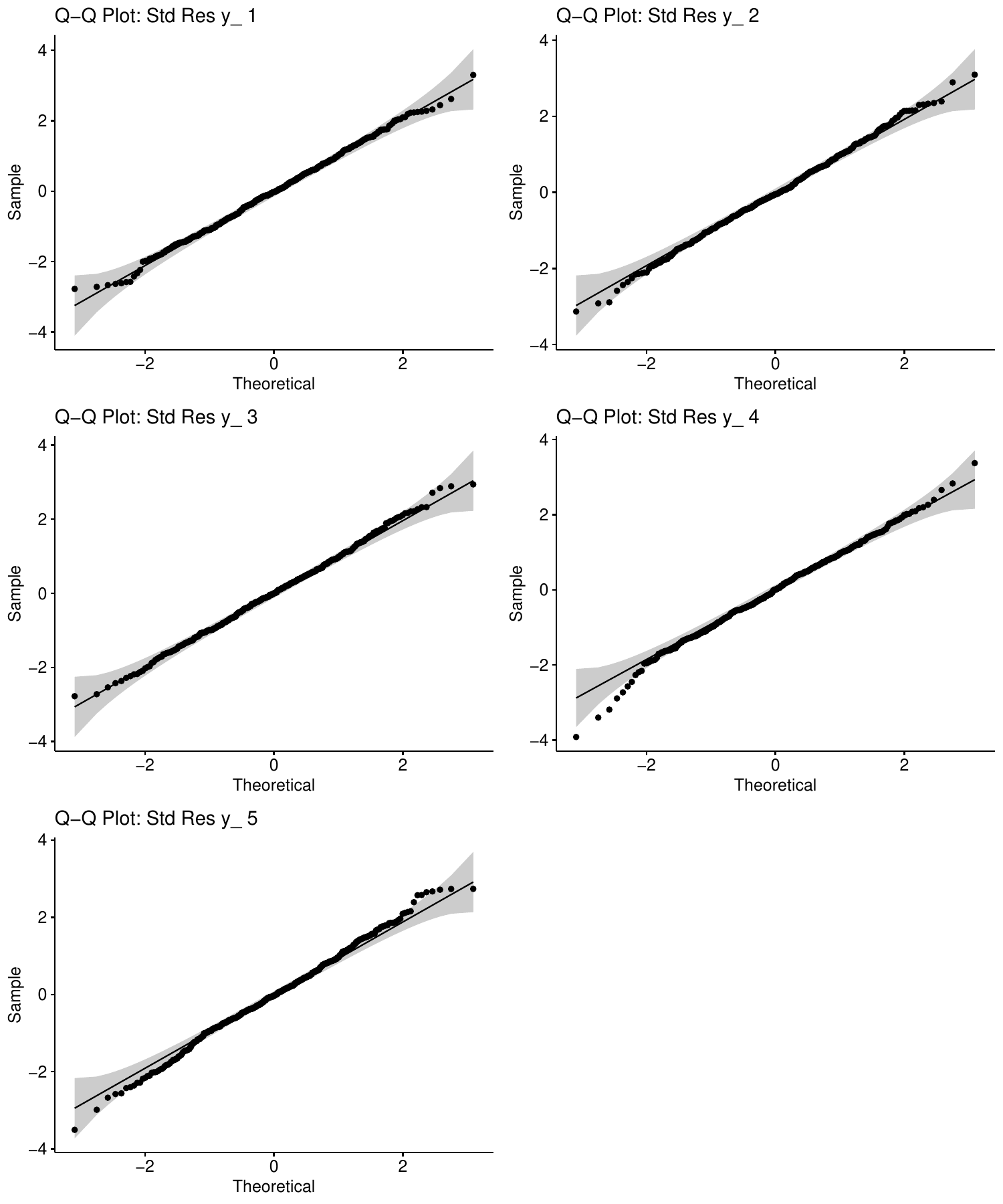}
        \caption{}
        \label{fig:b}
    \end{subfigure}

    \caption{\textbf{Misspecified Model - Dense VAR}. (a) standardized residuals; (b) Q-Q plots of standardized residuals. }
    \label{fig:residual_DENSE}
\end{figure}

In this setting, the posterior distribution for the number of states indicates a strong preference for three regimes, with \(P(M = 2 \mid \cdot) = 0.01\) and \(P(M = 3 \mid \cdot) = 0.99\), yielding an estimated mode at \(\hat{M} = 3\). Thus, under model misspecification with a dense autoregressive structure, our approach identifies multiple regimes, in contrast to the single-state solution found in the sparse case. Figure \ref{fig:residual_DENSE} presents the time-varying probability plot, illustrating the inferred state transitions over time. Additionally, Figure \ref{fig:matrixes_DENSE}(a) shows the estimated state-specific partial correlation matrices, which align closely with the true generating matrix, confirming that the model captures key structural dependencies.  

To further examine the factors driving the differentiation into three distinct states, we analyze the state-specific means and variances across dimensions, as shown in Figure \ref{fig:matrixes_DENSE}(d). Notably, the variance in state 2 for components 3 and 4 differs markedly from the other states, while component 4 also exhibits noticeable shifts in its mean. These distinctions suggest that, in the presence of a dense autoregressive structure, variability in both the mean and variance plays a critical role in defining latent regimes.  

Finally, to assess the model’s ability to capture the autocorrelation structure of the data, despite the misspecification, we examine the standardized residuals and QQ plots in Figure \ref{fig:residual_DENSE}. The results indicate an overall satisfactory goodness of fit, providing further evidence that our approach effectively extracts meaningful temporal patterns even when the assumed model deviates from the true data-generating process.

\section{Simulated fMRI data} \label{sec:suppl_simulatedfMRI}

In this section, we investigate the performance of our proposed model by evaluating its behavior under a controlled simulation framework. We generate synthetic fMRI data using the \texttt{neuRosim} package in R, which provides a principled approach to simulating BOLD signals, stimulus-driven activation, and multiple sources of structured noise. The simulation mimics a typical block-design fMRI experiment with region-specific activation patterns and realistic noise components.

\subsection{BOLD Signal Generation}  
The synthetic dataset consists of $ D= 10 $ brain regions and $ T = 200 $ time points, corresponding to a total scan duration of 400 seconds with a repetition time (TR) of 2 seconds. The experimental design follows a block paradigm where an external stimulus is periodically introduced every 80 seconds, lasting for 40 seconds. The stimulus function is generated using the \texttt{stimfunction()} function from \texttt{neuRosim}, creating a high-resolution representation of the experimental design. This function is subsequently downsampled to match the fMRI temporal resolution. For each brain region, the true underlying BOLD signal is simulated using \texttt{specifydesign()}, which convolves the stimulus function with a canonical double-gamma hemodynamic response function (HRF). The effect size of activation varies across regions, ensuring heterogeneous responses to the stimulus. Figure \ref{fig:BOLD_signal_generation} (middle) displays the generated true underlying BOLD signal. 

To model real-world fMRI data, we introduce multiple noise sources, each uniquely affecting individual brain regions. Specifically, (i) temporal noise is modeled as an auto-regressive (AR) process with parameters $\rho_1 = 0.5, \rho_2 = -0.3$, generated using \texttt{temporalnoise()}, accounting for slow signal drifts and temporal autocorrelation typical of fMRI data; (ii) physiological noise, induced by cardiac and respiratory cycles, is introduced using \texttt{physnoise()}, which generates region-specific fluctuations that exhibit structured oscillatory patterns influenced by the fMRI repetition time; and (iii) scanner noise is incorporated via \texttt{systemnoise()}, modeling Rician-distributed noise to account for thermal and hardware-induced variability in the fMRI acquisition process. Each region is assigned a noise level sampled from a uniform distribution between 1 and 3, reflecting spatially heterogeneous scanner noise. The final data representation is obtained by summing the clean BOLD signal with the noise components and standardizing each brain region by subtracting its mean and dividing by its standard deviation, as shown in Figure \ref{fig:BOLD_signal_generation} (top). By introducing region-specific variability across all noise sources, the simulated dataset closely resembles real fMRI time series, providing a robust testbed for evaluating the proposed methodology. 

\subsection{Results}

We apply our proposed sampler with the maximum allowable number of states set to 
$M_{\max} = 6$, while keeping the remaining parameter settings consistent with those described in Section~\ref{sec:simulation_studies}. The posterior distribution on the number of states exhibits a strong mode at 
$\hat{M} = 2$, with an overwhelming posterior probability of 
$p(M=1 \mid \cdot) = 0.99$, indicating that the model strongly favors a two-state structure. Similarly, the posterior distribution on the order of the discrete autoregressive (DAR) process was estimated at 
$\hat{P} = 2$, with posterior probabilities 
$p(P=1 \mid \cdot) = 0.15$, 
$p(P=2 \mid \cdot) = 0.79$, and 
$p(P=3 \mid \cdot) = 0.06$, confirming the adequacy of modeling dependencies up to two previous time points. Figure~\ref{fig:BOLD_signal_generation} (bottom) displays the time-varying probability plots inferred by the model. These plots reveal a clear alignment between the estimated state transitions and the periodic stimulus changes embedded in the true generated BOLD signal. This suggests that the proposed methodology effectively captures the latent switching dynamics in an unsupervised manner.

\begin{figure}
    \centering
    \includegraphics[width=0.5\linewidth]{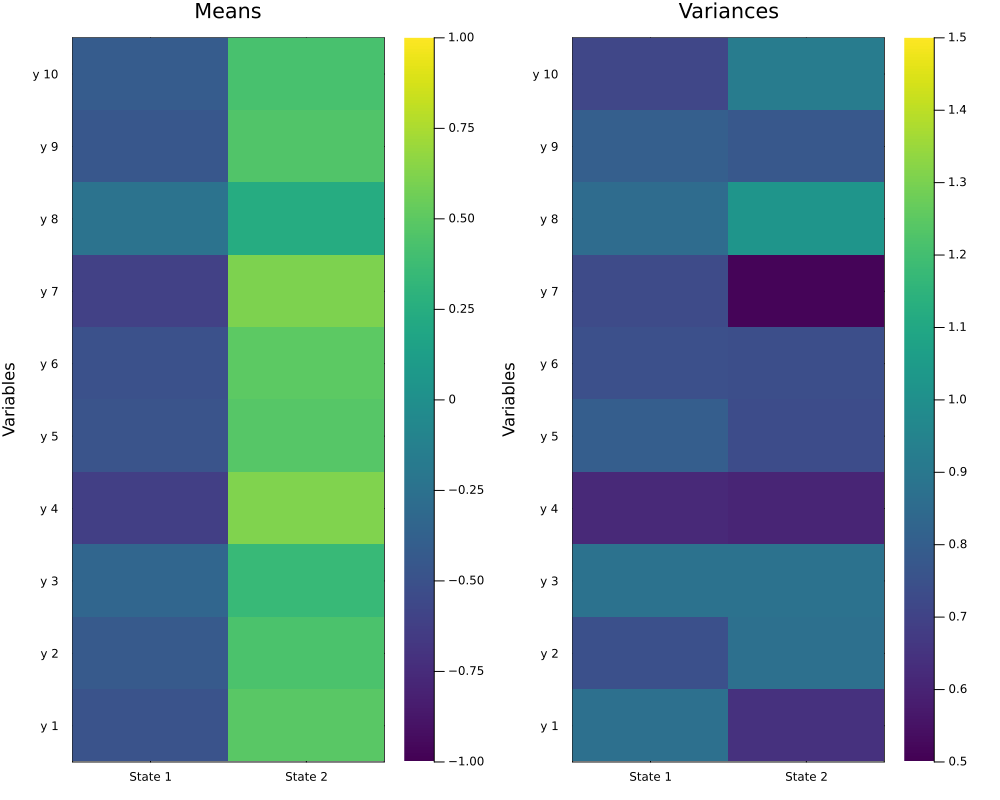}
    \caption{\textbf{Simulated fMRI data}. State-specific mean (left) and variance (right) values across different dimensions. }
    \label{fig:mean_variance_neuroSim}
\end{figure}

To further assess the characteristics of the inferred states, we examine the estimated state-specific means and variances, as shown in Figure~\ref{fig:mean_variance_neuroSim}. The heatmaps illustrate distinct mean activation patterns across the two states, with one state exhibiting consistently higher activation levels across most brain regions, while the other reflects lower or baseline activation. This separation of states suggests that the model successfully distinguishes between task-evoked and resting-state activity. The variance estimates, presented in the right panel of Figure~\ref{fig:mean_variance_neuroSim}, indicate that variability remains relatively stable across most regions, though with some fluctuations in specific regions, notably in region 
$y_7$. This might reflect localized differences in the neural response to the stimulus or variations in noise contributions across regions.

\begin{figure}[htbp]
    \centering
    \begin{subfigure}{0.42\textwidth} 
        \centering
        \includegraphics[width=\textwidth]{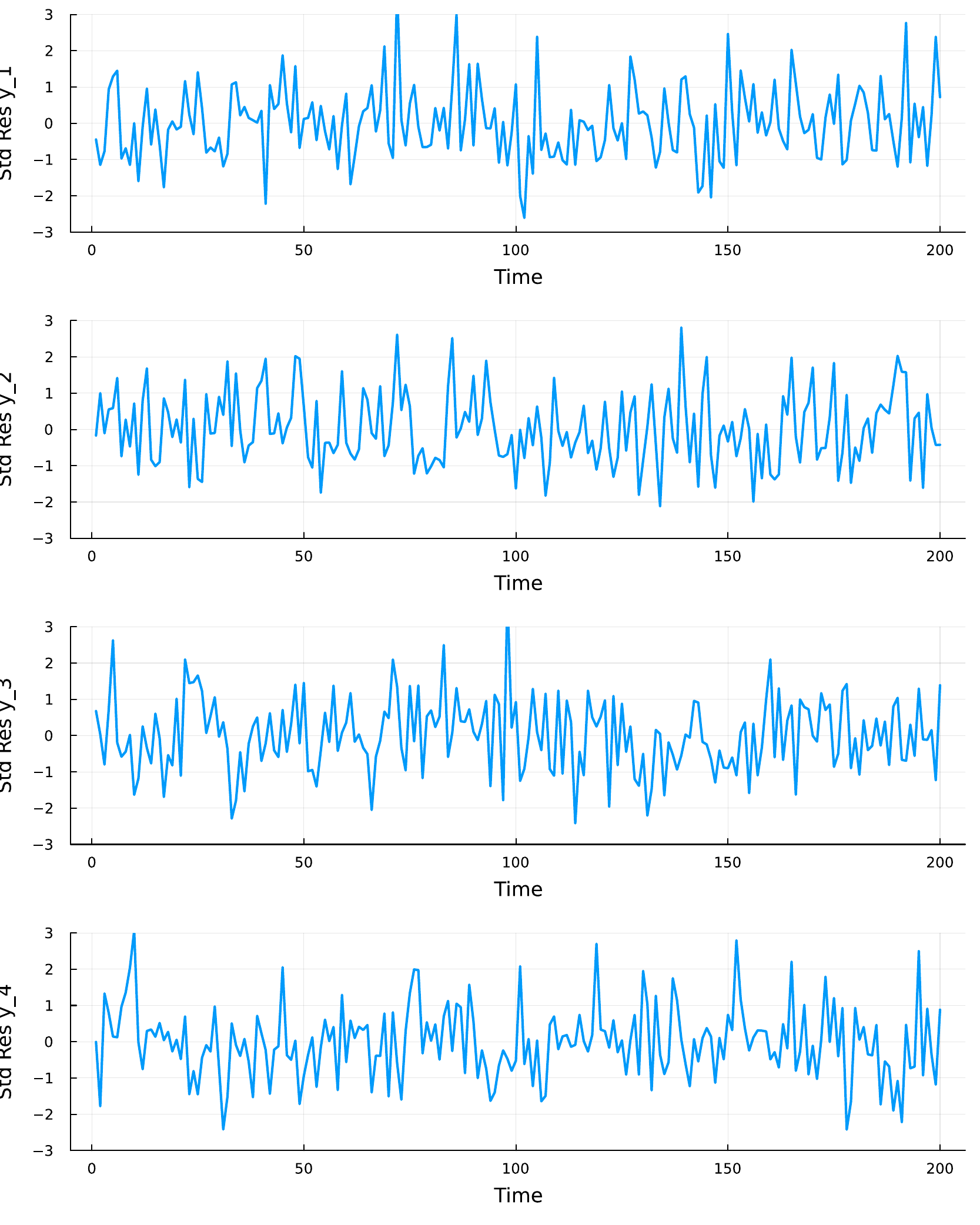}
        \caption{}
        \label{fig:a}
    \end{subfigure}
    \hspace{1em} 
    \begin{subfigure}{0.42\textwidth} 
        \centering
        \includegraphics[width=\textwidth]{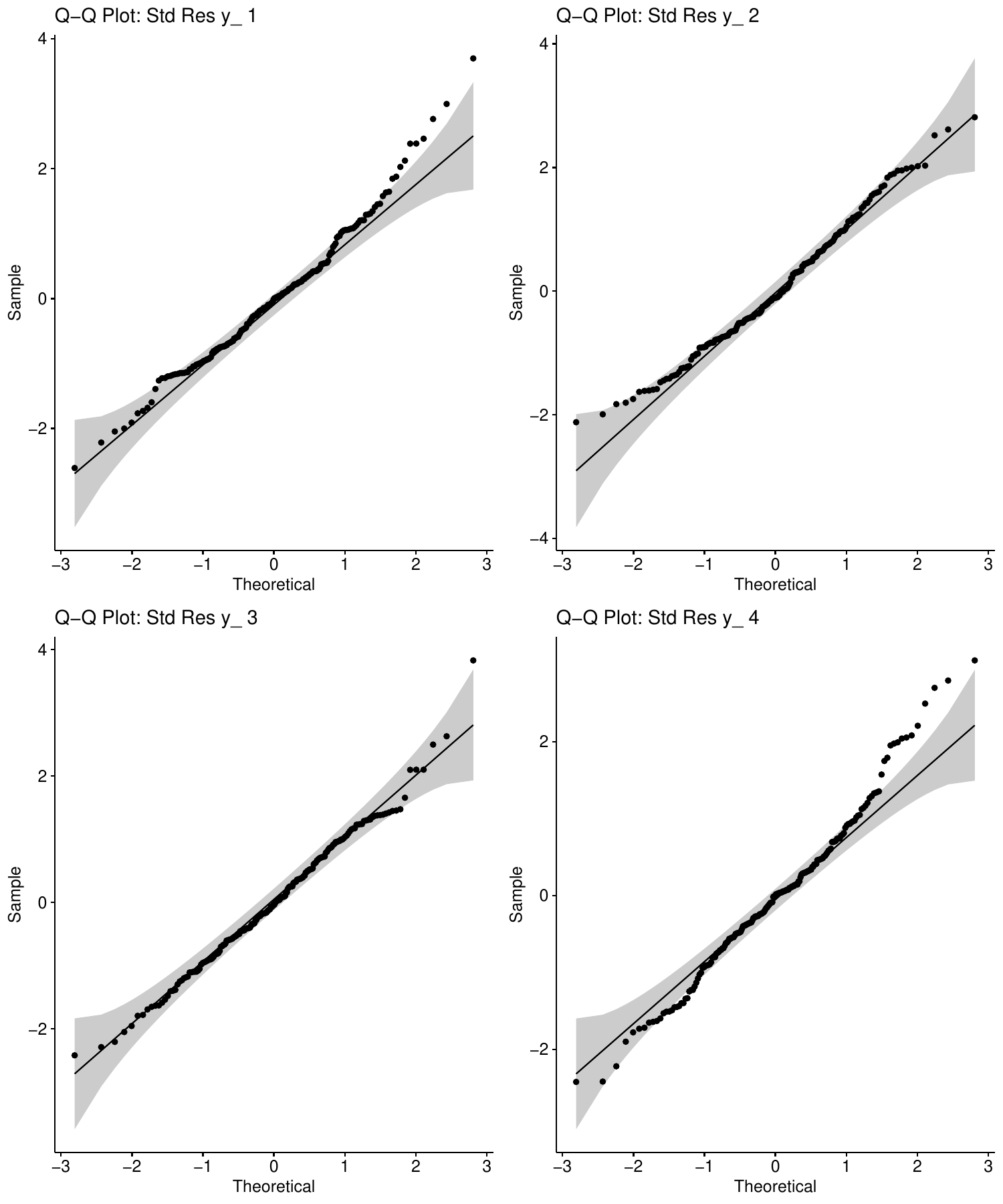}
        \caption{}
        \label{fig:b}
    \end{subfigure}

    \caption{\textbf{Simulated fMRI data}. (a) standardized residuals; (b) Q-Q plots of standardized residuals. }
    \label{fig:residual_neuroSim}
\end{figure}

Finally, we evaluate the model's goodness-of-fit by analyzing standardized residuals. Figure~\ref{fig:residual_neuroSim}(a) presents the residual time series for four selected brain regions, while the corresponding QQ plots in Figure~\ref{fig:residual_neuroSim}(b) assess their adherence to a normal distribution. While the residuals generally align well with normality, some deviations in the tails suggest minor departures, which may indicate room for further refinement in capturing noise structure. Nevertheless, the overall results confirm that the model successfully detects state transitions, accurately characterizes the stimulus-driven dynamics, and provides meaningful statistical summaries of the two inferred states.

\begin{figure}[htbp]
    \centering
    \includegraphics[width=0.6\textwidth]{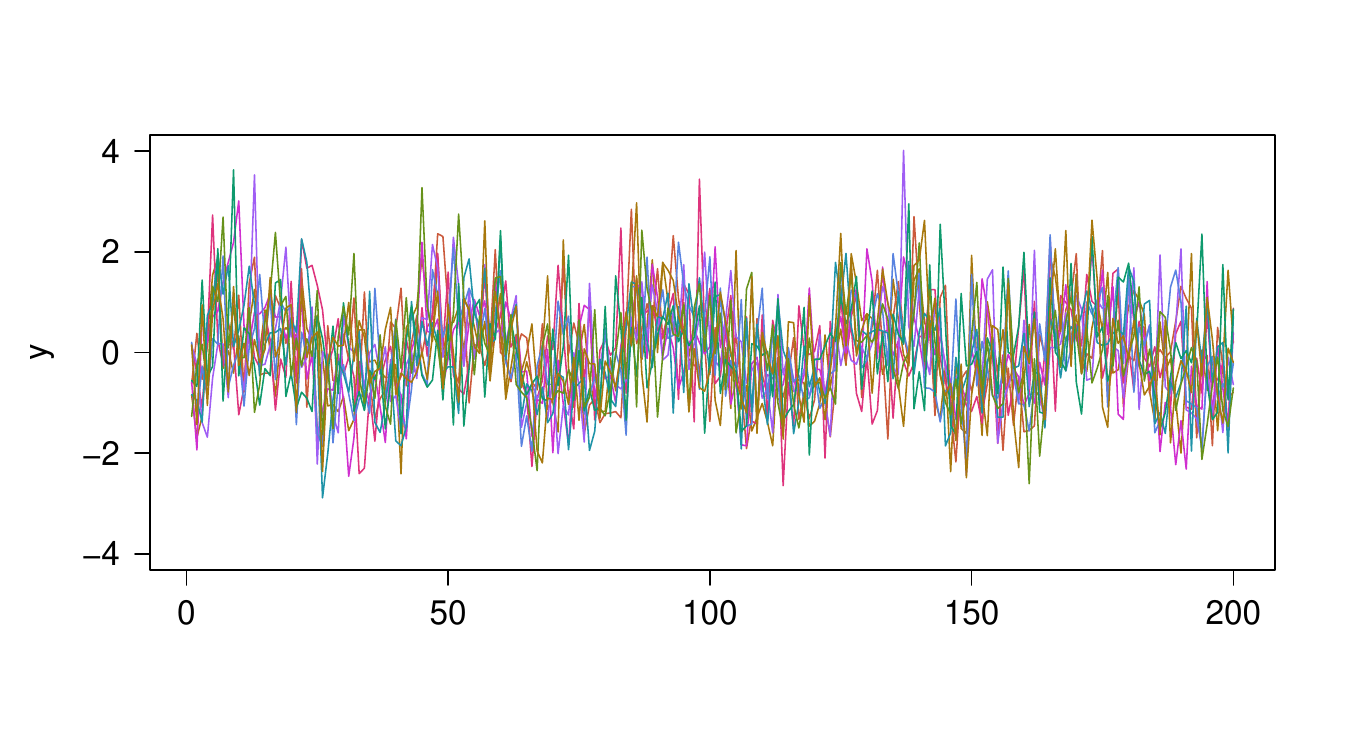}  
    \includegraphics[width=0.6\textwidth]{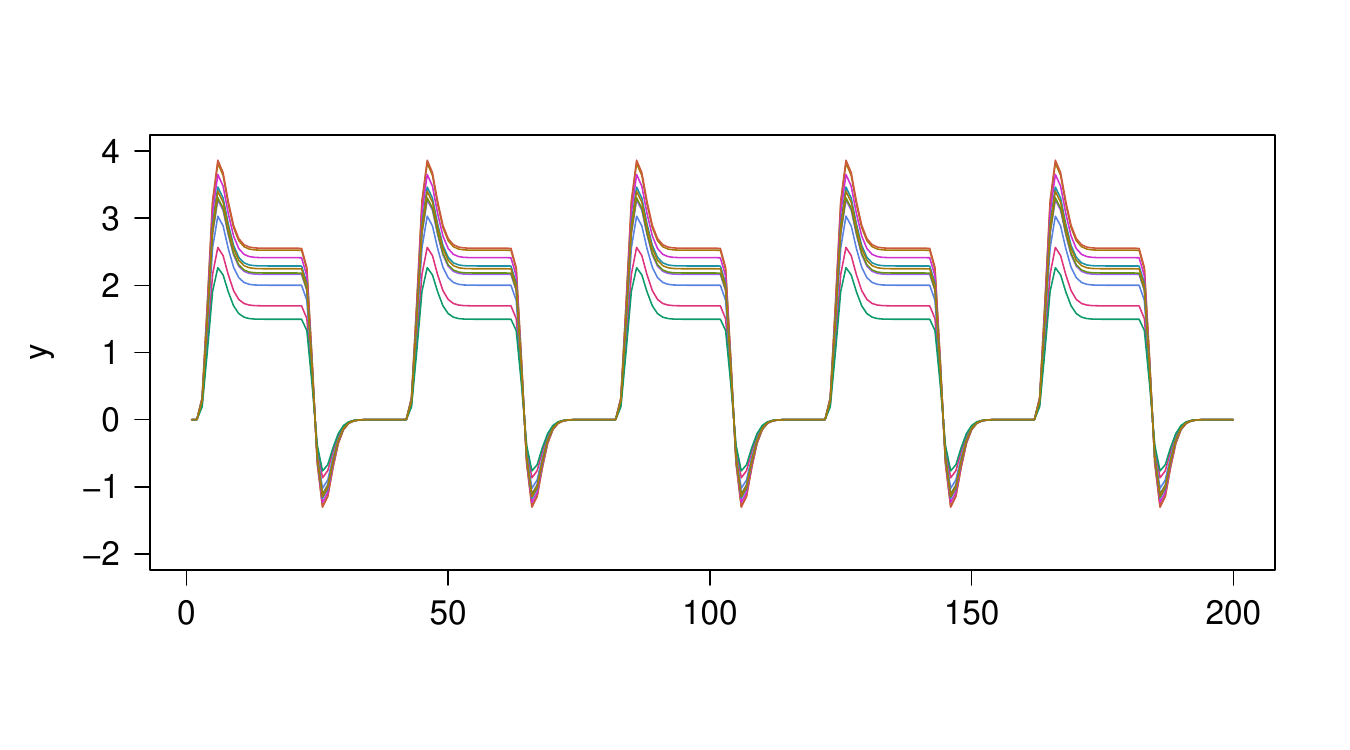}
    \includegraphics[width=0.6\textwidth]{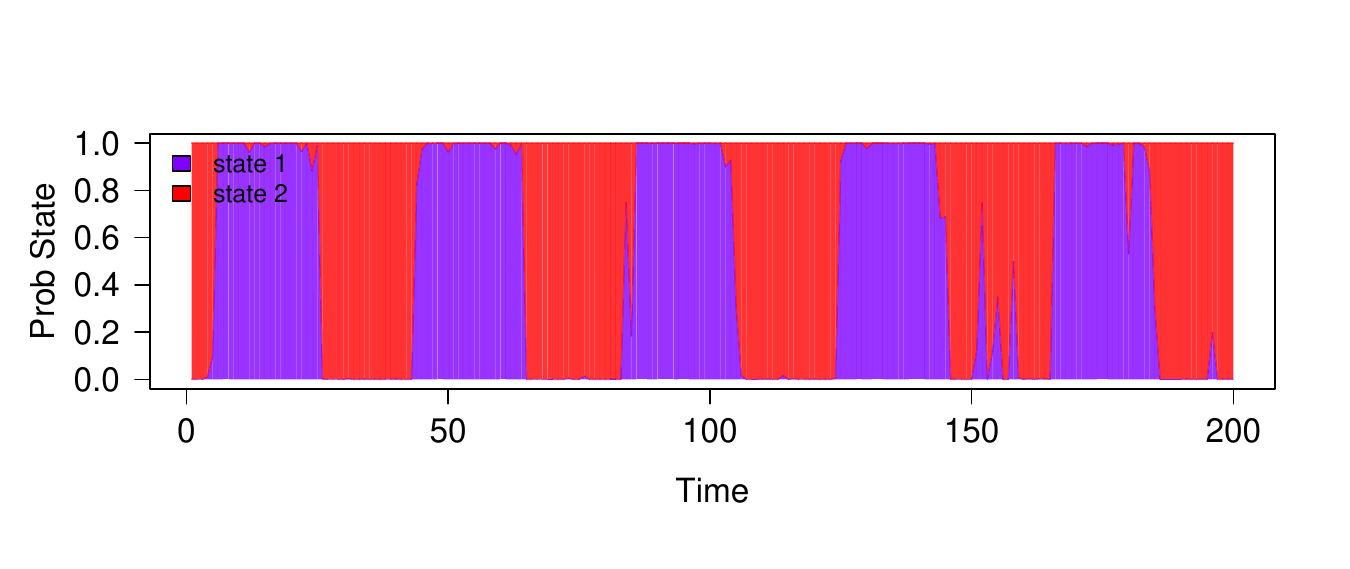}
\caption{\textbf{Simulated fMRI Data.} (a) Simulated fMRI data obtained by summing the clean BOLD signal with noise components and standardizing each brain region; (b) True underlying BOLD signal, generated by convolving the stimulus function with a canonical double-gamma HRF; (c) Estimated time-varying probability plot.}

    \label{fig:BOLD_signal_generation}
\end{figure}

\section{Sensitivity Analysis} \label{sec:suppl_sensitivity}

We carried out a sensitivity analysis study  by focusing on the impact of the hyperparameters of the  zero-inducing cumulative shrinkage prior that characterizes the DAR process formulated in Section \ref{sec:zero_inflated}.  In particular, we investigated the sensitivity of the hyperparameters of the Beta priors on the stick-breaking weights $v_0$ (i.e., $a_0$ and $b_0$), and $v_j$ (i.e., $a_j$ and $b_j$), recalling that the mixing probabilities $\{\phi_j\}_{j=0}^{P}$ are a by-product of the stick-breaking weights and that they determine the number of active DAR coefficients. We investigated four different scenarios: (i) $v_0 = v_j \sim \text{Beta}(0.5, 0.5)$, corresponding to a Jeffreys prior \citep{jeffreys1946invariant}; (ii) $v_0 = v_j \sim \text{Beta}(1,1)$, namely a uniform prior; (iii) $v_0 \sim \text{Beta}(1,10)$, $\,v_j \sim \text{Beta}(1, 10)$, so that the prior probability of autoregression is driven towards zero; (iv) $v_0 \sim \text{Beta}(1,10)$, $\,v_j \sim \text{Beta}(10, 1)$, i.e. the hyperparameter setting chosen as in Section \ref{sec:simulation_studies}. We simulated 30 time series from the same simulation setting of Section \ref{sec:simulation_studies}, for different sample sizes $T \in \{100, 500\}$. Table \ref{tab:sensitivity_analysis} reports posterior probabilities of the number of DAR parameters, $p (P = j | \cdot)$, over the 30 simulations, where we note that the true number of DAR parameters is  $P_{true} = 2$. It appears that cases (1) and (2) behave similarly, by identifying a posterior mode at 2 for both $T \in \{100,500 \}$, noting that as $T$ grows $p (P = 2 | \cdot)$ increases considerably. As expected, case (3) seems to penalize the probability of autoregression at higher lags, since for both sample sizes, large posterior probability is located at 1. Case (4) identifies the right number of lags for both sample sizes and seems to slightly favor probability mass to larger numbers of lags, as the probability of identifying a DAR process of order 1 is 0.045 and 0.002, while the probabilities of selecting 3 lags are 0.103 and 0.140, for $T=100$ and $T=500$, respectively. In our investigations, we also noted that as $T > 1000$ the sensitivity is not very noticeable. 


\begin{table}[htbp]
\centering
\footnotesize
\begin{tabular}{lccccc}
\hspace{-0.2cm}\\ \hline
       & \multicolumn{1}{l}{} & \multicolumn{2}{c}{$T = 100$} \\ \cmidrule{2-6}                 & \multicolumn{1}{l}{} & \multicolumn{1}{l}{}         \\ 
       &  $p \,(P = 1 | \, \cdot)$             &  $p \,(P = 2 | \, \cdot)$             &  $p \,(P = 3 | \, \cdot)$             &  $p \,(P = 4 | \, \cdot)$             & \multicolumn{1}{l}{ $p \,(P = 5 | \, \cdot)$ } \\ 
case 1 & 0.094                & 0.557                & 0.247                & 0.089                & 0.014                       \\
case 2 & 0.104                & 0.401                & 0.268                & 0.146                & 0.081                       \\
case 3 & 0.754                & 0.107                & 0.067                & 0.012                & 0.061                        \\
case 4 & 0.045                & 0.850                & 0.103                & 0.002                & 0.000                          \\
       & \multicolumn{1}{l}{} & \multicolumn{1}{l}{} & \multicolumn{1}{l}{} & \multicolumn{1}{l}{} & \multicolumn{1}{l}{}         \\  \cmidrule{2-6}   
       & \multicolumn{1}{l}{} & \multicolumn{2}{c}{$T = 500$}  \\ \cmidrule{2-6}                & \multicolumn{1}{l}{} & \multicolumn{1}{l}{}         \\
       &  $p \,(P = 1 | \, \cdot)$            &  $p \,(P = 2 | \, \cdot)$             &  $p \,(P = 3 | \, \cdot)$             &  $p \,(P = 4 | \, \cdot)$              & \multicolumn{1}{l}{ $p \,(P = 5 | \, \cdot)$ } \\
case 1 & 0.006                & 0.903                & 0.088                & 0.003               & 0.000                        \\
case 2 & 0.007                & 0.891                & 0.097                & 0.005                & 0.000                          \\
case 3 & 0.441                & 0.546                & 0.013                & 0.000                  & 0.000                          \\
case 4 & 0.002                  & 0.856                & 0.140                & 0.002                & 0.000    \\[0.1em] \hline                           
\end{tabular}
\caption{\textbf{Simulation study.} Sensitivity analysis on the DAR parameters for the cases (1) $v_0 \sim \text{Beta}(0.5, 0.5)$, $\,v_j \sim \text{Beta}(0.5, 0.5)$; (2) $v_0 \sim \text{Beta}(1,1)$, $\,v_j \sim \text{Beta}(1, 1)$; (3) $v_0 \sim \text{Beta}(1,10)$, $\,v_j \sim \text{Beta}(1, 10)$; (4) $v_0 \sim \text{Beta}(1,10)$, $\,v_j \sim \text{Beta}(10, 1)$. Note that $P_{true} = 2$. 
}
\label{tab:sensitivity_analysis}
\end{table}

We carried out further investigations on the impact of the hyperparameters of the zero-inducing cumulative shrinkage prior characterizing the DAR process. We studied the same four scenarios as above, by focusing on the posterior probability over the number of states. Table \ref{tab:sensitivity_analysis_Msample} reports, for each scenario, posterior probabilities of the number of states, $p (M = j | \cdot)$, over the 30 simulated replicates, where we note that the true number of states is  $M_{true} = 5$. It appears that for the smaller sample size (e.g. $T=100$) the sampler tends to estimate more states than necessary, for all combination of the hyperparameters. However, as the sample size increases,  the inference on the correct number of states increases considerably. In our investigations, we also noted that when $T > 1000$ the sensitivity is not very noticeable.

\begin{table}[htbp]
\centering
\footnotesize
\begin{tabular}{lrrrrrrr}
\hspace{-0.2cm}\\ \hline
       & \multicolumn{1}{c}{}          & \multicolumn{1}{c}{}         & \multicolumn{2}{c}{$T = 100$}    \\ \cmidrule{2-8}                             & \multicolumn{1}{l}{}       & \multicolumn{1}{l}{}       & \multicolumn{1}{l}{}       \\
       & \multicolumn{1}{c}{$p \,(M = 3 | \, \cdot)$ } & \multicolumn{1}{c}{$p \,(M = 4 | \, \cdot)$ } & \multicolumn{1}{l}{$p \,(M = 5 | \, \cdot)$ } & \multicolumn{1}{l}{$p \,(M = 6 | \, \cdot)$ } & \multicolumn{1}{l}{$p \,(M = 7 | \, \cdot)$ } & \multicolumn{1}{l}{$p \,(M = 8 | \, \cdot)$ } & \multicolumn{1}{l}{$p \,(M = 9 | \, \cdot)$ } \\
case 1 & 0.0                           & 0.1                          & 0.2                          & 0.333            & 0.267                      & 0.1                        & 0.0                        \\
case 2 & 0.033                         & 0.0                          & 0.167                        & 0.3              & 0.256                      & 0.21                       & 0.033                      \\
case 3 & 0.033                         & 0.133                        & 0.3                          & 0.333             & 0.1                        & 0.094                      & 0.006                      \\
case 4 & 0.033                         & 0.133                        & 0.267                        & 0.4              & 0.133                      & 0.033                      & 0.0                        \\
       & \multicolumn{1}{l}{}          & \multicolumn{1}{l}{}         & \multicolumn{1}{l}{}         & \multicolumn{1}{l}{}       & \multicolumn{1}{l}{}       & \multicolumn{1}{l}{}       & \multicolumn{1}{l}{}       \\ \cmidrule{2-8}
       & \multicolumn{1}{c}{}          & \multicolumn{1}{c}{}         & \multicolumn{2}{c}{$T = 500$}      \\ \cmidrule{2-8}                             & \multicolumn{1}{l}{}       & \multicolumn{1}{l}{}       & \multicolumn{1}{l}{}       \\
       & \multicolumn{1}{c}{$p \,(M = 3 | \, \cdot)$ } & \multicolumn{1}{c}{$p \,(M = 4 | \, \cdot)$ } & \multicolumn{1}{l}{$p \,(M = 5 | \, \cdot)$ } & \multicolumn{1}{l}{$p \,(M = 6 | \, \cdot)$ } & \multicolumn{1}{l}{$p \,(M = 7 | \, \cdot)$ } & \multicolumn{1}{l}{$p \,(M = 8 | \, \cdot)$ } & \multicolumn{1}{l}{$p \,(M = 9 | \, \cdot)$ } \\
case 1 & 0.0                           & 0.133                        & 0.867             & 0.0                        & 0.0                        & 0.0                        & 0.0                        \\
case 2 & 0.0                           & 0.133                        & 0.867             & 0.0                        & 0.0                        & 0.0                        & 0.0                        \\
case 3 & 0.0                           & 0.133                        & 0.833             & 0.033                      & 0.0                        & 0.0                        & 0.0                        \\
case 4 & 0.0                           & 0.133                        & 0.867             & 0.0                        & 0.0                        & 0.0                        & 0.0    \\[0.1em] \hline                     
\end{tabular} 
\caption{\textbf{Simulation study.} Sensitivity analysis of the DAR parameters for the following scenarios: (1) $v_0 \sim \text{Beta}(0.5, 0.5)$, $\,v_j \sim \text{Beta}(0.5, 0.5)$; (2) $v_0 \sim \text{Beta}(1,1)$, $\,v_j \sim \text{Beta}(1, 1)$; (3) $v_0 \sim \text{Beta}(1,10)$, $\,v_j \sim \text{Beta}(1, 10)$; (4) $v_0 \sim \text{Beta}(1,10)$, $\,v_j \sim \text{Beta}(10, 1)$. We report posterior probabilities of the number number of states, $p (M = j | \cdot)$, where we note that $M_{true} = 5$; results are shown for $T \in \{100, 500 \}$. }
\label{tab:sensitivity_analysis_Msample}
\end{table}

\section{Convergence Diagnostics} \label{sec:S_convergence}

\begin{figure}[htbp]
	\centering
	\includegraphics[width=0.6\linewidth]{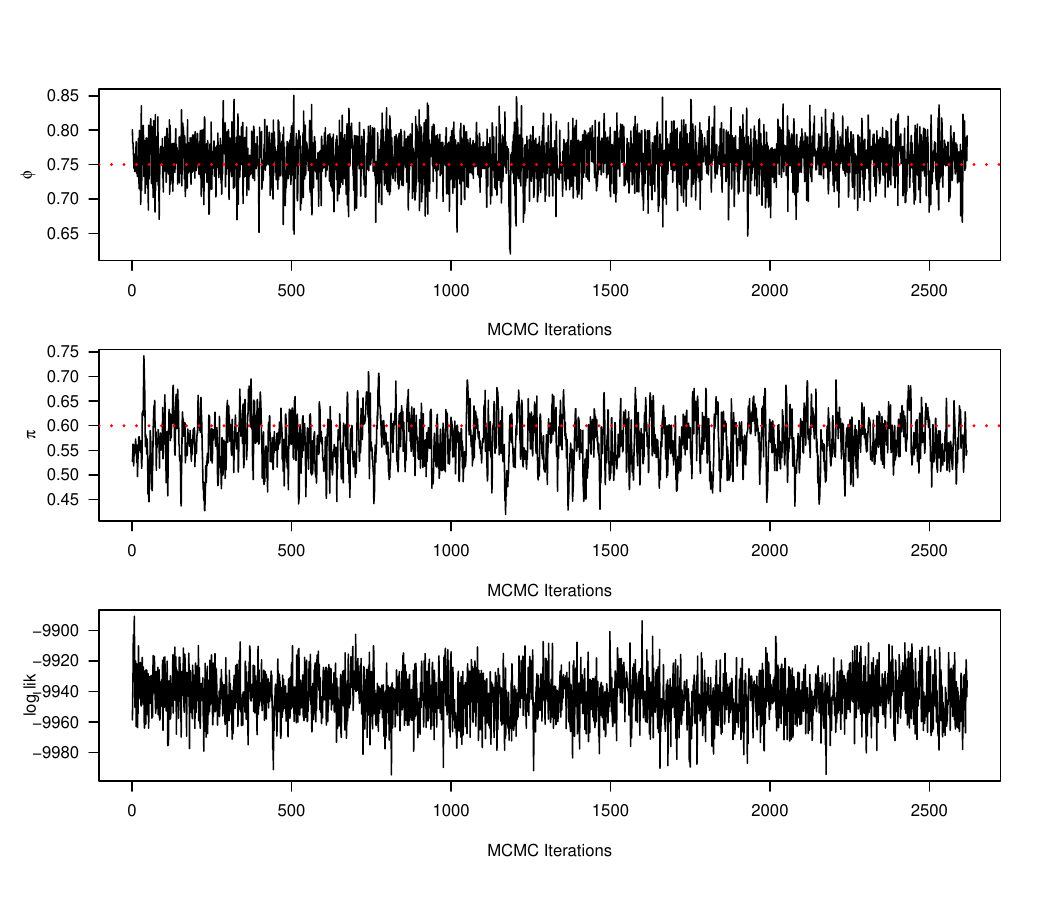}
\caption{Examples of trace plots of the DAR parameters ($\phi_j$ and $\pi_j$) and of the likelihood. Dotted red lines corresponds to true generating value of the DAR parameters.}
\label{fig:convergence_diagnostics}
\end{figure}

We verified convergence of the MCMC sampler by: (i) analyzing the trace plots of the parameters, e.g. the mean of the multivariate spiked Gaussian emissions, observing no pathological behavior (see Figure \ref{fig:convergence_diagnostics} for representative examples of trace plots of the DAR parameters); (ii) storing the values of the overall likelihood of the system (Eq. \eqref{eq:likelihood}) and plotting the corresponding trace, noting that it reached a stable regime (see Figure \ref{fig:convergence_diagnostics}, bottom plot, for a representative example of trace plot of the log likelihood);  (iii) verifying the Heidelberger and Welch’s convergence diagnostic test\citep{heidelberger1981spectral}: this diagnostic first tests the null hypothesis that the Markov Chain is in the stationary distribution and then tests whether the mean of a marginal posterior distribution can be estimated with sufficient precision, assuming that the Markov Chain is in the stationary distribution. In our experiments, this test was passed for every Markov chain we analyzed. 

\end{document}